\documentclass[12pt]{article}

\usepackage{verbatim}
\usepackage{natbib}
\usepackage{amsfonts}
\usepackage{amssymb}
\usepackage{amsmath}
\usepackage{amsthm}
\usepackage{mathrsfs}
\usepackage{mathtools}
\usepackage{nicefrac} 
\usepackage{enumerate}
\usepackage{hyperref}
\usepackage{todonotes}
\usepackage{algorithm}
\usepackage{algorithmic}
\usepackage{booktabs}
\usepackage{setspace}
\usepackage{etoc}

\newtheorem{theo}{Theorem}

\newtheorem{lem}{Lemma}

\newtheorem{cor}{Corollary}

\newcommand{\bs}{\boldsymbol}
\def\bal#1\eal{\begin{align}#1\end{align}}
\def\bals#1\eals{\begin{align*}#1\end{align*}}
\def\be#1\ee{\begin{equation}#1\end{equation}}
\newcommand{\lrb}[1]{\left(#1\right)}
\newcommand{\brb}[1]{\bigl(#1\bigr)}
\newcommand{\Brb}[1]{\Bigl(#1\Bigr)}
\newcommand{\bbrb}[1]{\biggl(#1\biggr)}
\newcommand{\labs}[1]{\left|#1\right|}
\newcommand{\babs}[1]{\bigl|#1\bigr|}

\DeclareMathOperator*{\argmax}{argmax\;}
\DeclareMathOperator*{\argmin}{argmin\;}

\newcommand{\R}{\mathbb{R}}
\newcommand{\ve}{\varepsilon}

\newcommand{\algexp}{Tempered Exp3 for Social Welfare}
\newcommand{\dyadic}{Dyadic Search for Social Welfare}

\newcommand{\pol}{x} %
\newcommand{\out}{y} %
\newcommand{\wtp}{v} %
\newcommand{\sw}{U} %
\newcommand{\swcum}{\mathbb \sw} %
\newcommand{\swexp}{{\bs \sw}} %
\newcommand{\dem}{G} %
\newcommand{\demcum}{\mathbb \dem} %
\newcommand{\demexp}{{\bs \dem}} %

\newcommand{\regadv}{\mathcal{R}_T(\{\wtp_i\}_{i=1}^T)} %
\newcommand{\regstoch}{\mathcal{R}_T(\demexp)} %

\newcommand{\constone}{\tilde{C}} %
\newcommand{\consttwo}{\tilde{c}_\delta} %
\newcommand{\constthree}{\tilde{n}} %
\newcommand{\ts}{\tau^\#} %

\newcommand{\wage}{w} 
\newcommand{\weight}{\omega} %

 \newcommand{\lsb}[1]{\left[#1\right]}

 \newcommand{\lcb}[1]{\left\{#1\right\}}
 \newcommand{\bcb}[1]{\bigl\{#1\bigr\}}

 \newcommand{\lfl}[1]{\left\lfloor#1\right\rfloor}
 \newcommand{\bfl}[1]{\bigl\lfloor#1\bigr\rfloor}

\usepackage[margin=3cm]{geometry}
\usepackage{fancyhdr}

\title{Adaptive maximization of social welfare}

\author{Nicolò Cesa-Bianchi\footnote{Università degli Studi di Milano and Politecnico di Milano. \href{mailto:nicolo.cesa-bianchi@unimi.it}{nicolo.cesa-bianchi@unimi.it}. NCB was supported by the MUR PRIN grant 2022EKNE5K (Learning in Markets and Society) and by the FAIR (Future Artificial Intelligence Research) project, funded by the NextGenerationEU program within the PNRR-PE-AI scheme.} \and  Roberto Colomboni\footnote{Politecnico di Milano and Università degli Studi di Milano. \href{mailto:roberto.colomboni@polimi.it}{roberto.colomboni@polimi.it}.} \and Maximilian Kasy\footnote{Department of Economics, University of Oxford. \href{mailto:maximilian.kasy@economics.ox.ac.uk}{maximilian.kasy@economics.ox.ac.uk}. Maximilian Kasy was supported by the Alfred P. Sloan Foundation, under the grant ``Social foundations for statistics and machine learning.''}}

\begin{document}
\maketitle

\onehalfspacing

\begin{abstract}
We consider the problem of repeatedly choosing policies to maximize social welfare.
Welfare is a weighted sum of private utility and public revenue.
Earlier outcomes inform later policies.
Utility is not observed, but indirectly inferred.
Response functions are learned through experimentation.

We derive a lower bound on regret, and a matching adversarial upper bound for a variant of the Exp3 algorithm.
Cumulative regret grows at a rate of $T^{2/3}$.
This implies that (i) welfare maximization is harder than the multi-armed bandit problem (with a rate of $T^{1/2}$ for finite policy sets), and (ii) our algorithm achieves the optimal rate.
For the stochastic setting, if social welfare is concave, we can achieve a rate of $T^{1/2}$ (for continuous policy sets), using a dyadic search algorithm. 

We analyze an extension to nonlinear income taxation, and sketch an extension to commodity taxation.
We compare our setting to monopoly pricing (which is easier), and price setting for bilateral trade (which is harder).

\end{abstract}

\clearpage
\section{Introduction}
\label{sec:introduction}

Consider a policymaker who aims to maximize social welfare, defined as a weighted sum of utility across individuals.
The policymaker can choose a policy parameter such as a sales tax rate, an unemployment benefit level, a health-insurance copay rate, etc. 
The policymaker does \textit{not} directly observe the welfare resulting from their policy choices.
They do, however, observe behavioral outcomes such as consumption of the taxed good, labor market participation, or health care expenditures.
They can revise their policy choices over time in light of observed outcomes.
How should such a policymaker act?
This is the question that we study.
To address this question, we bring together insights from welfare economics (in particular optimal taxation, \citealt{ramsey1927contribution, Mirrlees1971, baily1978some, Saez2001, chetty09suff}) with insights from machine learning (in particular online learning and multi-armed bandits, see \citealt{slivkins2019introduction, lattimore2020bandit} for recent reviews, and \cite{thompson1933likelihood,lai1985asymptotically} for classic contributions).

In our baseline model, individuals arrive sequentially and make a single binary decision.
In each period the policymaker chooses a tax rate that applies to this binary decision, and then observes the individual's response.
They do not observe the individual's private utility. 
Social welfare is given by a weighted sum of private utility and public revenue.
Later, we extend our model to nonlinear income taxation, where welfare weights vary as a function of individual earnings capacity, and sketch an extension to commodity taxation, where individual decisions involve a continuous consumption vector.

Our goal is to give guidance to the policymaker.
We propose algorithms to maximize cumulative social welfare, and we provide (adversarial and stochastic) guarantees for the performance of these algorithms.
In doing so, we also show that welfare maximization is a harder learning problem than reward maximization in the multi-armed bandit setting.
Private utility in our baseline model is equal to consumer surplus, which is given by the integral of demand. In order to learn this integral, we need to learn demand for counterfactual, suboptimal tax rates. This drives the difficulty of the learning problem.

\paragraph{Why welfare, why adversarial guarantees?}
Our algorithms are designed to maximize social welfare, which is not directly observable, rather than maximizing outcomes that are directly observable.
The definition of social welfare as an aggregation of individual utilities is at the heart of welfare economics in general, and of optimal tax theory in particular.
The distinction between utility and observable outcomes is important in practice.
To illustrate, consider the example of a policymaker who chooses the level of unemployment benefits, where the observable outcome is employment. 
The policymaker could use an algorithm that adaptively maximizes employment.
The problem with this approach is that employment might be maximized by making the unemployed as miserable as possible.
This is not normatively appealing. Such an algorithm would \textit{minimize} the utility of the unemployed, rather than maximizing social welfare. 
Similar examples can be given for many domains of public policy, including health, education, and criminal justice. In contrast to observable outcomes such as employment, welfare is improved by increasing the choice sets of those affected, not by reducing these choice sets.

Our theoretical analysis provides not only stochastic but also adversarial guarantees, which hold for arbitrary sequences of preference parameters.
Adversarial guarantees for algorithms promise robustness against deviations from the assumption that heterogeneity is independently and identically distributed over time. Possible deviations from this assumption include autocorrelation, trends, heteroskedasticity, more general non-stationarity, and other concerns of time-series econometrics. In the employment example, such deviations might for instance be due to the business cycle.
One might fear that adversarial robustness is achieved at the price of worsened performance for the i.i.d. setting, relative to less robust algorithms. That this is not the case follows from our theoretical characterizations.

\paragraph{Lower and upper bounds on regret}
Our main theorems provide lower and upper bounds on cumulative regret.
Cumulative regret is defined as the difference in welfare between the \textit{chosen} sequence of policies and the \textit{best} possible constant policy.
We consider both stochastic and adversarial regret. 
A lower bound on stochastic regret satisfies that, for any algorithm, there exists some stationary distribution of preference parameters for which the algorithm has to suffer at least a certain amount of regret.
An  upper bound on adversarial regret has to hold for a given algorithm and any sequence of preference parameters.

For a given algorithm, stochastic regret, averaged over i.i.d. sequences of preference parameters, is always less or equal than adversarial regret, for the worst-case sequence.
A lower bound on stochastic regret (for any algorithm) therefore implies a corresponding lower bound on adversarial regret, and an upper bound on adversarial regret (for a given algorithm) immediately implies an upper bound on stochastic regret.
When an adversarial upper bound coincides with a stochastic lower bound, in terms of rates of regret, it follows that the proposed algorithm is rate efficient, for both stochastic and adversarial regret. It follows, furthermore, that the bounds are sharp.

\paragraph{A lower bound on stochastic regret}
We first prove a stochastic (and thus also adversarial) lower bound on regret, for any possible algorithm in the welfare maximization problem.
Our proof of this bound constructs a family of possible distributions for preferences. 
This family is such that there are two candidate policies which are potentially optimal.
The difference in welfare between these two policies depends on the integral of demand over intermediate policy values.
In order to learn which of the two candidate policies is optimal, we need to learn behavioral responses for intermediate policies, which are strictly suboptimal.
Because of the need to probe these suboptimal policies sufficiently often, we obtain a lower bound on regret which grows at a rate of $T^{2/3}$, even if we restrict our attention to settings with finite, known support for preference parameters and policies. This rate is worse than the worst-case rate for bandits of $T^{1/2}$.

\paragraph{A matching upper bound on adversarial regret for modified Exp3}
We next propose an algorithm for the adaptive maximization of social welfare.
Our algorithm is a modification of the Exp3 algorithm \citep{auer2002nonstochastic}.
Exp3 is based on an unbiased estimate of cumulative welfare for each policy.
The probability of choosing a given policy is proportional to the exponential of this estimate of cumulative welfare, times some rate parameter. 
Relative to Exp3, we require two modifications for our setting. First, we need to discretize the continuous policy space. Second, and more interestingly, we need additional exploration of counterfactual policies, including some policies that are clearly sub-optimal, in order to learn welfare for the policies which are contenders for the optimum.
This need for additional exploration again arises because of the dependence of welfare on the integral of demand over counterfactual policy choices.
For our modified Exp3 algorithm, we prove an adversarial (and thus also stochastic) upper bound on regret. We show that, for an appropriate choice of tuning parameters, worst case cumulative regret over all possible sequences of preference parameters grows at a rate of $T^{2/3}$, up to a logarithmic term. The algorithm thus achieves the best possible rate.
Since the rates for our stochastic lower and adversarial upper bound coincide, up to a logarithmic term, we have a complete characterization of learning rates for the welfare maximization problem.

\paragraph{Improved stochastic bounds for concave social welfare}
The proof of our lower bound on regret is based on the construction of a distribution of preferences which delivers a non-concave social welfare function.
If we restrict attention to the stochastic setting, where preferences are i.i.d.\ over time, and if we assume that social welfare is concave, then we can improve upon this bound on regret.
We prove a lower bound on stochastic regret, under the assumption of concavity, which grows at the rate of $T^{1/2}$.
We then propose a dyadic search algorithm which achieves this rate, up to logarithmic terms.
This dyadic search algorithm maintains an ``active interval,'' containing the optimal policy with high probability, which is narrowed down over time.
Only policies within the active interval are sampled.

\paragraph{Extensions to non-linear income taxation and to commodity taxation}
Our discussion up to this point focuses on a minimal, stylized case of an optimal tax problem, where individual actions are binary, and the policy imposes a tax on this binary action.
Our arguments generalize, however, to more complicated and practically relevant settings.
This includes optimal nonlinear income taxation, as in \cite{Mirrlees1971} and \cite{Saez2001}, and commodity taxation for a bundle of goods, as in \cite{ramsey1927contribution}.
For nonlinear income taxation, different tax rates apply at different income levels, and welfare weights depend on individual earnings capacity.
In Section \ref{sec:income_taxation}, we discuss an extension of our tempered Exp3 algorithm to nonlinear income taxation, and characterize its regret.
For commodity taxation, different tax rates apply to different goods, and consumption decisions are continuous vectors.
In Section \ref{sec:commodity_taxation} we sketch an extension of our algorithm to commodity taxation, but leave its characterization for future research.

\paragraph{Roadmap}
The rest of this paper proceeds as follows.
We conclude this introduction with a discussion of some related work and relevant references.
Section \ref{sec:setup} introduces our setup, formally defines the adversarial and stochastic settings, and compares our setup to related learning problems.
Section \ref{sec:regret_bounds} provides lower and upper bounds on regret in the adversarial and stochastic settings.
Section \ref{sec:concave} restricts attention to the stochastic setting with concave social welfare, and provides improved regret bounds for this setting.
Section \ref{sec:income_taxation} discusses an extension of our baseline model to non-linear income taxation.
Section \ref{sec:commodity_taxation} sketches another extension of our baseline model to commodity taxation. 
Section \ref{sec:conclusion} concludes, and discusses some possible applications of our algorithm, as well as an alternative Bayesian approach to adaptive welfare maximization.
The proofs of Theorem \ref{theo:lower_stochastic} and Theorem \ref{theo:upper_exp3} can be found in Appendix \ref{sec:proofs}.
The proofs of our remaining theorems and proofs of technical lemmas are discussed in the Online Appendix.

\subsection{Background and literature}

To put our work in context, it is useful to contrast our framework with the standard approach in public finance and optimal tax theory, and with the frameworks considered in machine learning and the multi-armed bandit literature.

\paragraph{Optimal taxation}
Optimal tax theory, and optimal policy theory more generally, is concerned with the maximization of social welfare, where social welfare is understood as a (weighted) sum of subjective utility across individuals \citep{ramsey1927contribution, Mirrlees1971, baily1978some, Saez2001, chetty09suff}.
A key tradeoff in such models is between, first, \textit{redistribution} to those with higher welfare weights, and second, the efficiency cost of behavioral responses to tax increases. Such \textit{behavioral responses} might reduce the tax base.

Optimal tax problems are defined by normative parameters (such as welfare weights for different individuals), as well as empirical parameters (such as the elasticity of the tax base with respect to tax rates).
The typical approach in public finance uses historical or experimental variation to estimate the relevant empirical parameters (causal effects, elasticities).
These estimated parameters are then plugged into formulas for optimal policy choice, which are derived from theoretical models.
The implied optimal policies are finally implemented, without further experimental variation.

\paragraph{Multi-armed bandits}
The standard approach of public finance, which separates elasticity estimation from policy choice, contrasts with the adaptive approach that characterizes decision-making in many branches of AI, including online learning, multi-armed bandits, and reinforcement learning.
Multi-armed bandit algorithms, in particular, trade off \textit{exploration} and \textit{exploitation} over time \citep{RegretBandit2012, slivkins2019introduction, lattimore2020bandit}.
{Exploration} here refers to the acquisition of information for better future policy decisions, while {exploitation} refers to the use of currently available information for optimal policy decisions at the present moment.
The goal of bandit algorithms is to maximize a stream of {rewards}, which requires an optimal balance between exploration and exploitation.
Bandit algorithms for the stochastic setting are characterized by {optimism in the face of uncertainty}: Policies with uncertain payoff should be tried until their expected payoff is clearly suboptimal.

Bandit algorithms (and similarly, adaptive experimental designs for informing policy choice, as in \citealt{russo2020simple, adaptive2019}) are not directly applicable to social welfare maximization problems, such as those of optimal tax theory.
The reason is that bandit algorithms maximize a stream of \textit{observed} rewards. By contrast, social welfare as conceived in welfare economics is based on \textit{unobserved} subjective utility.

\paragraph{Adversarial decision-making}
Adversarial models for sequential decision-making find their roots in repeated game theory \citep{hannan1957approximation}, while related settings were independently studied in information theory \citep{cover1966behavior} and computer science \citep{vovk1990aggregating,littlestone1994weighted,cesa1997use}. Regret minimization, also in a bandit setting, was investigated as a tool to prove convergence of uncoupled dynamics to equilibria in $N$-person games \citep{hart2000simple,hart2001general} -- the exponential weighting scheme used by Exp3 is also known as \textsl{smooth fictitious play} in the game-theoretic literature \citep{fudenberg1995consistency}. 
Recent works \citep{seldin2014one,zimmert2021tsallis} show that simple variants of Exp3 simultaneously achieve essentially optimal regret bounds in adversarial, stochastic, and contaminated settings, without prior knowledge of the actual regime. This suggests that algorithms designed for adversarial environments can behave well in more benign settings, whereas the opposite is provably not true.

\paragraph{Bandit approaches for economic problems}
Bandit-type approaches have been applied to a number of other economic and financial scenarios in the literature where rewards \textit{are} observable. These include monopoly pricing \citep{kleinberg2003value} (see also the survey \citealt{den2015dynamic}), second-price auctions \citep{cesa2015regret,weed2016online,cesa2017algorithmic}, first-price auctions \citep{han2020optimal,han2020learning,achddou2021fast,cesa2023transparency}---see also \citep{kolumbus2022auctions,feng2018learning,feng2021convergence}, and combinatorial auctions \citep{daskalakis2022learning}. 
Bandit-type approaches have also been applied to some settings where rewards are not directly observable, including bilateral trading \cite{cesa2021regret,cesa2023smooth,cesa2023bilateral}, and the newsvendor problem \citep{lugosi2023hardness}.

Bandit algorithms are widely used in online advertising and recommendation. Online learning methods are successfully used for tuning the bids made by autobidders (a service provided by advertising platforms) \citep{lucier2024autobidders}. While these algorithms are analyzed in adversarial environments, the extent to which they are deployed in commercial products remains unclear.

\section{Setup}
\label{sec:setup}

At each time $i = 1,2, \ldots, T$, one individual arrives who is characterized by an unknown willingness to pay $\wtp_i \in [0,1]$.
This individual is exposed to a tax rate $\pol_i$, and makes a binary decision $\out_i = \bs 1(\pol_i \leq \wtp_i)$.
The implied public revenue is $\pol_i  \cdot \out_i$.
The implied private welfare is $\max(\wtp_i  - \pol_i, 0)$.
We define social welfare as a weighted sum of public revenue and private welfare, with a weight $\lambda \in (0,1)$ for the latter.
Social welfare for time period $i$ is therefore given by
\be
  \sw_i(\pol_i) = \underbrace{\pol_i  \cdot \bs 1(\pol_i \leq \wtp_i)}_{\textrm{Public revenue}}\quad + \quad \lambda  \cdot \underbrace{\max(\wtp_i  - \pol_i, 0)}_{\textrm{Private welfare}}.
\ee
After period $i$, we observe $\out_i$ and the tax rate $
\pol_i$, but nothing else.
In particular, we do \textit{not} observe welfare $\sw_i(\pol_i)$.

We can rewrite social welfare $\sw_i(\pol)$ as follows. Denote $\dem_i(\pol)  = \bs  1(\wtp_i \geq \pol)$, so that $\out_i = \dem_i(\pol_i)$. This is the individual demand function.
Then private welfare can be written as
$
  \max(\wtp_i  - \pol, 0) = 
  \int_{\pol}^1 \dem_i(\pol') \mathrm{d} \pol'.
$
That is, private welfare is given by integrated demand.\footnote{This reflects the absence of income effects in our model, which implies that private utility, consumer surplus, compensating variation, and equivalent variation all coincide.} 
This representation of private welfare implies
\be
  \sw_i(\pol) = \underbrace{\pol \cdot  \dem_i(\pol)}_{\textrm{Public revenue}}  + \quad \lambda  \cdot \underbrace{\int_\pol^1 \dem_i(\pol') \mathrm{d} \pol'}_{\textrm{Private welfare}}.
\ee
We consider algorithms for the choice of $\pol_i$ which might depend on the observable history $(\pol_j, \out_j)_{j=1}^{i-1}$, as well as possibly a randomization device.

\paragraph{Notation}
For the \textit{adversarial} setting, we will consider cumulative demand and welfare, denoted by blackboard bold letters, summing across $j=1, \ldots, i$.
In particular,
\bals
  \demcum_i(\pol) &= \sum_{j\leq i} \dem_i(\pol),&
  \swcum_i(\pol) &= \sum_{j\leq i} \sw_i(\pol),&
  \swcum_i &= \sum_{j\leq i} \sw_j(\pol_j).
\eals
$\demcum_i(\pol)$ and $\swcum_i(\pol)$ are cumulative demand and welfare for a counterfactual, fixed policy $\pol$.
$\swcum_i$, without an argument, is the cumulative welfare for the policies $\pol_j$ actually chosen.

For the \textit{stochastic} setting, we will analogously consider expected demand and expected welfare, denoted by boldface letters. The expectation is taken across some stationary distribution $\mu$ of $\wtp_i$, where $\wtp_i$ is statistically independent of $\pol_i$, and of $\wtp_j$ for $j\neq i$.
In particular,
\bals
  \demexp(\pol) &= E[\dem_i(\pol)],&
  \swexp(\pol) &= E[\sw_i(\pol)].
\eals

\subsection{Regret}

\paragraph{The adversarial case}
Following the literature, we consider regret for both the adversarial and the stochastic setting.
In the adversarial setting, we allow for arbitrary sequences of willingness to pay, $\{\wtp_i\}_{i=1}^T$.
We compare the expected performance of any given algorithm for choosing $\{\pol_i\}_{i=1}^T$ to the performance of the best possible constant policy $\pol$.
This comparison yields cumulative expected regret, which is given by
\be
  \regadv = \sup_{\pol} E\left [ \swcum_T(\pol)- \swcum_T \Big| \{\wtp_i\}_{i=1}^T \right ].
\ee
The expectation in this expression is taken over any possible randomness in the tax rates $\pol_i$ chosen by the algorithm; there is no other source of randomness.

\paragraph{The stochastic case}
We also consider the stochastic setting.
In this setting, we add structure by assuming that the $\wtp_i$ are i.i.d.\ draws from some distribution $\mu$ on $[0,1]$, 
with implied demand function $\demexp(x) = P(\wtp_i \geq x)$.
This demand function is identified by the regression
$$
  \demexp(\pol) = E[\out_i | \pol_i = \pol].
$$
The expectation in this expression is taken over the distribution of $\wtp_i$, which is presumed to be statistically independent of the tax rate $\pol_i$.
Expected welfare for this distribution of $\wtp_i$ is given by 
$$
 \swexp(\pol) = 
  \pol \cdot \demexp(\pol) + \lambda \int_{\pol}^1 \demexp(\pol') \mathrm{d}\pol'.
$$
Cumulative expected regret in the stochastic case equals 
\bal
  \regstoch &= \sup_{\pol} E\left [\swcum_T(\pol)-\swcum_T \right ]
  = T \cdot \sup_{\pol}  \swexp(\pol)- E\left[\sum_{i\leq T}\swexp(\pol_i) \right ].
\eal
The expectation in this expression is taken over both any possible randomness in the tax rates $\pol_i$, and the i.i.d.\ draws $\wtp_i$.

\subsection{Comparison to related learning problems}
\label{sec:comparison}

Before proceeding with our analysis of regret, we take a step back, and compare our learning problem to two related problems that have received some attention in the literature. 
The first of these is the adaptive \textbf{monopoly pricing} problem; see for instance \cite{kleinberg2003value}.
This problem is equivalent to our setting when we set $\lambda = 0$, interpret $\pol$ as a price, and $\sw^{\mathrm{MP}}_i$ as monopolist profits (neglecting production costs):
\bal
  \sw_i^{\mathrm{MP}}(\pol) &= \pol_i  \cdot \bs 1(\pol_i \leq \wtp_i)
  =  \underbrace{\pol \cdot  \dem_i(\pol).}_{\textrm{Monopolist revenue}}
\eal
As in our adaptive taxation setting, the feedback received at the end of period $i$ is 
$$\out_i = \dem_i(\pol_i) = \bs 1(\pol_i \leq \wtp_i).$$

Another related problem is price setting for \textbf{bilateral trade}, see for instance \cite{cesa2023bilateral}.
In this problem, welfare $\sw_i^{\mathrm{BT}}(\pol)$ is given by the sum of seller and buyer welfare.
Trade happens if and only if both sides agree to transact at the proposed price. Buyer willingness to pay is given by $\wtp^b_i$, while the seller is willing to trade at prices above $\wtp^s_i$.
\be
\begin{array}{lllll}
	\sw_i^{\mathrm{BT}}(\pol)
&=&
	\bs 1(\wtp^b_i  \geq \pol) \cdot  \max(\pol - \wtp^s_i, 0)
&+&
	\bs 1(\wtp^s_i  \leq \pol) \cdot  \max(\wtp^b_i  - \pol, 0)
\\&=&
	\dem^b_i(\pol) \cdot \underbrace{\int_0^x \dem^s_i(\pol') \mathrm{d}\pol'}_{\textrm{Seller welfare}}
&+&
	\dem^s_i(\pol) \cdot \underbrace{\int_x^1 \dem^b_i(\pol') \mathrm{d}\pol'}_{\textrm{Buyer welfare}}.
\end{array}
\ee
Feedback in this case is a little richer: We observe both whether the buyer $b$ would have accepted the posted price,  and whether the seller would have accepted this price, 
\[
	\out^b_i =\dem^b_i(\pol_i) = \bs 1(\pol_i \leq \wtp^b_i) \qquad\textrm{and}\qquad \out^s_i = \dem^s_i(\pol_i) = \bs 1(\pol_i \geq \wtp^s_i).
\]

\begin{table}
  \caption{Regret rates for different learning problems}
  \label{tab:regret_comparison}
  \begin{center}
    \begin{tabular}{lllcc}
      \toprule
      Model & \multicolumn{2}{c}{Policy space} & \multicolumn{2}{c}{Objective function} \\
      & Discrete & Continuous & Pointwise & One-sided Lipschitz\\
      \midrule
      Monopoly price setting &  $T^{1/2}$ & $T^{2/3}$  &Yes & Yes\\
      Optimal taxation  & $T^{2/3}$ &  $T^{2/3}$ & No & Yes\\
      Bilateral trade & $T^{2/3}$ & $T$ & No & No\\
      \bottomrule
    \end{tabular}
  \end{center}
  \footnotesize
  \textit{Notes:}
  This table shows the efficient rates of regret for different learning problems.
  Rates are up to logarithmic terms, and apply to both the stochastic and the adversarial setting.
  Regret rates are shown for the discrete case, where the space of policies $\pol$ is restricted to a finite set, and the continuous case, where $\pol$ can take any value in $[0,1]$.
  The columns on the right describe the properties of the objective function in each problem, which drive the differences in regret rates.  
  
  Rates for the optimal taxation case are proven in this paper.
  Rates for the continuous monopoly price setting case are from \cite{kleinberg2003value}; the discrete case reduces to a standard bandit problem.
  Rates for the continuous bilateral trade case are from \cite{cesa2023bilateral}; the discrete case can be deduced by adapting the arguments in the same paper (for the stochastic i.i.d.\ case with independent sellers' and buyers' valuations), or by adapting the techniques in \cite{cesa2023smooth} (for the adversarial case, allowing the learner to use weakly budget balanced mechanisms).
\end{table}

\paragraph{Lipschitzness and information requirements}
The difficulty of the learning problem in each of these models critically depends on (i) the Lipschitz properties of the welfare function, and (ii) the information required to evaluate welfare at a point.

We say that a generic welfare function $W:[0,1]\to\R$ is one-sided Lipschitz if $W(\pol + \ve) \le W(\pol) + \ve$ for all $0 \le \pol \le 1$ and all $0 \le \ve \le 1 - \pol$.
One-sided Lipschitzness allows us to bound the approximation error of a learning algorithm operating on a finite subset of the set of policies.
One-sided Lipschitzness is an intrinsic property of both the monopoly pricing and the optimal taxation problem; it is not an assumption that is additionally imposed.
To see this for monopoly pricing, note that, for $\epsilon \ge 0$, $\sw_i^{\mathrm{MP}}(\pol +\ve) = (\pol +\ve) \cdot \bs 1(\pol + \ve \leq \wtp_i) \le \pol \cdot \bs 1(\pol \leq \wtp_i) +\ve = \sw_i^{\mathrm{MP}}(\pol)  +\ve$.
For social welfare, $\sw_i(\pol) = (\pol_i+\ve) \cdot \bs 1(\pol_i+\ve \leq \wtp_i) + \lambda  \cdot \max(\wtp_i  - \pol_i-\ve, 0) \le \pol \cdot \bs 1(\pol \leq \wtp_i)+\ve  + \lambda\cdot \max(\wtp_i  - \pol_i, 0) = 
\sw_i(\pol)  +\ve$.

We say that learning $W( \cdot )$ requires only pointwise information if $W( \pol )$ is a function of $\dem(\pol)$, and does not depend on $\dem( \cdot )$ otherwise.
Pointwise information allows us to avoid exploring policies that are clearly suboptimal, when we aim to learn the optimal policy.

\autoref{tab:regret_comparison} summarizes the Lipschitz properties and information requirements in each of the three models; the following justifies the claims made in \autoref{tab:regret_comparison}:
\begin{enumerate}
  \item For \textbf{monopoly pricing}, welfare $\sw_i^{\mathrm{MP}}(\pol)$ is one-sided Lipschitz and only depends on $\dem_i(\pol)$ pointwise.
  
  \item For \textbf{optimal taxation}, welfare $\sw_i(\pol)$ is one-sided Lipschitz and depends on both $\dem_i(\pol)$ at the given $\pol$ (pointwise), and on an integral of $\dem_i(\pol')$ for a range of values of $\pol'$ (non-pointwise).
  
  \item For \textbf{bilateral trade}, welfare $\sw_i^{\mathrm{BT}}(\pol)$ is not one-sided Lipschitz and depends on both $\dem_i^b(\pol)$ and $\dem_i^s(\pol)$ (pointwise), as well as the integrals of $\dem_i^b(\pol')$ and $\dem_i^s(\pol')$ (non-pointwise).
\end{enumerate}
These properties suggest a ranking in terms of the difficulty of the corresponding learning problems, and in particular in terms of the rates of divergence of cumulative regret: The information requirements of optimal taxation are stronger than those of monopoly pricing, but its continuity properties are more favorable than those of bilateral trade.
This intuition is correct, as shown by \autoref{tab:regret_comparison}. The rates for monopoly pricing and for bilateral trade are known (or can be easily adapted) from the literature. In this paper we prove corresponding rates for optimal taxation.

In comparing optimal taxation and monopoly pricing to conventional multi-armed bandits, it is worth emphasizing that there are two distinct reasons for the slower rate of convergence. 
First, the continuous support of $\pol$, as opposed to a finite number of arms, which is shared by optimal taxation and monopoly pricing.
Second, the requirement of additional exploration of sub-optimal policies for the optimal tax problem.
As shown in \autoref{tab:regret_comparison}, the continuous support alone is enough to slow down convergence, with no extra penalty for the additional exploration requirement, in terms of rates.
If, however, we restrict our attention to a discrete set of feasible policies $\pol$, then monopoly pricing reduces to a multi-armed bandit problem, with a minimax regret rate of $T^{1/2}$.
The optimal tax problem, by contrast, still has a rate of $T^{2/3}$, even if we restrict our attention to the case of finite known support for $\wtp$ and $\pol$, as shown by the proof of Theorem \ref{theo:lower_stochastic} below.

\paragraph{Hannan consistency}
The cumulative regret of any non-adaptive algorithm necessarily grows at a rate of $T$. This includes, in particular, randomized experiments where the policy is chosen uniformly at random, from a fixed policy set, in every period. Algorithms for which adversarial regret (and thus also stochastic regret) grows at a rate less than $T$, so that per-period regret goes to $0$ as $T$ increases, are known as \textit{Hannan consistent}. Non-adaptive algorithms are not Hannan consistent.
Table \ref{tab:regret_comparison} implies that Hannan consistent algorithms exist in all settings considered, with the exception of Bilateral trade and continuous policy spaces.

\section{Stochastic and adversarial regret bounds}
\label{sec:regret_bounds}

We now turn to our main theoretical results, lower and upper bounds on stochastic and adversarial regret for the problem of social welfare maximization.
We first prove a lower bound on stochastic regret, which applies to any algorithm, and which immediately implies a lower bound on adversarial regret.
We then introduce the algorithm \algexp.
We show that, for an appropriate choice of tuning parameters, this algorithm achieves the rates of the lower bound on regret, up to a logarithmic term.
Formal proofs of these bounds can be found in \autoref{sec:proofs}.

\subsection{Lower bound}

\begin{theo}[Lower bound on regret]
  \label{theo:lower_stochastic}
  Consider the setup of Section \ref{sec:setup}.
  There exists a constant $C>0$ such that, for any randomized algorithm for the choice of $\pol_1, \pol_2, \dots$ and any time horizon $T \in \mathbb{N}$, the following holds.  
  \begin{enumerate}
  \item There exists a distribution $\mu$ on $[0,1]$ with associated demand function $\demexp$ for which the stochastic cumulative expected regret $\regstoch$ is at least $C \cdot T^{2/3}$.
  \item There exists a sequence $(\wtp_1, \ldots, \wtp_T)$ for which the adversarial cumulative expected regret $\regadv$ is at least $C \cdot T^{2/3}$.
  \end{enumerate}
\end{theo}

\begin{figure}[t]
  \caption{Construction for proving the lower bound on regret}
  \label{fig:lower_bound}
  \begin{center}
    \includegraphics[width=.8\textwidth]{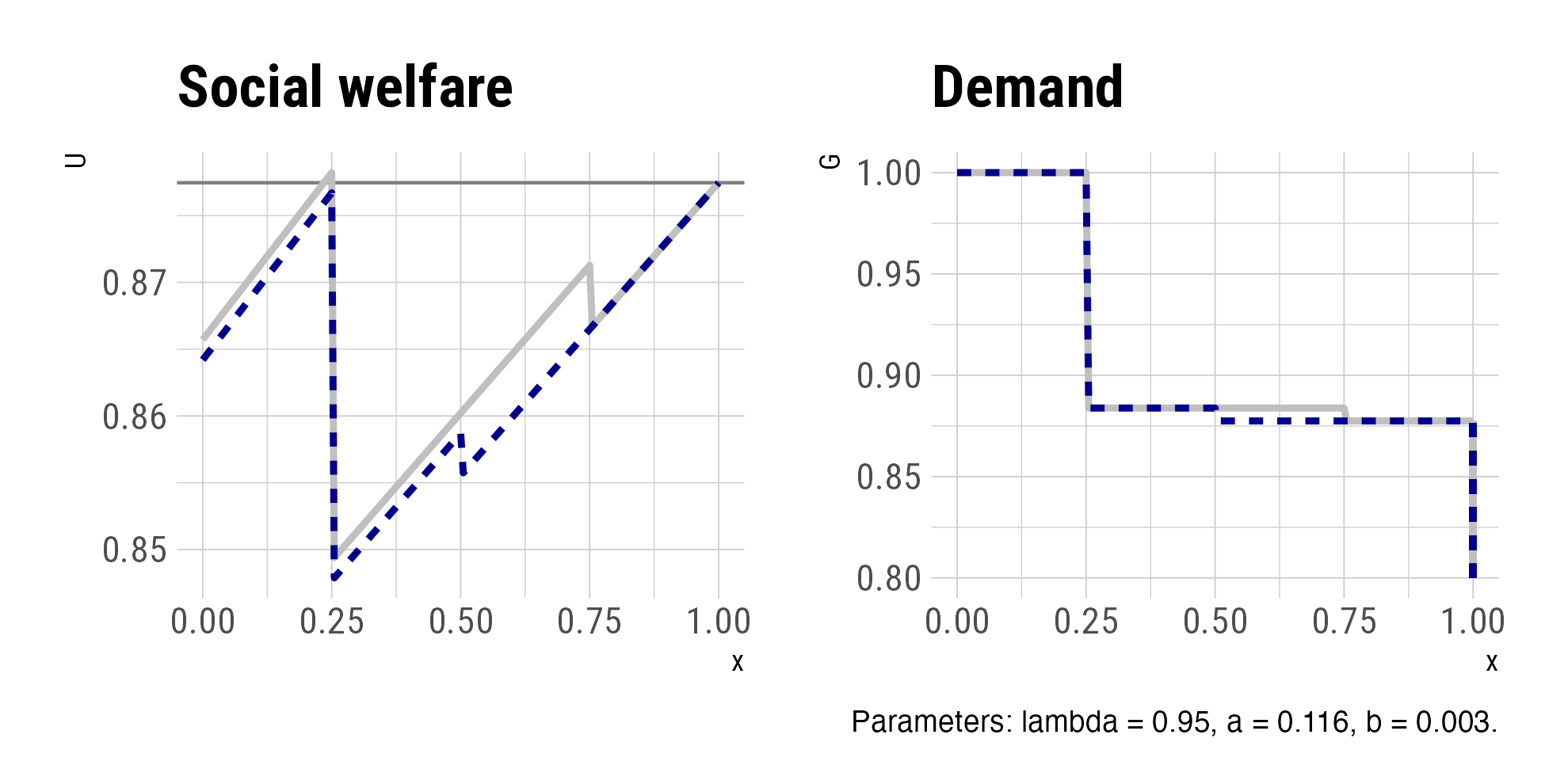}
  \end{center}
  \footnotesize
  \textit{Notes:} This figure illustrates our construction for proving the lower bound on regret. The relative social welfare of policies $1$ and $.25$ depends on the sign of $\epsilon$. The bright line corresponds to $\epsilon=-1$, the dark, dashed line to $\epsilon=1$. In order to distinguish between these two, we must learn demand in the intermediate interval $[.5,.75]$.
\end{figure}

The proof of Theorem \ref{theo:lower_stochastic} 
can be found in \autoref{sec:proofs}.
The adversarial lower bound follows immediately from the stochastic lower bound, since worst case regret (over possible sequences of $\wtp_i$) is bounded below by average regret (over i.i.d.\ draws of $\wtp_i$), for any distribution of $\wtp_i$.

\paragraph{Sketch of proof}
To prove the stochastic lower bound we construct a family of distributions $\{\mu^\epsilon\}_{\epsilon \in [-1,1]}$ for $\wtp_i$, indexed by a parameter $\epsilon \in [-1,1]$.
The distributions in this family have four points of support, $(\nicefrac{1}{4}, \nicefrac{1}{2}, \nicefrac{3}{4}, 1)$.
The probability of these points is given by 
$$\left ( a, (1+\epsilon)b, (1-\epsilon)b, 1- a - 2b \right ).$$
The values of $a$ and $b$ are chosen such that (i) the two middle points $\nicefrac{1}{2}, \nicefrac{3}{4}$  are far from optimal, for any value of $\epsilon$, and (ii) learning which of the two end points $(\nicefrac{1}{4}, 1)$ is optimal requires sampling from the middle.\footnote{Specifically, 
    $a \coloneqq \frac{(1-\lambda) \cdot (136 - 99\cdot \lambda)}{2 \cdot (4-3\cdot\lambda)\cdot(24-17\cdot\lambda)}$,   and
    $b \coloneqq \frac{1-\lambda}{2 \cdot(24 - 17 \cdot \lambda)}$.
These two constants are strictly greater than zero, and satisfy $1 - a - 2\cdot b > 0$.} 
For each $\epsilon \in [-1,1]$, denote the demand function associated to $\mu^\epsilon$ by $\demexp^\epsilon$, and the expected social welfare associated to $\demexp^\epsilon$ by $\swexp^\epsilon$.
Property (ii) holds because of the integral term $\int_{\tfrac14}^1 \demexp^\epsilon(\pol') \mathrm{d}\pol'$, which shows up in $\swexp^\epsilon(1)-\swexp^\epsilon(\nicefrac{1}{4})$.
This construction is illustrated in \autoref{fig:lower_bound}. This figure shows plots of $\demexp^\epsilon$ and of $\swexp^\epsilon$ for $\lambda = .95$ and $\epsilon \in \{\pm 1\}$.

The difference in welfare $\swexp^\epsilon(1)-\swexp^\epsilon(\nicefrac{1}{4})$ of the two candidates optimal policies $\nicefrac{1}{4}$ and $1$ depends on the sign of $\epsilon$. 
In order not to suffer expected regret that grows as $|\epsilon| \cdot T$, 
any learning algorithm needs to sample policies from points that are informative about the sign of $\epsilon$. 
The only  points that are informative are those in the region $(\nicefrac{1}{2},\nicefrac{3}{4}]$, where welfare is bounded away from optimal welfare. 

More specifically, the learning algorithm has to sample on the order of $|\epsilon|^{-2}$
times from the region $(\nicefrac{1}{2},\nicefrac{3}{4}]$, to be able to detect the sign of $\epsilon$, incurring regret on the order of $|\epsilon|^{-2}$ in the process.
Any learning algorithm therefore incurs regret on the order of $\min\brb{|\epsilon|^{-2},|\epsilon| \cdot  T}$, which, for $\epsilon \propto T^{-1/3})$, leads to the conclusion.

\subsection{An algorithm that achieves the lower bound}

\begin{algorithm}[t]
  \caption{\algexp}
  \label{alg:tempered_exp3}
\begin{algorithmic}[1]
  \REQUIRE Tuning parameters $K$, $\gamma$ and $\eta$.
  
  \STATE Calculate evenly spaced grid-points $\tilde \pol_k = (k-1)/K$,\\
  and initialize $\widehat \demcum_{1k} = 0$ and $\widehat \swcum_{1k} = 0$ for $k=1,\ldots, K+1$.

  \FOR{individual $i=1,2,\ldots, T$} 

    \STATE For all $k=1,2,\ldots, K+1$, set \hfill \COMMENT{Assignment probabilities}
      \be      
        p_{ik} =(1-\gamma)  \cdot \frac{ \exp(\eta \cdot  \widehat \swcum_{ik})}{\sum_{k'}\exp(\eta  \cdot \widehat \swcum_{ik'})} + \frac{\gamma }{K+1}.
      \ee
          
    \STATE Choose $k_i$ at random according to the probability distribution $(p_{i,1},\ldots, p_{i,K+1})$.\\
    Set $\pol_i = \tilde \pol_{k_i}$, and query $\out_i$ accordingly.
      
    \STATE For all $k=1,2,\ldots, K+1$, set \hfill \COMMENT{Estimated demand}
      \be
        \widehat \demcum_{i+1, k}= \widehat \demcum_{i, k} + y_i \cdot \frac{\bs 1(k_i = k)}{p_{ik}}.
      \ee

    \STATE For all $k=1,2,\ldots, K+1$, set \hfill \COMMENT{Estimated welfare}
    \be
      \widehat \swcum_{i+1, k} = \tilde \pol_k  \cdot  \widehat \demcum_{i+1, k} + \tfrac{\lambda}{K}  \cdot  \sum_{k'> k} \widehat \demcum_{i+1, k'}.
    \ee
  \ENDFOR
  
\end{algorithmic}
\end{algorithm}

We next introduce Algorithm \ref{alg:tempered_exp3}, which allows us to essentially achieve the lower bound on regret, in terms of rates.

\paragraph{Conventional Exp3}
Algorithm \ref{alg:tempered_exp3} is a modification of the Exp3 algorithm.
Conventional Exp3 \citep{auer2002nonstochastic} is designed to maximize the standard bandit objective, $\sum_{i \leq T} \out_i$.
Exp3 maintains an unbiased running estimate of the cumulative payoff of each arm $k$, calculated using inverse probability weighting, $\widehat \demcum_{i,k} = \sum_{j<i} \out_i  \cdot \frac{\bs 1(k_i = k)}{p_{ik}}$. 
In period $i$, arm $k$ is chosen with probability $p_{ik} =(1-\gamma)  \cdot \tfrac{ \exp(\eta \cdot  \widehat \demcum_{ik})}{\sum_{k'}\exp(\eta  \cdot \widehat \demcum_{ik'})} + \frac{\gamma }{K+1}$, where $\eta$ and $\gamma$ are tuning parameters.
$p_{ik}$ is thus increasing in the estimated average performance $\tfrac{\widehat \demcum_{i,k}}{i}$ of arm $k$ in prior periods.
Because $\widehat \demcum_{i,k}$ is \emph{not} normalized by the number of time periods $k$, more weight is given to the best-performing arms over time, as estimation uncertainty for average performance decreases.
In both these aspects, Exp3 is similar to the popular Upper Confidence Bound algorithm (UCB) for stochastic bandit problems \citep{lai1987adaptive,agrawal1995sample,auer2002finite}.
In contrast to UCB, Exp3 is a randomized algorithm.
Randomization is required for adversarial performance guarantees. This is analogous to the necessity of mixed strategies in Nash equilibrium for zero-sum games.

\paragraph{Modifications relative to conventional Exp3}
Relative to this algorithm, we require three modifications.
First, we discretize the continuous support $[0,1]$ of $\pol$, restricting attention to the grid of policy values $\tilde \pol_k = (k-1)/K$.
Second, since welfare $\sw_i(\pol)$ is not directly observed for the chosen policy $\pol$, we need to estimate it indirectly. 
In particular, we first form an estimate $\widehat \demcum_{i k}$ of cumulative demand for each of the policy values $\tilde \pol_k$, using inverse probability weighting.
We then use this estimated demand, interpolated using a step-function, to form estimates of cumulative social welfare, 
$
\widehat \swcum_{ik} ={ \tilde \pol_k  \cdot  \widehat \demcum_{i k} + \tfrac{\lambda}{K}  \cdot  \sum_{k'> k} \widehat \demcum_{i k'}}.
$
Third, we require additional exploration, relative to Exp3.
Since social welfare depends on demand for counterfactual policy choices, we need to explore policies that are away from the optimum, in order to learn the relative welfare of approximately optimal policy choices.
The mixing weight $\gamma$, which determines the share of policies sampled from the uniform distribution, needs to be larger relative to conventional Exp3, to ensure sufficient exploration away from the optimum.

\begin{theo}[Adversarial upper bound on regret of \algexp]
  \label{theo:upper_exp3}
  Consider the setup of Section \ref{sec:setup}, and Algorithm \ref{alg:tempered_exp3}.
  Assume that $(K+1) \eta < \gamma$.\\
  Then for any  sequence $(\wtp_1, \ldots, \wtp_T)$ expected regret $\regadv$ is bounded above by
  \be
    \left ( \gamma   + \eta  \cdot (e-2) 
     \tfrac{K+1}{K} \cdot \left (\tfrac{2K+1}{6} + \tfrac{\lambda^2}{\gamma}  \right )
    + \tfrac{\lambda}{K} \right ) \cdot T
    + \tfrac{\log(K+1)}{\eta}.
    \label{eq:upper_exp3}
  \ee
  Suppose additionally that $c_1,c_2,c_3>0$ are constants.
  Then, there exists a constant $c_4$ such that, if we set
  $\gamma = c_1  \cdot \left(\tfrac{\log(T)}{T}\right)^{1/3}$, 
  $\eta = c_2  \cdot \gamma^2$,  and
  $ K = \lfloor c_3 / \gamma \rfloor $,
  the expected regret $\regadv$ is bounded above by 
  \be
    c_4 \cdot \log(T)^{1/3} T^{2/3}.
  \ee
\end{theo}

\begin{cor}[Stochastic upper bound on regret of \algexp]
  Under the assumptions of Theorem \ref{theo:upper_exp3}, suppose additionally that $\wtp_i$ is drawn i.i.d.\ from some distribution with associated demand function $\demexp$.
  Then expected regret $\regstoch$ is bounded above by the same expressions as in Theorem \ref{theo:upper_exp3}.
\end{cor}

The proof of Theorem \ref{theo:upper_exp3} can again be found in \autoref{sec:proofs}.

\paragraph{Tuning}
The statement of the theorem leaves the constants $c_1,c_2,c_3$ in the definition of the tuning parameters unspecified.
Suppose we wish to choose the tuning parameters so as to optimize the upper bound obtained in Theorem \ref{theo:upper_exp3}.
Ignoring the rounding of $K$, an approximate solution to this problem is given by
\bals
\eta  &= 1/a \cdot (\log(T)/T)^{2/3}\\
\gamma  &=\lambda  \sqrt{(e-2)/a} \cdot (\log(T)/T)^{1/3}\\
K &=\sqrt{3\lambda a/(e-2)} \cdot (T/\log(T) )^{1/3}
\eals
where
$$a = \left(9(e-2)\right)^{1/3} (\sqrt{\lambda /3}+\lambda )^{2/3}.$$
This solution is obtained by taking the upper bound in Equation \eqref{eq:upper_exp3}, approximating $(K+1) / K \approx 1$ and $(2K+1) / 6 \approx K/3$, and solving the first order conditions with respect to the three tuning parameters.
This approximation, and the tuning parameters specified above, yield an approximate upper bound on regret of $6  \cdot \log(T)^{1/3} T^{2/3}$.

\paragraph{Unknown time horizon}
Note that the proposed tuning depends crucially on knowledge of the time horizon $T$ at which regret is to be evaluated. 
In order to extend our rate results to the case of unknown time horizons, we can use the so-called doubling trick; cf. Section 2.3 of \cite{cesa2006prediction}: 
Consider a sequence of epochs (intervals of time-periods) of exponentially increasing length, and re-run Algorithm \ref{alg:tempered_exp3} for each time-period separately, tuning the parameters over the current epoch length.
This construction converts Algorithm \ref{alg:tempered_exp3} into an ``anytime algorithm'' which enjoys the same regret guarantees of Theorem \ref{theo:upper_exp3}, up to a multiplicative constant factor.
Another more efficient strategy to achieve the same goal is to modify Algorithm \ref{alg:tempered_exp3}, allowing the parameters $\eta$ and $\gamma$ to change at each iteration, and splitting each bin associated with the discretization parameter $K$ whenever more precision is required.

\paragraph{The extra $\log(T)^{1/3}$ term}
There is a rate discrepancy between our our upper and lower bounds on regret, corresponding to the $\log(T)^{1/3}$ term in our upper bound.
We conjecture the existence of an alternative algorithm that can eliminate this extra logarithmic term, albeit at the cost of reduced computational efficiency and a less transparent theoretical analysis.
Our conjecture is based on known results for the standard multi-armed bandit problem with $K$ arms.
The Exp3 algorithm achieves an upper bound of order $\sqrt{ K\log(K)T }$ for this problem, which includes an extra logarithmic term compared to the known lower bound of order $\sqrt{KT}$. Exp3 is an instance of the Follow-The-Regularized-Leader (FTRL) algorithm with importance weighting and the negative entropy as the regularizer. It is known that using the $\frac{1}{2}$-Tsallis entropy as the regularizer in the FTRL algorithm with importance weighting results in regret guarantees of order $\sqrt{KT}$ for the bandit problem \citep{lattimore2020bandit}. 
However, unlike Exp3, the FTRL with Tsallis entropy involves a more complex proof and requires solving an optimization problem in each period.
Analogous statements might be true for our setting.

\paragraph{Numerical example}
\begin{figure}[t]
  \caption{\algexp -- Numerical example}
  \label{fig:numerical}
  \begin{center}
    \includegraphics[width=.7\textwidth]{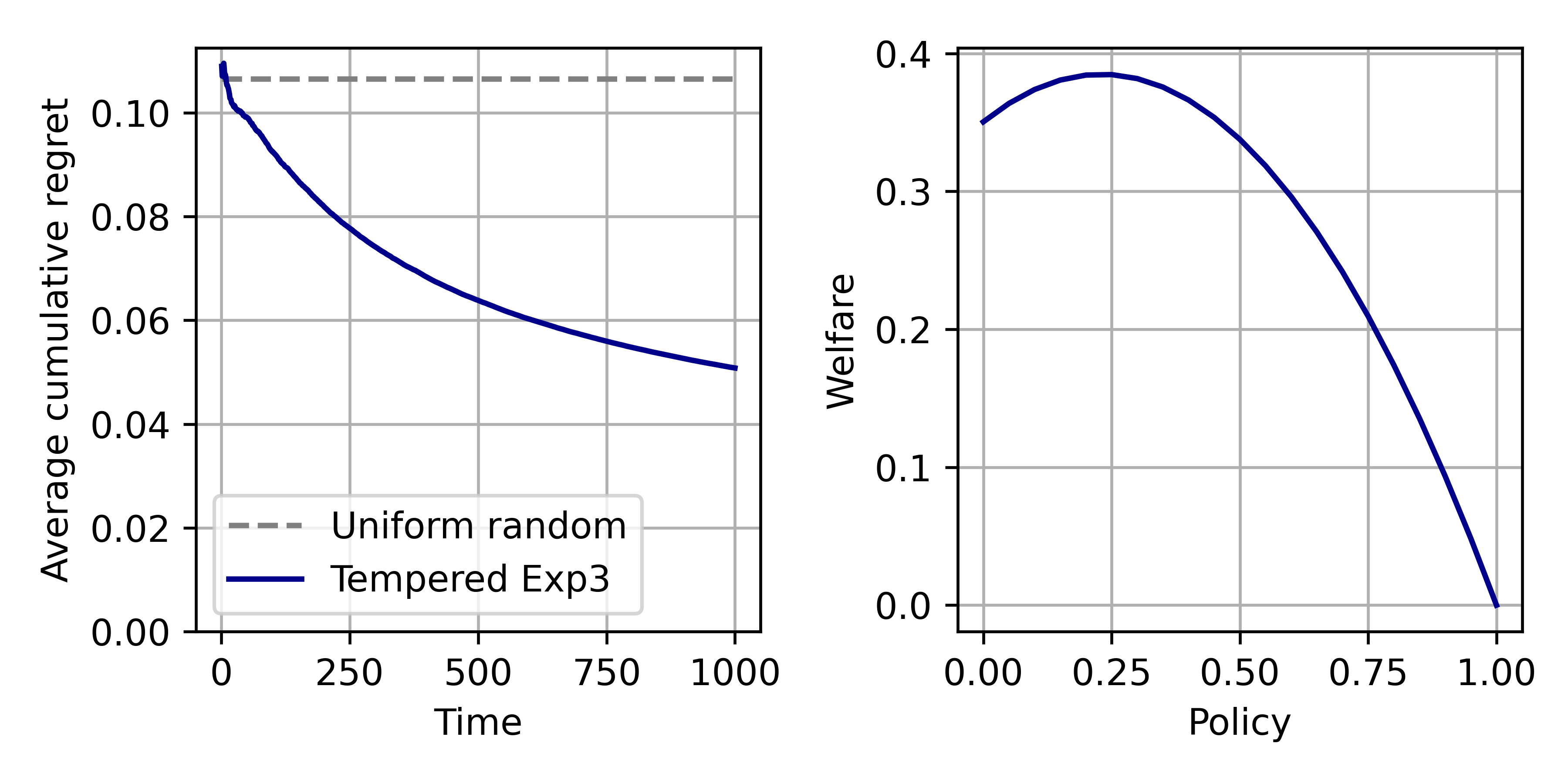}
  \end{center}
  \footnotesize
  \textit{Notes:} This figure illustrates the performance of our algorithm for the stochastic case, where $\wtp_i$ is drawn uniformly at random from $[0,1]$ for all $i$, the weight $\lambda$ equals $.7$, and the tuning parameters are 
  $K = 20,\;\eta = .025,\;\gamma = .1$.
  The left plot shows the cumulative average regret of our algorithm, averaged across 4000 simulations.
  The right plot shows expected social welfare $\swexp(\pol)$ as a function of the policy $\pol$.
\end{figure}
For illustration, Figure \ref{fig:numerical} plots the cumulative average regret of \algexp~for the case where $\wtp_i$ is sampled uniformly at random from $[0,1]$ each time period. Initially, the performance of our algorithm is, by construction, equal to the performance of choosing a policy uniformly at random. Over time, however, the average regret of our algorithm drops by more than half, in this numerical example.
Note that the rate at which cumulative regret declines in Figure \ref{fig:numerical} (for i.i.d. sampling from a fixed distribution) is unrelated to the regret rate of Theorem \ref{theo:upper_exp3} (for the worst case sequence of $\wtp_i$, for each time horizon $T$).

\section{Stochastic regret bounds for concave social welfare}
\label{sec:concave}

Theorem \ref{theo:lower_stochastic} in Section \ref{sec:regret_bounds} provides a lower bound proportional to $T^{2/3}$ for adversarial and stochastic regret in social welfare maximization.
The proof of this lower bound constructs a distribution for the $\wtp_i$.
This distribution is such that expected social welfare $\swexp(\pol)$ is non-concave, as a function of $\pol$;
two global optima are separated by a region of lower welfare.
In order to learn which of two candidates for the globally optimal policy is actually optimal, it is necessary to sample policies in between. These intermediate policies yield lower welfare, and sampling them contributes to cumulative regret. This construction is illustrated in \autoref{fig:lower_bound}.

Given that the construction relies on non-concavity of expected social welfare, could we achieve lower regret if we knew that social welfare is actually concave?
The answer turns out to be yes, for the stochastic setting (in the adversarial setting, cumulative welfare is necessarily non-concave). 
One reason is that concavity ensures that the function is unimodal. 
To estimate the difference in social welfare between two policies it therefore suffices to sample policies that lie in the interval between them.
These in-between policies yield social welfare exceeding the minimum of the two boundary policies. 
A second reason is that concavity prevents unexpected spikes in social welfare. This property allows us to test carefully chosen triples of points for extended periods, to ensure that one of them is suboptimal, without incurring significant regret. 

For the stochastic setting with concave social welfare, we present an algorithm that achieves a bound on regret of order $T^{1/2}$, up to logarithmic terms.
Before describing our proposed algorithm, \dyadic, let us formally state the improved regret bounds.
The proofs of these lower and upper bounds can be found in Online Appendix \ref{sec:onlineproofs}.

\begin{theo}[Lower bound on regret for the concave case]
    \label{theo:lower_concave}
    Consider the setup of Section \ref{sec:setup}.
    There exists a constant $C>0$ such that, for any randomized algorithm for the choice of $\pol_1, \pol_2, \dots$ and any time horizon $T \in \mathbb{N}$, the following holds:  

    There exists a distribution $\mu$ on $[0,1]$ with associated demand function $\demexp$ and \emph{concave} social welfare function $\swexp$, for which the stochastic cumulative expected regret $\regstoch$ is at least $C \cdot T^{1/2}$.
\end{theo}

\begin{theo}[Stochastic upper bound on regret of \dyadic]
    \label{theo:upper_concave}
    Consider the stochastic setup of Section \ref{sec:setup} and time horizon $T \in \mathbb{N}$. If Algorithm \ref{alg:dyadic} is run with confidence parameter $\delta = \frac{1}{T^{5/2}}$, 
    and if the social welfare function $\swexp$ is concave, then, the expected regret $\regstoch$ is of order at most $T^{1/2}$, up to logarithmic terms.
\end{theo}

\begin{algorithm}[t]
    \caption{\dyadic}
    \label{alg:dyadic}
    \begin{algorithmic}[1]
        \REQUIRE A confidence parameter $\delta \in (0,1)$.
        \STATE ${I}_1 = [0,1]$, $t_0 = 0$, $k = 0$
        \FOR{epochs $\tau=1,2,\dots$}
            \STATE Let $c= (\sup I_\tau + \inf I_\tau) / 2$, and $d=\sup I_\tau -  \inf I_\tau$.
            \hfill \COMMENT{Subinterval for sampling}
            \IF{$\tau$ is odd} 
            \STATE Let $l = c - \tfrac14 d$, $r = c + \tfrac14 d$.\\
             \ELSE
             \STATE Let $l = c - \tfrac16 d$, $r = c + \tfrac16 d$.
            \ENDIF
            \FOR{$t=t_{\tau-1}+1,t_{\tau-1}+2,\dots$}
                \STATE Select $w \in \argmax_{w' \in \{l,c,r, (l,c), (c,r) \}} \Gamma_{t-1}(w')$,\hfill \COMMENT{Sampling}\\ 
                \quad breaking ties following the order $l,c,r,(l,c),(c,r)$ 
                
                \IF{$w \in \{l,c,r\}$}
                        \STATE Set $\pol_t = w$.
                \ELSE
                \STATE Set $\pol_t = w_1 + (w_2-w_1) \cdot \frac{k+\nicefrac{1}{2}}{ n_{t-1}(w_1,w_2) + 1 }$, and
                $k = (k + 1) \mod n_{t-1}(w_1,w_2)+1$.
                \ENDIF

                \STATE Calculate $J_t(l,c)$, $J_t(c,r)$,  and $J_t(l,r)$, as in Equations \eqref{eq:CI} and \eqref{eq:CI_bis}.\\
                \hfill \COMMENT{Inference}
            
                \IF{$\inf \brb{J_t(l,c)} \ge 0$  or  $\inf \brb{J_t(l,r)} \ge 0$}
                    \STATE let $ {I}_{\tau+1} = {I}_{\tau} \cap [l , 1 ]$ and $t_\tau = t$ and \textbf{break} \hfill \COMMENT{Shrinking the active interval}
                \ELSIF{$\sup\brb{J_t(c,r)} \le 0$  or  $\sup\brb{J_t(l,r)}\le 0$}
                    \STATE let $ {I}_{\tau+1} = {I}_{\tau} \cap [ 0 , r ]$ and $t_\tau = t$ and \textbf{break}
                \ENDIF
            \ENDFOR
        \ENDFOR 
    \end{algorithmic}
\end{algorithm}

\paragraph{Dyadic search}
Our algorithm is based on a modification of dyadic search, as discussed in \citep{bachoc2022aNearOptimal,bachoc2022regret}.
At any point in time, this algorithm maintains an active interval $I_\tau$, which contains the optimal policy with high probability.
Only policies within this interval are sampled going forward.
As evidence accumulates, this interval is trimmed down, by excluding policies that are sub-optimal with high probability.

The algorithm proceeds in epochs $\tau$.
At the start of each epoch, a sub-interval $[l,r] \subset I_\tau$ is formed, with mid-point $c = (l+r) / 2$. The points $l,c,r$ are in a dyadic grid, that is, they are of the form $k/2^m$.
After sampling from $[l,r]$, we calculate confidence intervals $J_t(l,c)$, $J_t(c,r)$, and $J_t(l,r)$ for the welfare differences 
$\Delta(l,c)$, $\Delta(c, r)$, and $\Delta(l, r)$, where $\Delta(\pol,\pol') = \swexp(\pol') - \swexp(\pol)$.

If the confidence interval $J_t(l,c)$ or $J_t(l,r)$ lies above $0$, concavity implies that the optimal policy cannot lie to the left of $l$; we can thus trim the active interval $I_\tau$ by dropping all points to the left of $l$.
Symmetrically, if the confidence interval $J_t(c,r)$ or $J_t(l,r)$ lies below $0$, we can trim $I_\tau$ by dropping all points to the right of $r$.

\paragraph{Confidence intervals for welfare differences}
This procedure requires the construction of confidence intervals for welfare differences of the form
\be
  \Delta(\pol,\pol') = \swexp(\pol') - \swexp(\pol) = \pol' \cdot \demexp(\pol') - \pol \cdot \demexp(\pol) - \lambda \int_{\pol}^{\pol'} \demexp(\pol'') \mathrm{d}\pol''.
\ee
At time $t$, we estimate demand $\demexp(\pol)$, for policies $\pol$ chosen in previous periods, as\footnote{We use the convention $0/0 = 0$ and $a / 0 = +\infty$ whenever $a>0$. Furthermore, every summation over an empty set of indices is understood to have value $0$.}
\bals
\widehat \demexp_t(\pol) &= \frac{1}{n_t(\pol)} \sum_{i \leq t} \out_i  \cdot \bs 1(\pol_i = \pol), &
n_t(\pol) &= \sum_{i \leq t} \bs 1(\pol_i = \pol).
\eals
We similarly estimate integrated demand $\int_{\pol}^{\pol'} \demexp(\pol'') \mathrm{d}\pol''$ by $(\pol' - \pol)$ times the average of realized demand $\out_i$ for observations $\pol_i$ in the open interval $(\pol, \pol')$. We have to be careful, however, to use a sample of $\pol_i$ that is (approximately) uniformly distributed over this interval. This can be  achieved for our dyadic search procedure, as specified in Algorithm \ref{alg:dyadic}, by truncating the time index used to estimate this average.\footnote{The sampling procedure in Algorithm \ref{alg:dyadic} samples sequentially from the dyadic grid in the active interval, refining the grid in subsequent iterations.
$s(\pol,\pol',t)$ provides a truncation of the time index such that one round of such dyadic sampling has been completed.}
Let 
$$s(\pol,\pol',t) =  \max \left\{s \leq t:\; \log_2\left(1 + \sum_{i\leq s} \bs 1(\pol_i \in (\pol, \pol'))\right) \in \mathbb{N}\right\}.$$
We define
\bals
\widehat \demexp_t(\pol, \pol') &= \frac{1}{n_t(\pol, \pol') + 1} \sum_{i \le s(\pol,\pol',t)} \out_i  \cdot \bs 1(\pol_i \in (\pol, \pol')), &
n_t(\pol, \pol') &= \sum_{i \le s(\pol,\pol',t)} \bs 1(\pol_i \in (\pol, \pol')).
\eals
At each round, Algorithm \ref{alg:dyadic} maintains estimates for welfare differences among three points $l,c,r$ (for left, center and right, respectively).
The estimate of the welfare difference between $\pol'=c$ and $\pol = l$ (or between $\pol'=r$ and $\pol = c$) is given by
\be
    \widehat \Delta_t(\pol,\pol') = \pol' \cdot \widehat \demexp_t(\pol') - \pol \cdot \widehat \demexp_t(\pol) - \lambda  \cdot (\pol' - \pol)  \cdot \widehat \demexp_t(\pol, \pol').
\ee
while the estimate of the welfare difference between $r$ and $l$ is given by
\be
    \widehat \Delta_t(l,r) = \widehat \Delta_t(l, c) + \widehat \Delta_t(c, r).
\ee
To construct confidence intervals for $\Delta(\pol,\pol')$, we also need to quantify the uncertainty of our demand estimates. 
We use the following interval half-lengths for confidence intervals for tax revenue at $\pol$, and for the private welfare difference between $\pol'$ and $\pol$:
\bals
  \Gamma_t(\pol) &= \pol \cdot \sqrt{ \tfrac{1}{ 2 n_t(\pol)} \log \lrb{\tfrac{2}{\delta}}},
  & \Gamma_t(\pol,\pol') &=  \lambda  \cdot (\pol' - \pol)  \cdot \lrb{\sqrt{ \tfrac{1}{ 2 \brb{n_t(\pol, \pol') + 1}} \log \lrb{\tfrac{2}{\delta}}} + \tfrac{2}{ n_t(\pol, \pol') + 1}} . 
\eals
Using the shorthand $a\pm b = [a-b,a+b]$, our confidence interval for $\Delta(\pol,\pol')$, where $\pol'=c$ and $\pol=l$ (or $\pol'=r$ and $\pol=c$) is given by
\be
    J_t(\pol,\pol') =
    \widehat \Delta_t(\pol,\pol') \pm \left(\Gamma_t(\pol') +  \Gamma_t(\pol) +  \Gamma_t(\pol,\pol') \right),
    \label{eq:CI}
\ee
while our confidence interval for $\Delta(l,r)$ is given by
\be
    J_t(l,r) =
    \widehat \Delta_t(l,r) \pm \left(\Gamma_t(r) +  \Gamma_t(l) +  \Gamma_t(l,c) + \Gamma_t(c,r) \right).
    \label{eq:CI_bis}
\ee
With these preliminaries, we are now ready to state our algorithm, \dyadic, in Algorithm \ref{alg:dyadic}.

Before concluding this section, we highlight two features of Algorithm \ref{alg:dyadic}. 
First, two of the three points $l,c,r$, and the corresponding estimates of demand, are kept from each epoch to the next. 
Second, estimation of the integral term is performed by querying points following a fixed and balanced design on the dyadic grid -- instead of, for example, using a randomized Monte Carlo procedure which may lead to unbalanced exploration. 
This implies that the points queried to estimate the integral terms can be easily reused to obtain other integral estimates from each epoch to the next. These two features combined ensure that Algorithm \ref{alg:dyadic} recycles information very efficiently to prune the active interval as quickly as possible, which leads to better regret.

\section{Income taxation}
\label{sec:income_taxation}

We discuss two extensions of the baseline model of optimal taxation that we introduced in Section \ref{sec:setup}.
These extensions incorporate features that are important in more realistic models of optimal taxation.
For both of these extensions, we propose a properly modified version of Algorithm \ref{alg:tempered_exp3}.
The first extension, discussed in this section, is a variant of the Mirrlees model of optimal income taxation \citep{Mirrlees1971, Saez2001, saez2002optimal}. 
The second extension, discussed in Section \ref{sec:commodity_taxation} is a variant of the Ramsey model of commodity taxation \citep{ramsey1927contribution}.

Our model of income taxation generalizes our baseline model by allowing for heterogeneous wages $\wage_i$,
welfare weights $\weight(\wage_i)$, 
extensive-margin labor supply responses determined by the cost of participation $\wtp_i$, 
and non-linear income taxes $\pol_i = \mathbf{\pol}(\wage_i)$.
Two simplifications are maintained in this model, relative to a more general model of income taxation. First, only extensive margin responses (participation decisions) by individuals are allowed; there are no intensive margin responses (hours adjustments).
Second, as in the baseline model of Section \ref{sec:setup}, there are no income effects. 
In imposing these assumptions, our model mirrors the model of optimal income taxation discussed in Section II.2 of \cite{saez2002optimal}.

\paragraph{Setup}
At each time $i = 1,2, \ldots, T$, one individual arrives who is characterized by
(i) a potential wage $\wage_i \in [0,1]$, and
(ii) an unknown cost of participation $\wtp_i \in [0,1]$.
This individual makes a binary labor supply decision $\out_i$.
If they participate in the labor market ($\out_i=1$), they earn $\wage_i$, but pay a tax according to the tax rate $\pol_i = \mathbf{\pol}(\wage_i)$ on their earnings $\wage_i$. They furthermore incur a non-monetary cost of participation $\wtp_i$.

Their optimal labor supply decision is therefore given by 
$\out_i = \bs 1( \wtp_i\leq \wage_i \cdot (1 - \pol_i))$, and private welfare equals $\max(\wage_i\cdot (1 - \pol_i) - \wtp_i , 0)$.
The implied public revenue is equal to the tax on earnings $\pol_i\cdot \wage_i$ if $\out_i =1$, and $0$ otherwise.

We define social welfare as a weighted sum of public revenue and private welfare, with a weight $\weight(\wage_i)$ for the latter.
Typically, $\weight$ is a decreasing function of $\wage$, reflecting a preference for redistribution towards those with lower earnings potential, cf. \cite{saez2013generalized}.
Social welfare for time period $i$, as a function of the tax schedule $\mathbf{\pol}( \cdot )$, is therefore given by
\be
  \sw_i(\mathbf{\pol}(\cdot)) = \underbrace{\mathbf{\pol}(\wage_i)\cdot \wage_i  \cdot \bs 1( \wtp_i \leq \wage_i  \cdot (1 - \mathbf{\pol}(\wage_i)))}_{\textrm{Public revenue}}\quad + \quad \weight(\wage_i)  \cdot \underbrace{\max(\wage_i  \cdot (1  - \mathbf{\pol}(\wage_i)) - \wtp_i  , 0)}_{\textrm{Private welfare}}.
\ee
After period $i$, we observe $\out_i$ and the tax schedule $
\mathbf{\pol}_i( \cdot )$.
If $\out_i = 1$, we also observe $\wage_i$. Nothing else is observed.\footnote{It should be noted that in this model we take the transfer $\pol_0$ for individuals without other income as given. The effective tax owed by an employed individual equals $\mathbf{\pol}(\wage_i)\cdot \wage_i-\pol_0$. The ``unconditional basic income'' $\pol_0$ does not affect labor supply, given our assumption that there are no income effects, and it enters social welfare additively. It is therefore without loss of generality to omit $\pol_0$ from our model.}

\paragraph{Piecewise constant tax schedules}
We next construct a generalization of Algorithm \ref{alg:tempered_exp3} based on piecewise constant tax schedules, with tax rates changing at the grid-points $\mathcal W$, where $0 \in \mathcal{W}\subset [0,1]$.
Formally, define $\left\lfloor \wage\right\rfloor  = \max \{w' \in \mathcal{W}:\; w'\leq \wage\} $, rounding the wage $\wage$ down to the nearest grid-point in $\mathcal W$,\footnote{Here we use slightly non-standard notation, where $\left\lfloor  \cdot \right\rfloor $ denotes rounding down to the nearest grid-point, rather than the nearest integer.} 
Denote $H = |\mathcal W|$, and let
$$
  \mathcal {X_W} = \{\mathbf{\pol}( \cdot ):\;\forall \wage \in [0,1],\; \mathbf{\pol}(\wage) = \mathbf{\pol}(\left\lfloor \wage\right\rfloor)\}.
$$
For $\wage \in \mathcal W$ and any $\pol \in [0,1]$, denote
$$
  \dem_i(\wage, \pol)  = \wage_i \cdot  \bs  1( \wtp_i\leq \wage_i \cdot (1 - \pol)) \cdot \bs 1(\left\lfloor \wage_i\right\rfloor = \wage),
$$
so that $\out_i \cdot \wage_i = \dem_i(\wage_i, \mathbf{\pol}_i(\wage_i))$. $\dem_i(\wage, \pol)$ is the individual labor supply function, in monetary units, interacted with an indicator for whether the wage $\wage_i$ falls into the tax bracket starting at $\wage$.
With this notation, we can rewrite
$$
  \max(\wage_i\cdot (1 - \pol) - \wtp_i , 0) = 
  \int_{\pol}^{1} \dem_i(\left\lfloor \wage_i\right\rfloor, \pol') \mathrm{d}\pol'.
$$
For piecewise constant tax rates $\mathbf{\pol}( \cdot )$ we get
\bal
  \sw_i(\mathbf{\pol}(\cdot)) &= 
  \sum_{\wage \in \mathcal W} 
  \left [\mathbf{\pol}(\wage)  \cdot \dem_i(\wage, \mathbf{\pol}(\wage)) + 
  \weight(\wage_i) \cdot \int_{\mathbf{\pol}(\wage)}^{1} \dem_i(\wage, \pol') \mathrm{d}\pol'\right ].
  \label{eq:swf_income}
\eal
Cumulative social welfare is given by $\swcum_i= \sum_{j\leq i} \sw_i(\mathbf{\pol}_i( \cdot ))$, and we correspondingly define cumulative expected regret, in the adversarial setting, as
$$
  \mathcal R_T = \sup_{\mathbf{\pol}( \cdot ) \in \mathcal{X_W}} E\left [ \swcum_T(\mathbf{\pol}( \cdot ))- \swcum_T \Big| \{\wtp_i\}_{i=1}^T, \{\wage_i\}_{i=1}^T \right ].
$$
The supremum here is taken over all tax schedules $\mathbf{\pol}( \cdot )$ that are piecewise constant between the gridpoints $\wage \in \mathcal W$.

\paragraph{Algorithm}
Algorithm \ref{alg:tempered_exp3_income} generalizes Algorithm \ref{alg:tempered_exp3} to this setting.
As before, we form an unbiased estimate $\widehat \dem_i$ of $\dem_i$ using inverse probability weighting, map this estimate into a corresponding estimate $\widehat \sw_i$ of $\sw_i$, based on Equation \eqref{eq:swf_income}, and cumulate across time periods to obtain $\widehat \swcum_i$. 
Note that $\wage_i$ is observed whenever $\out_i = 1$.
This implies that the estimate $\widehat \dem_i$ is in fact a function of observables, and the same holds for $\widehat \sw_i$.

Algorithm \ref{alg:tempered_exp3_income} keeps track of estimated demand and social welfare for each bin (``tax bracket''), as defined by the gridpoints $\wage \in \mathcal W$.
The algorithm then constructs a distribution $p_i(\pol| \wage)$ over tax rates $\pol \in \mathcal X$ given $\wage$, using the tempered Exp3 distribution.
The tax schedule $\mathbf{\pol}( \cdot )$ is sampled according to these (marginal) distributions of tax rates for each bracket.
Though immaterial for the following theorem, we choose the perfectly correlated coupling, across brackets, of these marginal distributions, which is implemented using the random variable $A_i$ in Algorithm \ref{alg:tempered_exp3_income}.

\begin{algorithm}[t]
  \caption{Tempered Exp3 for Optimal Income Taxation}
  \label{alg:tempered_exp3_income}
\begin{algorithmic}[1]
  \REQUIRE Tuning parameters $K$, $\gamma$ and $\eta$, and set of gridpoints $\mathcal W \subset [0,1]$.
  
  \STATE   Calculate evenly spaced grid-points $\mathcal X = \{0, \tfrac1K,\tfrac2K,\ldots,1\}$.
  
  \STATE Initialize $\widehat \demcum_{1}(\wage, \pol) = 0$ 
  and $\widehat \swcum_{1}(\wage, \pol) = 0$ 
  for all $\wage \in \mathcal X$ and all $\pol \in \mathcal X$.

  \FOR{individual $i=1,2,\ldots, T$}
     
    \STATE For all $\pol, \wage \in \mathcal X$, set  $\left\lfloor \wage\right\rfloor = \max \{\wage' \in \mathcal{W}:\; \wage'\leq \wage\} $, and\\ \hfill \COMMENT{Assignment probabilities}
      \be   
        p_{i}(\pol | \wage) = (1-\gamma)  \cdot 
        \frac{ \exp(\eta \cdot  \widehat \swcum_{i}(\pol, \left\lfloor \wage\right\rfloor))}{\sum_{\pol' \in \mathcal X} \exp(\eta  \cdot \widehat \swcum_{i}(\pol', \left\lfloor \wage\right\rfloor))} 
        + \frac{\gamma }{K + 1}.
      \ee

    \STATE Draw $A_i \sim U[0,1]$.
    For all $\wage \in [0,1]$, set 
      \be
        \mathbf{\pol}_i(\wage) = \max \left \{\pol \in \mathcal X: 
        \sum_{\pol'\in \mathcal X, \pol'<\pol} p_{i}(\pol' | \wage) \leq A_i \right \},
      \ee    
      and query $\out_i$ accordingly.
    \STATE For all $\wage \in \mathcal W$ and $\pol \in \mathcal X$, set \hfill \COMMENT{Estimated labor supply}
      \be
        \widehat \dem_i(\pol, \wage) = \out_i \cdot \wage_i  \cdot \frac{\bs 1(\left\lfloor \wage_i\right\rfloor = \wage, \mathbf{\pol}_i(\wage_i) = \pol)}{p_{i}(\pol | \wage)}.
      \ee
    \STATE For all $\wage \in \mathcal W$ and $\pol \in \mathcal X$, set \hfill \COMMENT{Estimated welfare}
    \be
      \widehat \swcum_{i+1}(\pol, \wage) = \widehat \swcum_{i}(\pol, \wage)
      + \pol \cdot  \widehat \dem_i(\pol, \wage)
      + \frac{\weight(\wage_i)}{K}  \cdot \sum_{\pol' \in \mathcal X, \pol'>\pol } \widehat \dem_{i}(\pol', \wage).
    \ee
  \ENDFOR
  
\end{algorithmic}

\end{algorithm}

\begin{theo}[Adversarial upper bound on regret of Tempered Exp3 for Optimal Income Taxation]
  \label{theo:upper_exp3_income}
  Consider the setup of Section \ref{sec:income_taxation}, and Algorithm \ref{alg:tempered_exp3_income}.
  Assume that $(K+1) \eta < \gamma$, and that $\weight(\wage) \leq 1$ for all $\wage$.\\
  Then for any  sequence $(\wtp_1, \ldots, \wtp_T)$ expected regret $\regadv$ is bounded above by
  \be
    \left ( \gamma   + \eta  \cdot (e-2) 
     \tfrac{K+1}{K} \cdot \left (\tfrac{2K+1}{6} + \tfrac{1}{\gamma}  \right )
    + \tfrac{1}{K} \right ) \cdot T
    + \tfrac{H\log(K+1)}{\eta}.
    \label{eq:upper_exp3_sw}
  \ee
  Suppose additionally\footnote{for simplicity, we assume that in the following tuning $K$ is an integer. If not, round $K$ to the closest integer.} that
  $ K = c_1  \cdot  (T / H)^{1/3}$,
  $\gamma = c_2  /(K+1)$, and
  $\eta = c_3 / (K+1)^2$,  
  for some constants $c_1,c_2,c_3$.
  Then expected regret $\regadv$ is bounded above by 
  \be
    c_4 \cdot H^{1/3} \cdot  \log(T)^{1/3} T^{2/3},
  \ee
  for some constant $c_4$.
\end{theo}

\section{Commodity taxation}
\label{sec:commodity_taxation}

In this section, we generalize our baseline model of optimal taxation to a model of commodity taxation with multiple goods $j \in \{1,\ldots,k\}$ and continuous demand functions $\out_i(\pol) \in [0,1]^k$, where $\pol \in [0,1]^k$ is a vector of tax rates.
We again assume that there are no income effects. Our setup is a version of the classic Ramsey model \citep{ramsey1927contribution}.
We propose a generalization of \algexp~to this setting. 
In the following, we use $\langle \pol, \out \rangle$ to denote the Euclidean inner product between $\pol$ and $\out$.

\paragraph{Setup}
At each time $i = 1,2, \ldots, T$, one individual arrives who is characterized by
a utility function $u_i: [0,1]^k \rightarrow \mathcal R$.  
This individual is exposed to a tax vector $\pol_i \in [0,1]^k$, and makes a continuous consumption decision $\out_i$.
Public revenue is given by $\langle \pol_i, \out_i \rangle$.
Private utility is given by $u_i(\out_i)$ plus their consumption of a numeraire good, which has price normalized to $1$ and enters utility additively.
The individual consumption choice $\out_i$ costs $\langle \pol_i + p, \out \rangle$, where $p$ is the (exogenously given) vector of pre-tax prices.
This cost of purchasing $\out_i$ reduces the consumption of the numeraire good.
The optimal individual decision is therefore given by
\be
  \out_i = \dem_i(\pol_i)=  
  \argmax_{\out \in [0,1]^k} \left [  u_i(\out) - \langle \pol_i + p, \out \rangle  \right]. 
\ee  
The implied private welfare is 
$$
  v_i(\pol) = v_0 + \max_{\out \in [0,1]^k} \left [ u_i(\out) - \langle \pol + p, \out \rangle \right ],
$$
where we have added a constant $v_0$, chosen such that $v_i(0) = 0$; this is just a normalization to simplify notation below.

We define social welfare as a weighted sum of public revenue and private welfare, with a weight $\lambda$ for the latter.
Social welfare for time period $i$, as a function of the tax vector $\pol$, is therefore given by
\be
  \sw_i(\pol_i) = \underbrace{\langle \pol_i, \out_i \rangle}_{\textrm{Public revenue}}\quad + \quad \lambda  \cdot \underbrace{v_i(\pol_i)}_{\textrm{Private welfare}}.
\ee
After period $i$, we observe $\out_i$ and the tax vector $\pol_i$.
Nothing else is observed.

\paragraph{Mapping demand to welfare}
By the envelope theorem \citep{milgrom2002envelope}, 
$$
  \nabla_\pol v_i(\pol) = \dem_i(\pol).
$$
Let $\mathcal V$ be the set of differentiable functions $v$ on $[0,1]^k$ such that $\nabla_\pol v \in L^2$, and such that $v(0)=0$.
Consider the following operator, mapping the demand function $\dem$ into the corresponding indirect utility function $v$.
\be
  \Pi (\dem( \cdot )) \in \argmin_{v( \cdot ) \in \mathcal V}
  \int_{[0,1]^k} \left \| \nabla_\pol v(\pol) - \dem(\pol) \right \|^2 \mathrm{d}\pol
\ee
We can think of the operator $\Pi$ as combining two operators.
First, the function $\dem$ is projected on the subspace of functions on $[0,1]^k$ which can be written as the gradient of some function $v$.
Second, the projected $\dem$ is then integrated to get $v(\pol)$ for any $\pol$. Integration here is along some curve in $[0,1]^k$ from $0$ to $\pol$. Given the first projection, the choice of curve does not matter for the resulting function $v$.
A formal analysis of Tempered Exp3 for Commodity Taxation would need to prove existence of the projection. We leave such a formal analysis, including lower and upper regret bounds, for future research.

\begin{algorithm}[t]
  \caption{Tempered Exp3 for Commodity Taxation}
  \label{alg:tempered_exp3_commodity}
\begin{algorithmic}[1]
  \REQUIRE Tuning parameters $K$, $\gamma$ and $\eta$.
  
  \STATE Calculate the set of evenly spaced grid-points 
  $\mathcal X = \{0,\tfrac1K,\ldots,1\}^k$\\
  and initialize $\widehat \demcum_{1}(\pol) = 0$ for all grid points.

  \FOR{individual $i=1,2,\ldots, T$}
    
    \STATE For all $\pol \in \mathcal X$, set \hfill \COMMENT{Estimated welfare}
    \be
      \widehat \swcum_{i}(\pol) = \langle \pol_i, \widehat \demcum_i\rangle + \lambda  \cdot \widehat v_i(\pol_i).
    \ee  
    
    \STATE For all $\pol \in \mathcal X$, set   \hfill \COMMENT{Assignment probabilities}
    \be      
      p_{i} =(1-\gamma)  \cdot \frac{ \exp(\eta \cdot  \widehat \swcum_{i}(\pol))}{\sum_{\pol'}\exp(\eta  \cdot \widehat \swcum_{i}(\pol'))} + \frac{\gamma }{(K+1)^k}.
    \ee

    \STATE Choose $\pol_i$ at random according to the probability distribution $p_i$, and query $\out_i$ accordingly.
      
    \STATE For all $\pol \in [0,1]^k$, set \hfill \COMMENT{Estimated demand}
      \bal
        \widetilde \demcum_{i+1}(\pol)&= \widehat \demcum_{i}(\pol) + y_i \cdot \frac{\bs 1(\pol_i = \lfloor \pol \rfloor)}{p_{i}}\\
        \widehat v_{i+1}(\pol) &= \Pi(\widetilde \demcum_{i+1})\\
        \widehat \demcum_{i+1}(\pol) &= \nabla_\pol \widehat v_{i+1}(\pol).
      \eal
      
  \ENDFOR
  
\end{algorithmic}
\end{algorithm}

\section{Conclusion}
\label{sec:conclusion}

\paragraph{Possible applications}
The setup introduced in Section \ref{sec:setup} was deliberately stylized, to allow for a clear exposition of the conceptual issues that arise when adaptively maximizing social welfare.
The algorithm that we proposed for this setup, and the generalizations discussed later in the paper, are nonetheless directly practically relevant. They remain appropriate in economic settings that are considerably more general than the setting described by our model.

The reasons for this generality have been elucidated by the public finance literature, cf. \cite{chetty09suff}, building on the generality of the envelope theorem, cf. \cite{milgrom2002envelope,sinander2022converse}.
By the envelope theorem, the welfare impact of a marginal tax change on private welfare can be calculated ignoring any behavioral responses to the tax change.
This holds in generalizations of our setup that allow for almost arbitrary action spaces (including discrete and continuous, multi-dimensional, and dynamic actions), and for arbitrary preference heterogeneity.
The expressions for social welfare that justify our algorithms remain unchanged under such generalizations.
That said, the validity of these expressions does require the absence of income effects and of externalities.
If there are income effects or externalities, the algorithms need to be modified.

Our approach is motivated by applications of algorithmic decisionmaking for public policy, where a policymaker cares about welfare, but also faces a government budget constraint.
Possible application domains of our algorithm include the following.
In \textit{public health and development economics}, field experiments such as \cite{cohen2010free} vary the level of a subsidy for goods such as insecticide-treated bed nets (ITNs), estimating the impact on demand. Our algorithm could be used to find the optimal subsidy level quickly and apply it to experimental participants. A term capturing positive externalities of the use of ITNs could be added to social welfare, leaving the algorithm otherwise unchanged.
In \textit{educational economics}, many studies evaluate the impact of financial aid on college enrollment \citep{dynarski2023college}.
An adaptive experiment might vary the level of aid provided, where aid is conditional on college attendance and conditional on pre-determined criteria of need or merit.
In such an experiment, a variant of our algorithm for optimal income taxation might be used, where the welfare weights $\weight$ are a function of need or merit, and the outcome $\out$ is college attendance.
In \textit{environmental economics}, many experiments (e.g., \citealt{lee2020experimental}) study the impact of electricity pricing on household electricity consumption. Once again, our baseline algorithm (for binary household decisions about connecting to the grid) or our algorithm for commodity taxation (for continuous household decisions about consumption levels) could be applied, to learn optimal prices, taking into account both distributional considerations and externalities.

These examples are all drawn from public policy, where there is an intrinsic concern for social welfare.
This contrasts with commercial applications, where the goal is typically to maximize (directly observable) profits by monopolist pricing \citep{den2015dynamic}, or more generally by reserve price setting in auctions \citep{nedelec2022learning}. Adaptive pricing algorithms are used in applications such as online ad auctions.
A concern for welfare might enter in such commercial settings if there is a participation constraint that needs to be satisfied for consumers.
Suppose for example that consumers or service providers need to first sign up for a platform, say for e-commerce or for gig work, and then repeatedly engage in transactions on this platform.
To sign up in the first place, their expected welfare needs to exceed their outside option. This constraint might then enter the platform provider's objective, in Lagrangian form, adding a term for welfare, and leading to objectives such as those maximized by our algorithms.

\paragraph{An alternative approach: Thompson sampling}
The main algorithm proposed in this paper, \algexp, is designed to perform well in the adversarial setting.
In the construction of this algorithm, no probabilistic assumptions were made about the distribution of $\wtp_i$.
In the stochastic framework, a sampling distribution is assumed, for instance that the $\wtp_i$ be i.i.d. over time.
The Bayesian framework completes this by assuming a prior distribution over the parameters which govern the sampling distribution.

One popular heuristic for adaptive policy choice in the Bayesian framework is Thompson sampling \citep{thompson1933likelihood, Thompson2018}, also know as probability matching, which assigns a policy with probability equal to the posterior probability that this policy is optimal.
In our setting, Thompson sampling could be implemented as follows.
First, form a posterior for the demand function $\demexp(\pol) = E[\out| \pol]$, based on all the data available from previous periods $j < i$. Sample one draw $\widetilde \demexp( \cdot )$ from this posterior. Map this draw into a draw $\widetilde \swexp( \cdot )$ from the posterior for $\swexp( \cdot )$ via
$\widetilde \swexp(\pol) = \pol \cdot  \widetilde \demexp(\pol) + \lambda  \cdot \int_\pol^1 \widetilde \demexp(\pol') \mathrm{d} \pol'$.
Find the maximizer $\pol_i = \argmax_\pol \widetilde \swexp(\pol)$. This is the policy recommended by Thompson sampling.
We conjecture that this algorithm will outperform random assignment, but will under-explore relative to the optimal algorithm. 
Adding further forced exploration to this algorithm might improve cumulative welfare.
A formal analysis of algorithms of this type is left for future research.

A natural class of priors for $\demexp$ are Gaussian process priors \citep{williams2006gaussian}. If outcomes $\out$ are conditionally normal (rather than binary, as in our baseline model), then the posterior for demand is available in closed form, and the posterior mean is equal to the best linear predictor given past outcomes $\out_j$. Furthermore, since social welfare is a linear transformation of demand, the posterior for $\swexp$ is then also linear and available in closed form. For details, see \cite{kasy2017taxation}.

\bibliographystyle{apalike}
\bibliography{adaptive_social_welfare,ncb}

\appendix
\section{Proofs}
\label{sec:proofs}

\subsection{Theorem \ref{theo:lower_stochastic} (Lower bound on regret)}

\paragraph{Defining a family of distributions for $\wtp$}
Recall that, for each $\epsilon \in [-1,1]$, the probability distribution $\mu^\epsilon$ is defined as the probability measure supported on $(\nicefrac{1}{4}, \nicefrac{1}{2}, \nicefrac{3}{4}, 1)$ with masses $\big( a, (1+\epsilon) \cdot b, (1-\epsilon) \cdot b, 1 - a - 2 \cdot b \big)$, where
\begin{equation*}
        a \coloneqq \frac{(1-\lambda) \cdot (136 - 99\cdot \lambda)}{2 \cdot (4-3\cdot\lambda)\cdot(24-17\cdot\lambda)}\;, \qquad
        b \coloneqq \frac{1-\lambda}{2 \cdot(24 - 17 \cdot \lambda)}\;.
    \end{equation*} 
Furthermore, for each $\epsilon \in [-1,1]$, recall that $\demexp^\epsilon$ and $\swexp^\epsilon$ are respectively the demand function and the expected social welfare associated to $\mu^\epsilon$ (see Figure \ref{fig:lower_bound} for an illustration).
     Let $v_1, v_2, \dots \in [0,1]$ be the sequence of individual valuations.
For each $\epsilon \in [-1,1]$, consider a distribution $P^\epsilon$ such that the individual valuations $v_1, v_2, \dots$ form a $P^\epsilon$-i.i.d. sequence (independent of the randomization used by the algorithm) with common distribution $\mu^\epsilon$.

\paragraph{Explicit lower bound on regret that will be proven}
Define
\begin{equation*}
	c_1 \coloneqq \frac{\lambda}{4} \cdot b\;, \quad
        c_2 \coloneqq \frac{1}{8} \cdot \frac{1-\lambda}{4 - 3 \cdot \lambda}\;, \quad
        c_3 \coloneqq b \cdot \sqrt{\frac{2}{a \cdot (1 - a - 2 \cdot b)}}\;.
\end{equation*}
We will prove that, for any randomized algorithm and any time horizon $T \in \mathbb{N}$, there exists $\epsilon \in [-1,1]$ such that
\begin{equation*}
	\mathcal{R}_T(\demexp^\epsilon) \ge C\cdot T^{2/3} \;,
\end{equation*}
where
\bal
\label{eq:constants}
	C &\coloneqq
	\min \bigg( \frac{c_1^2 \cdot c_3^2}{c_2},\; \frac{c_2}{2},\; \frac{1}{16} \cdot \sqrt[3]{\frac{c_1^2 \cdot c_2}{c_3^2}} \bigg)
\\
&=\min \bigg( \frac{\lambda^2 \cdot (4 - 3 \cdot \lambda)^3}{8 \cdot (136-99 \cdot \lambda) \cdot ( 26 - 19 \cdot \lambda  )} , \;
	 \frac{\lambda^{\nicefrac{2}{3}} \cdot (1-\lambda)^{\nicefrac{4}{3}} \cdot (136 - 99 \cdot \lambda)^{\nicefrac{1}{3}} \cdot (26 - 19 \cdot \lambda)^{\nicefrac{1}{3}}}{128 \cdot (4 - 3 \cdot \lambda) \cdot (24 - 17 \cdot \lambda)^{\nicefrac{4}{3}}} \bigg)  > 0 \nonumber
\eal

Fix a randomized algorithm to choose the policies $x_1,x_2,\dots$, and fix a time horizon $T \in \mathbb{N}$.

\paragraph{Number of mistakes and lower bound on regret}
We need to count the random number of times the algorithm has played in the regions $(\nicefrac{1}{2}, \nicefrac{3}{4}]$, $[0, \nicefrac{1}{2}]$ and $(\nicefrac{3}{4}, 1]$ up to time $T$. This can be done relying on the following random variables:
\begin{equation*}
   n_1 \coloneqq \sum_{i=1}^T \bs 1_{(\nicefrac{1}{2}, \nicefrac{3}{4}]} ( x_i ) \;, \qquad
   n_2 \coloneqq \sum_{i=1}^T \bs 1_{[0, \nicefrac{1}{2}]} ( x_i ) \;, \qquad
   n_3 \coloneqq \sum_{i=1}^T \bs 1_{(\nicefrac{3}{4}, 1]} ( x_i ) \;. \qquad
\end{equation*}
Notice that since the intervals $(\nicefrac{1}{2}, \nicefrac{3}{4}], [0, \nicefrac{1}{2}]$ and $(\nicefrac{3}{4}, 1]$ form a partition of $[0,1]$, we have that
\begin{equation}
	\label{eq:counters}	
	n_1 + n_2 + n_3 = T
\end{equation}
For each $\epsilon \in [-1,1]$, denote by $E^\epsilon$ the expectation taken with respect to the distribution $P^\epsilon$.
	Notice that, for each $\epsilon \in [-1,1]$, the expected regret when the underlying distribution is $P^\epsilon$ equals
\begin{equation}
\label{eq:explicit_regret}
   \mathcal{R}_T(\demexp^\epsilon)
   = T \cdot \sup_{x \in [0,1]} \swexp^\epsilon (x) - \sum_{i=1}^T E^\epsilon \big( \swexp^\epsilon(x_i) \big) \;.
\end{equation}
	Algebraic calculations show that, for each $\epsilon \in [-1,1]$
\begin{equation}
\label{eq:algebraic_calculation_1}
	\max\limits_{x \in (\nicefrac{1}{2}, \nicefrac{3}{4}] } \swexp^\epsilon (x) = \swexp^\epsilon (\nicefrac{3}{4}) \;,\quad	\max\limits_{x \in [0, \nicefrac{1}{2}] } \swexp^\epsilon (x) = \swexp^\epsilon (\nicefrac{1}{4}) \; ,\quad \max\limits_{x \in (\nicefrac{3}{4}, 1] } \swexp^\epsilon (x) = \swexp^\epsilon (1) \;,
\end{equation}
\begin{equation}
\label{eq:algebraic_calculation_2}
	\qquad \text{and} \qquad \swexp^\epsilon (1) - \swexp^\epsilon (\nicefrac{1}{4}) = c_1 \cdot \epsilon\;.  \qquad\qquad
\end{equation}
Further calculations show also that
\begin{equation}
\label{eq:algebraic_calculation_3}
	\min\limits_{\epsilon \in [-1,1] }\min \brb{ \swexp^\epsilon (\nicefrac{1}{4}), \swexp^\epsilon (1) } = \swexp^1 (\nicefrac{1}{4}) \;, \quad
	\max\limits_{\epsilon \in [-1,1] }\max\limits_{x \in (\nicefrac{1}{2}, \nicefrac{3}{4}]} \swexp^\epsilon (x) = \swexp^{-1} (\nicefrac{3}{4})\;,
\end{equation}
\begin{equation}
\label{eq:algebraic_calculation_4}
	\qquad \text{and} \qquad \swexp^1 (\nicefrac{1}{4}) - \swexp^{-1} (3/4) = c_2 \;. \qquad\qquad
\end{equation}
Equations \eqref{eq:algebraic_calculation_1}, \eqref{eq:algebraic_calculation_2}, \eqref{eq:algebraic_calculation_3}, and \eqref{eq:algebraic_calculation_4} imply that
\begin{equation}
\label{eq:optimum_epsilon>0}
	\sup_{x \in [0,1]} \swexp^\epsilon (x) = \swexp^\epsilon(1) \;, \qquad \text{if } \epsilon \in [0,1] \;.
\end{equation}
It follows that, if $\epsilon \in [0,1]$,
\begin{align}
\label{eq:first_regret_bound}
	\mathcal{R}_T(\demexp^\epsilon)
&\overset{\eqref{eq:explicit_regret}}{=}
	T \cdot \sup_{x \in [0,1]} \swexp^\epsilon (x) - \sum_{i=1}^T E^\epsilon \big( \swexp^\epsilon(x_i) \big)
\nonumber
\\
&\overset{\eqref{eq:optimum_epsilon>0}}{=}
	T \cdot \swexp^\epsilon (1)
- \sum_{i=1}^T E^\epsilon \Big( \swexp^\epsilon(x_i) \cdot \big( \bs 1_{(\nicefrac{1}{2}, \nicefrac{3}{4}]} ( x_i ) + \bs 1_{[0, \nicefrac{1}{2}]} ( x_i ) + \bs 1_{(\nicefrac{3}{4}, 1]} ( x_i ) \big) \Big) 
\nonumber
\\
&\overset{\eqref{eq:algebraic_calculation_1}}{\ge}
	T \cdot \swexp^\epsilon (1) - \sum_{i=1}^T E^\epsilon \Big( \swexp^\epsilon(\nicefrac{3}{4}) \cdot \bs 1_{(\nicefrac{1}{2}, \nicefrac{3}{4}]} (x_i)
+ \swexp^\epsilon(\nicefrac{1}{2}) \cdot \bs 1_{[0, \nicefrac{1}{2}]} ( x_i ) + \swexp^\epsilon(1) \cdot \bs 1_{(\nicefrac{3}{4}, 1]}  ( x_i )  \Big) 
\nonumber
\\
&\overset{\eqref{eq:counters}}{=}
	\big( \swexp^\epsilon (1) - \swexp^\epsilon (\nicefrac{3}{4}) \big) \cdot E^\epsilon (n_1) + \big( \swexp^\epsilon (1) - \swexp^\epsilon (\nicefrac{1}{4}) \big) \cdot  E^\epsilon (n_2) 
\nonumber
\\
&\overset{\eqref{eq:algebraic_calculation_3}}{\ge}
	\big( \swexp^1 (\nicefrac{1}{4}) - \swexp^{-1} (\nicefrac{3}{4}) \big) \cdot E^\epsilon (n_1) + \big( \swexp^\epsilon (1) - \swexp^\epsilon (\nicefrac{1}{4}) \big) \cdot  E^\epsilon (n_2) 
\nonumber
\\
&\overset{\eqref{eq:algebraic_calculation_4}}{=}
	c_2 \cdot E^\epsilon(n_1) + \big( \swexp^\epsilon (1) - \swexp^\epsilon (\nicefrac{1}{4}) \big) \cdot  E^\epsilon (n_2)
\nonumber
\\
&\overset{\eqref{eq:algebraic_calculation_2}}{=}
	c_2 \cdot E^\epsilon(n_1) + c_1 \cdot \epsilon \cdot E^\epsilon (n_2) \;
\end{align}
Notice that inequality \eqref{eq:first_regret_bound} quantifies how much regret the algorithm is going to suffer in terms of the expected number of times it plays in the wrong regions, when the demand function is $\demexp^\epsilon$ and $\epsilon>0$. 

In the same way inequality \eqref{eq:first_regret_bound} was proven, we can prove that, if $\epsilon \in [0,1]$,
\begin{equation}
	\label{eq:second_regret_bound}	
	\mathcal{R}_T(\demexp^{-\epsilon})
\ge
	c_2 \cdot E^{-\epsilon}(n_1) + c_1 \cdot \epsilon \cdot E^{-\epsilon} (n_3) \ge c_1 \cdot \epsilon \cdot E^{-\epsilon} (n_3) \;,
\end{equation}
which again quantifies how much regret the algorithm is going to suffer in terms of the expected number of times it plays in the wrong regions, when the demand function is $\demexp^{-\epsilon}$ and $\epsilon>0$.

\paragraph{Intuition for the remainder of the proof}
At high level, inequalities \eqref{eq:first_regret_bound} and \eqref{eq:second_regret_bound} tell us that, if $|\epsilon|$ is not negligible, the algorithm has to play a substantially different number of times in the region $(\nicefrac{3}{4},1]$, depending on the sign of $\epsilon$, not to suffer significant regret when the demand function is $\demexp^{\epsilon}$. The crucial idea is that the only way for the algorithm to present this different behavior is by playing in the only informative region about the sign of $\epsilon$, i.e., the region $(\nicefrac{1}{2},\nicefrac{3}{4}]$. However, as shown in \eqref{eq:first_regret_bound}, selecting policies in this region comes at a cost in terms of regret. To relate quantitatively the number of times the algorithm has to play in this costly region with the difference in the expected number of times the algorithm selects policies in the region $(\nicefrac{3}{4},1]$ is the last missing ingredient that we can obtain relying on information theoretic techniques: It can be proved (and a formal proof is provided in the online appendix, in Section \ref{sec:pinsker_proof}) that, for each $\epsilon \in [0,1]$,
\begin{equation}
\label{eq:Pinsker}
	E^{-\epsilon}(n_3) \ge E^{\epsilon}(n_3) - c_3 \cdot \epsilon \cdot T \cdot \sqrt{E^\epsilon(n_1)} \;.
\end{equation}
Now, if the algorithm is going to suffer low regret when $\epsilon>0$, then by \eqref{eq:first_regret_bound} we have an upper bound on the number of times the algorithm plays in the region $(\nicefrac{1}{2},\nicefrac{3}{4}]$ and a lower bound on the number of times it plays in the region $(\nicefrac{3}{4},1]$, whenever $\epsilon>0$. In turn, by \eqref{eq:Pinsker}, this gives a lower bound on the number of times the algorithm plays in the sub-optimal region $(\nicefrac{3}{4},1]$ when $\epsilon<0$. Then, relying on \eqref{eq:second_regret_bound}, we have an explicit lower bound on how much regret the algorithm is going to suffer when $\epsilon<0$.
We will now carry out this plan ---and prove the theorem--- as follows.

\paragraph{Low regret cannot be achieved for both positive and negative $\epsilon$}
To get a contradiction, suppose that
\begin{equation}
\label{eq:contradiction}
	\forall \epsilon \in [-1,1]\, \qquad \mathcal{R}_T(\demexp^{\epsilon})
< C \cdot T^{2/3} \;.
\end{equation}
It follows from \eqref{eq:first_regret_bound} that, for each $\epsilon \in [0,1]$,
\begin{equation}
\label{eq:third_regret_bound}
	E^\epsilon(n_1) \overset{\eqref{eq:first_regret_bound}}{\le} \frac{\mathcal{R}_T(\demexp^{\epsilon})}{c_2} \overset{\eqref{eq:contradiction}}{\le} \frac{C}{c_2} \cdot T^{2/3} \;, \qquad
	E^\epsilon(n_2) \overset{\eqref{eq:first_regret_bound}}{\le} \frac{\mathcal{R}_T(\demexp^{\epsilon})}{c_1 \cdot \epsilon} \overset{\eqref{eq:contradiction}}{\le} \frac{C}{c_1 \cdot \epsilon} \cdot T^{2/3} \;.
\end{equation}
This implies, relying also on \eqref{eq:second_regret_bound} and \eqref{eq:Pinsker}, that for each $\epsilon \in [0,1]$ we have
\begin{align}
\label{eq:fourth_regret_bound}
	\mathcal{R}_T(\demexp^{-\epsilon})
&\overset{\eqref{eq:second_regret_bound}}{\ge}
	c_1 \cdot \epsilon \cdot E^{-\epsilon} (n_3)
\overset{\eqref{eq:Pinsker}}{\ge}
	c_1 \cdot \epsilon \cdot \big( E^{\epsilon}(n_3) - c_3 \cdot \epsilon \cdot T \cdot \sqrt{E^\epsilon(n_1)} \big)
\nonumber
\\
&\overset{\eqref{eq:counters}}{=}
	c_1 \cdot \epsilon \cdot \brb{ T - E^{\epsilon} (n_1) - E^{\epsilon} (n_2) - c_3 \cdot \epsilon \cdot T \cdot \sqrt{E^\epsilon(n_1)} }
\nonumber
\\
&\overset{\eqref{eq:third_regret_bound}}{\ge}
	c_1 \cdot \epsilon \cdot \bbrb{ T - \frac{C}{c_2} \cdot T^{2/3} - \frac{C}{c_1 \cdot \epsilon} \cdot T^{2/3} - c_3 \cdot \epsilon \cdot T \cdot \sqrt{\frac{C}{c_2} \cdot T^{2/3}} }
\nonumber
\\
&\overset{\phantom{\eqref{eq:counters}}}{=}
	c_1 \cdot \epsilon \cdot \bbrb{ 1 - \frac{C}{c_2} \cdot T^{-1/3} - \frac{C}{c_1 \cdot \epsilon} \cdot T^{-1/3} - c_3 \cdot \epsilon \cdot T^{1/3} \cdot \sqrt{\frac{C}{c_2}} } \cdot T \;.
\end{align}
Pick $\epsilon \coloneqq T^{-1/3} \cdot \sqrt{\frac{\sqrt{ C \cdot c_2}}{c_1 \cdot c_3}}$. First, note that since $0 < C \overset{\eqref{eq:constants}}{\le} \frac{c_1^2 \cdot c_3^2}{c_2}$ we have that $\epsilon \in (0,1]$. Plugging this value of $\epsilon$ in \eqref{eq:fourth_regret_bound} leads to
\begin{align}
\label{eq:final_regret_bound}
	C \cdot T^{2/3}
&\overset{\eqref{eq:contradiction}}{>}
	\mathcal{R}_T(\demexp^{-\epsilon})
\nonumber
\\
&\overset{\eqref{eq:fourth_regret_bound}}{\ge}
	\sqrt{\frac{ \sqrt{C \cdot c_2} \cdot c_1 }{c_3}} \cdot \lrb{ 1- \frac{C}{c_2} \cdot T^{-1/3} - 2 \cdot \sqrt{\frac{c_3}{c_1 \cdot \sqrt{c_2}}} \cdot C^{3/4} } \cdot T^{2/3}
\nonumber
\\
&\overset{\eqref{eq:constants}}{\ge}
	\frac{1}{2} \cdot \sqrt{\frac{ \sqrt{C \cdot c_2} \cdot c_1 }{c_3}} \cdot \lrb{ 1- 4 \cdot \sqrt{\frac{c_3}{c_1 \cdot \sqrt{c_2}}}\cdot C^{3/4} } \cdot T^{2/3}
\nonumber
\\
&\overset{\eqref{eq:constants}}{\ge}
	\frac{1}{4} \cdot \sqrt{\frac{ \sqrt{C \cdot c_2} \cdot c_1 }{c_3}} \cdot T^{2/3} \;,
\end{align}
where the second to last inequality follows from $C \le \frac{c_2}{2}$, while the last inequality follows from $C \le \frac{1}{16} \sqrt[3]{\frac{c_1^2 \cdot c_2}{c_3^2}}$.
Rearranging inequality \eqref{eq:final_regret_bound} leads to the contradiction
\begin{equation*}
	C \overset{\eqref{eq:final_regret_bound}}{>} \lrb{\frac{1}{4} \cdot \sqrt{\frac{c_1 \cdot \sqrt{c_2}}{c_3}} }^{4/3} = \frac{1}{8} \cdot \sqrt[3]{\frac{2 \cdot c_1^2 \cdot c_2}{c_3^2}} > \frac{1}{16} \cdot \sqrt[3]{\frac{c_1^2 \cdot c_2}{c_3^2}} \overset{\eqref{eq:constants}}{\ge} C \;. 
\end{equation*}
Since \eqref{eq:contradiction} leads to a contradiction, it follows that there exists $\epsilon \in [-1,1]$ such that $\mathcal{R}_T(\demexp^{\epsilon}) \ge C \cdot T^{2/3}$. Given that the time horizon $T$ and the randomized algorithm were arbitrarily fixed, the theorem is proved.

\subsection{Theorem \ref{theo:upper_exp3} (Adversarial upper bound on regret)}

The proof of this theorem builds upon the proof of Theorem 6.5 in \cite{cesa2006prediction}.
Relative to this theorem, we need to additionally consider the discretization error introduced by Algorithm \ref{alg:tempered_exp3}, and explicitly control the variance of estimated welfare.

  Recall our notation $\swcum$ and $\swcum(\pol)$ for realized cumulative welfare, and for cumulative welfare for the counterfactual, fixed policy $\pol$.
  We further abbreviate  $\swcum_{Tk} = \swcum(\tilde \pol_k)$.
  Throughout this proof, the sequence $\{\wtp_i\}_{i=1}^T$ is given and conditioned on in any expectations.

  \begin{enumerate}
    \item \textbf{Discretization}\\
    Recall that  $\sw_i(\pol) = \pol  \cdot \bs 1(\pol \leq \wtp_i) + \lambda  \cdot \max(\wtp_i  - \pol, 0)$.
    Let 
    $$\tilde \wtp_i = \max_k \{\tilde \pol_k:\; \tilde \pol_k\leq \wtp_i\}$$ 
    (this is $\wtp_i$ rounded down to the next gridpoint $\tilde \pol_k$), and denote 
    \bals
      \tilde \sw_i(\pol) &= \pol  \cdot \bs 1(\pol \leq \wtp_i) + 
      \lambda  \cdot \max(\tilde \wtp_i  - \pol, 0),\\
      \tilde \swcum_{i}(\pol) &= \sum_{j \leq i}\tilde \sw_j(\pol),
    \eals
    as well as $\tilde \swcum_{ik} = \tilde \swcum_{i}(\tilde \pol_k)$.
    Then it is immediate that $\tilde \sw_i(\pol) \leq \sw_i(\pol)$,    
    $$\sup_\pol |\tilde \sw_i(\pol) - \sw_i(\pol)| \leq \frac{\lambda}{K},$$
    and $\argmax_{\pol} \tilde \swcum_i(\pol) \in \{\tilde \pol_1,\dots,\pol_{K+1}\}$,
    and therefore
    $$
      \max_k \tilde \swcum_{ik} \geq
      \sup_{\pol} \swcum_i(\pol)  -  i \cdot \frac{\lambda}{K}
    $$

    \item \textbf{Unbiasedness}\\
    At the end of period $i$, $\widehat \dem_k$ is an unbiased estimator of $\sum_{j \leq i} \bs 1(\tilde x_k \leq \wtp_j)$ for all $k$.
    Therefore, $E\left [ \widehat \swcum_{ik} \right ] = \tilde \swcum_{ik}$ for all $i$ and $k$.
    
    \item \textbf{Upper bound on optimal welfare}\\    
      Define $W_i =\sum_{k}\exp(\eta  \cdot \widehat \swcum_{ik})$, and       
      $q_{ik} = { \exp(\eta \cdot  \widehat \swcum_{ik})}/{W_i}$.
      
      It is immediate that,
      $$
         E[\log W_T] \geq \eta  \cdot E[\max_k \widehat \swcum_{Tk}]\geq \eta  \cdot \max_k E[ \widehat \swcum_{Tk}] = \eta  \cdot \max_k \tilde \swcum_{Tk} .
      $$
      Furthermore 
      $$
        E[\log W_T] = \sum_{0\leq i < T} E\left [  \log\left(\frac{W_{i+1}}{W_{i}}\right) \right ] + \log(W_0).
      $$
      Given our initialization of the algorithm, $\log(W_0) = \log(K+1)$.
      
    \item \textbf{Lower bound on estimated welfare}\\
    Denote $\widehat \sw_{ik} = \tilde \pol_k  \cdot  \widehat H_{k} + \tfrac{\lambda}{K}  \cdot  \sum_{k'> k} \widehat H_{k'}$,
     where $\widehat H_k = \frac{\out_i}{p_{ik}}  \cdot \bs 1(k_i = k)$,\\
     so that 
     $\widehat \swcum_{ik} =\sum_{j < i} \widehat \sw_{jk}$, and $E[\widehat \sw_{jk}] = \sw_i(\tilde \pol_k)$.

    By definition of $W_i$,
    $$
      \log\left(\frac{W_{i+1}}{W_{i}}\right) = \log\left ( \sum_k q_{ik}  \cdot \exp (\eta  \cdot \widehat \sw_{ik})\right ).
    $$
    Since $p_k \geq \gamma / (K+1)$ for all $k$,  $\widehat \sw_{ik} \in [0, (K+1)/\gamma]$ for all $i$ and $k$, and therefore $\eta  \cdot \widehat \sw_{ik} \leq (K+1)  \cdot \eta / \gamma \leq 1$ (where the last inequality holds by assumption).   
    Using $\exp(a) \leq 1 + a + (e-2) a^2$ for any $a \leq 1$ yields
    \bals
      \exp\left ( \eta  \widehat \sw_{ik} \right ) 
      &\leq  1 + \eta  \cdot \widehat \sw_{ik} + (e-2) \cdot \left (\eta  \cdot \widehat \sw_{ik}  \right )^2.
    \eals    
    Therefore,
    \bals
       \log\left(\frac{W_{i+1}}{W_{i}}\right) 
      \leq & 
      \log\left ( \sum_k q_{ik}  \cdot \left ( 1 + \eta  \cdot \widehat \sw_{ik} + (e-2)\cdot \left (\eta  \cdot \widehat \sw_{ik}  \right )^2 \right ) \right)\\
      \leq & \eta  \cdot \sum_k q_{ik} \cdot  \widehat \sw_{ik} +
      (e-2) \cdot \eta^2   \cdot  \sum_k q_{ik}  \cdot  \widehat \sw_{ik}^2
    \eals
    The second inequality follows from $\log(1+ x) \leq x$.
    
    \item \textbf{Connecting the first order term to welfare}\\
    Note that, by definition, $q_{ik} = \left ( p_{ik} - \tfrac{\gamma}{K+1} \right ) \big / (1-\gamma)$.
    Therefore
    \bals
      \sum_k q_{ik} \cdot  \widehat \sw_{ik} &=
       \frac{1}{1 - \gamma}\sum_k p_{ik} \cdot  \widehat \sw_{ik} - \tfrac{\gamma}{(1-\gamma)(K+1)}  \cdot \sum_k \widehat \sw_{ik},
    \eals
    and thus
    \bals
      E\left [ \sum_k q_{ik} \cdot  \widehat \sw_{ik} \right ] &\leq \frac{1}{1 - \gamma} E\left [ \tilde \sw_i(\pol_i) \right ],
    \eals
    where we have used the fact that $0\leq \tilde \sw_k \leq 1$  for all $k$, given our definition of $\tilde \sw$, and the fact that $k_i$ is distributed according to $p_{ik}$, by construction.

    \item \textbf{Bounding the second moment of estimated welfare}\\
    It remains to bound the term $E\left [  \sum_k q_{ik}  \cdot  \widehat \sw_{ik}^2 \right ]$.
    As in the preceding item, we have
    \bals
      \sum_k q_{ik} \cdot \widehat \sw_{ik}^2 &\leq
       \frac{1}{1 - \gamma}\sum_k p_{ik} \cdot  \widehat \sw_{ik}^2.
    \eals
    
    We can rewrite
    \bals
      \widehat \sw_{ik} &=
      \left ( \tilde \pol_k  \cdot \bs 1(k_i = k)  + \tfrac{\lambda}{K}  \cdot \bs 1(k_i > k) \right )  \cdot \frac{\out_{i}}{p_{ik_i}}.
    \eals
    
    Bounding $y_i \leq 1$ immediately gives
    \bals
      E_i\left [ \widehat \sw_{ik} ^2 \right ] 
      &\leq  \frac{\tilde \pol_k^2 }{p_{ik}} 
    +\left ( \tfrac{\lambda}{K} \right )^2 \cdot \sum_{k'>k} \frac{1}{p_{ik'}},   
    \eals
    and therefore
    \bals
      E_i\left [ \sum_k p_{ik} \cdot  \widehat \sw_{ik}^2 \right ] &\leq \sum_k \tilde \pol_k^2 + \left ( \tfrac{\lambda}{K} \right )^2 \cdot \sum_k\sum_{k'>k} \frac{p_{ik}}{p_{ik'}}\\
      & \leq \sum_k \left ( \tfrac{k}{K} \right ) ^2 + \left ( \tfrac{\lambda}{K} \right )^2 \cdot \sum_k p_{ik} \sum_{k'\neq k} \tfrac{K+1}{\gamma}\\
      & = \tfrac{K(K+1)(2K+1)}{6 K^2} + \tfrac{\lambda^2}{\gamma} \tfrac{K+1}{K}\\
      & = \tfrac{K+1}{K} \cdot \left (\tfrac{2K+1}{6} + \tfrac{\lambda^2}{\gamma}  \right ).
    \eals
    
    \item \textbf{Collecting inequalities}\\
    Combining the preceding items, we get
    \bals
    &\eta  \cdot\left ( \sup_{\pol} \swcum(\pol)-  T \cdot \frac{\lambda}{K} \right ) \\
    \leq & \eta  \cdot \max_k \tilde \swcum_{Tk}  
    \leq  E[\log W_T]& (\textrm{Item 1})\\
    =& \sum_{0 \leq i < T} E\left [  \log\left(\frac{W_{i+1}}{W_{i}}\right) \right ] + \log(K+1) & (\textrm{Item 3})\\
    \leq &  \frac{\eta}{1 - \gamma} \cdot  E\left [\tilde \swcum \right ] + (e-2) \cdot 
    \frac{\eta^2}{1-\gamma} \sum_{1\leq i\leq T}  \sum_k  E\left [ p_{ik}  \cdot  \widehat \sw_{ik}^2 \right ]  + \log(K+1)& (\textrm{Item 4 and 5})\\
     \leq&   \frac{\eta}{1 - \gamma} \cdot  E\left [\tilde  \swcum \right ] + (e-2) \cdot 
     \frac{\eta^2}{1-\gamma}T \cdot \tfrac{K+1}{K} \cdot \left (\tfrac{2K+1}{6} + \tfrac{\lambda^2}{\gamma}  \right )  + \log(K+1). & (\textrm{Item 6})
    \eals

    Multiplying by $(1-\gamma)$ and dividing by $\eta$, adding $\gamma \sup_{\pol} \swcum(\pol) + T \frac{\lambda}{K}$ to both sides and subtracting $E\left [\tilde \swcum \right ]$, bounding $\sup_{\pol} \swcum(\pol) \leq  T$, and $E\left [\tilde \swcum \right ]\leq E\left [ \swcum \right ]$ (from Item 1), yields
    \bal
      &\sup_{\pol} \swcum(\pol) - E\left [ \swcum \right ]\nonumber\\ \leq &
      \left ( \gamma   + \eta  \cdot (e-2) 
       \tfrac{K+1}{K} \cdot \left (\tfrac{2K+1}{6} + \tfrac{\lambda^2}{\gamma}  \right )
      + \tfrac{\lambda}{K} \right ) \cdot T
      + \tfrac{\log(K+1)}{\eta}.
      \label{eq:regret_bound}
    \eal
    This proves the first claim of the theorem.
    
    \item \textbf{Optimizing tuning parameters}\\
    Suppose now that we choose the tuning parameters as follows:
    \bals
      \gamma &= c_1  \cdot \left(\tfrac{\log(T)}{T}\right)^{1/3}, &
      \eta &= c_2  \cdot \gamma^2, &
      K &= c_3 / \gamma.
    \eals
    Plugging in we get
    \bals
      &\sup_{\pol} \swcum(\pol) - E\left [ \swcum \right ]\nonumber\\ \leq &
      \left ( \gamma   + c_2  \cdot \gamma^2  \cdot (e-2) 
       \tfrac{K+1}{K} \cdot \left (\tfrac{2c_3 / \gamma+1}{6} + \tfrac{\lambda^2}{\gamma}  \right )
      + \lambda \cdot \gamma / c_3  \right ) \cdot T
      + \tfrac{\log(K+1)}{c_2  \cdot \gamma^2}\\
      &=  \log(T)^{1/3} T^{2/3} \cdot \left ( c_1 + (e-2) 
       \tfrac{K+1}{K}  \cdot c_1 c_2 \left ( \tfrac{c_3}{3} + \lambda^2 +\tfrac{\gamma}{6} \right )   +\lambda \tfrac{c_1}{c_3} + \frac{\log(T^{1/3} \log(T)^{-1/3} c_3 / c_1 + 1 )}{c_1^2\log(T) }  \right ) \\
       &= \log(T)^{1/3} T^{2/3} \cdot \left ( c_1 + (e-2) \cdot c_1 c_2 \left ( \tfrac{c_3}{3} + \lambda^2 \right )   +\lambda \tfrac{c_1}{c_3} + \frac{1}{3 c_1^2}  + o(1)\right ).
    \eals
    The second claim of the theorem follows.

  \end{enumerate}

\clearpage
\Huge
\begin{center}
    Online Appendix
\end{center}
\normalsize

\pagestyle{fancy}
\fancyhead[L]{\textbf{Online Appendix}}

\section{Additional proofs}
\label{sec:onlineproofs}
\localtableofcontents

\subsection{Claim \eqref{eq:Pinsker} (Relating choice probabilities for positive and negative $\epsilon$)}
\label{sec:pinsker_proof}

\begin{proof}[Proof of Claim \eqref{eq:Pinsker}]$\;$\\
Let $w_1, w_2, \dots \in [0,1]$ be the randomization seeds to be used by the algorithm. In the light of the Skorokhod representation theorem \cite[Section 17.3]{williams1991probability}, we may assume without (much) loss of generality that, for each $\epsilon \in [-1,1]$, these seeds form a sequence of $P^\epsilon$-i.i.d. $[0,1]$-valued uniform random variables. In particular, this implies,
\begin{equation}
\label{eq:iid_seeds}
   P^\epsilon_{(w_i)_{i \in \mathbb{N}}} = P^{-\epsilon}_{(w_i)_{i \in \mathbb{N}}}\;, \qquad \forall \epsilon \in [0,1] \;.
\end{equation}

Recall that a sequence of functions $\alpha \coloneqq (\alpha_i)_{i \in \mathbb{N}}$ is called a randomized algorithm if
	\begin{equation*}
	\alpha_1 \colon	[0,1] \to [0,1] \;, \qquad \forall i \in \mathbb{N}, \quad \alpha_{i+1} \colon [0,1]^{i+1} \times \{0,1\}^i \to [0,1] \;. 
	\end{equation*}
The feedback function associated to our problem is
	\begin{equation*}
		\varphi \colon [0,1] \times \{\nicefrac{1}{4}, \nicefrac{1}{2}, \nicefrac{3}{4}, 1\} \to \{0,1\}\;, \qquad
		(x,v) \mapsto \bs 1 (x \le v) \;. 
	\end{equation*}
Now, a randomized algorithm $\alpha$ generates a sequence of choices $x_1, x_2, \dots $ using the randomization seeds $w_1,w_2,\dots$ and the received feedback $z_1, z_2, \dots \in \{0,1\}$ in the following inductive way on $i \in \mathbb{N}$
\begin{align*}
	x_1 &\coloneqq \alpha_1(w_1) \;, \qquad & z_1 &\coloneqq \varphi(x_1,v_1) \;,
	\\
	x_{i+1} &\coloneqq \alpha_{i+1}(w_1, \dots, w_{i+1}, z_1, \dots, z_i) \;, \qquad & z_{i+1} & \coloneqq \varphi(x_{i+1},v_{i+1})\;.
\end{align*} 
For each $a \in [0,1]$, fix a binary representation $0.a_1 a_2 a_3\dots$ and define $\xi(a) \coloneqq 0.a_1 a_3 a_5\dots$ and $\zeta(a)\coloneqq 0.a_2 a_4 a_6\dots$. Notice that $\xi,\zeta \colon [0,1] \to [0,1]$ are independent with respect to the Lebesgue measure on $[0,1]$ and that their (common) distribution is a uniform on $[0,1]$.
For each $x \in [0,1]$, define $\psi_x \colon [0,1] \to \{0,1\}, u \mapsto \bs 1_{[0, \nicefrac{1}{4}]} (x) + \bs 1_{(\nicefrac{1}{4}, \nicefrac{1}{2}]} (x) \cdot \bs 1_{[0, 1-a]} (u) + \bs 1_{(\nicefrac{3}{4}, 1]} (x) \cdot \bs 1_{[0, 1-a-2\cdot b]} (u)$. Define by induction on $i \in \mathbb{N}$ the following process
\begin{align*}
	\tilde{x}_1 &\coloneqq \alpha_1\brb{\zeta(w_1)} \;,
\\
	\tilde{z}_1 &\coloneqq \varphi \Brb { \tilde{x}_1, \psi_{\tilde{x}_1} \brb{ \xi(w_1) } } \;,
\\
	\tilde{x}_{i+1} &\coloneqq \alpha_{i+1}\brb{ \zeta(w_1), \dots, \zeta(w_{i+1}), \tilde{z}_1, \dots, \tilde{z}_i },
\\
	\tilde{z}_{i+1} & \coloneqq 
\begin{cases}
                    \varphi(\tilde{x}_{i+1},v_{i+1}), & \tilde{x}_{i+1} \in (\nicefrac{1}{2}, \nicefrac{3}{4}]\\
                    \varphi\Brb{\tilde{x}_{i+1},\psi_{\tilde{x}_{i+1}} \brb{ \xi (w_{i+1}) }}, & \textrm{otherwise} .
                    \end{cases}	
\end{align*}
Since, for each $\epsilon \in [-1,1]$ and each $i \in \mathbb{N}$,
\begin{align*}
	P^\epsilon(z_i = 1 \mid x_i) &= 
                    \begin{cases}
                    1 & x_i \in [0, \frac{1}{4}]\\
                    1-a & x_i \in (\frac{1}{4}, \frac{1}{2}]\\
                    1-a-(1+\epsilon)\cdot b  & x_i \in (\frac{1}{2}, \frac{3}{4}]\\
                    1-a-2 \cdot b & x_i \in (\frac{3}{4}, 1]
                    \end{cases}
,\;
\\
	P^\epsilon( \tilde{z}_i = 1 \mid \tilde{x}_i) &= 
                    \begin{cases}
                    1 & \tilde{x}_i \in [0, \frac{1}{4}]\\
                    1-a & \tilde{x}_i \in (\frac{1}{4}, \frac{1}{2}]\\
                    1-a-(1+\epsilon)\cdot b  & \tilde{x}_i \in (\frac{1}{2}, \frac{3}{4}]\\
                    1-a-2 \cdot b & \tilde{x}_i \in (\frac{3}{4}, 1]
                    \end{cases}
\end{align*}

it follows that, for each $\epsilon \in [-1,1]$ and each $i \in \mathbb{N}$, the random variable $\tilde{x}_i$ has the same distribution as the random choice $x_i$ made by the randomized algorithm $\alpha$ at time $i$ when the underlying distribution is $P^\epsilon$, i.e.,

\begin{equation}
\label{eq:simulation}
P^\epsilon_{\tilde{x}_i} = P^\epsilon_{x_i} \;.
\end{equation}
As with the process $x_1, x_2, \dots$, we have to count the number of times the process $\tilde{x}_1, \tilde{x}_2, \dots$ lands in the regions $(\nicefrac{1}{2}, \nicefrac{3}{4}]$, $[0, \nicefrac{1}{2}]$ and $(\nicefrac{3}{4}, 1]$ up to the time $T$. This can be done relying on the following random variables
\begin{equation*}
   \tilde{n}_{1} \coloneqq \sum_{i=1}^T \bs 1_{(\nicefrac{1}{2}, \nicefrac{3}{4}]} ( \tilde{x}_i ) \;, \qquad
   \tilde{n}_{2} \coloneqq \sum_{i=1}^T \bs 1_{[0, \nicefrac{1}{2}]} ( \tilde{x}_i ) \;, \qquad
   \tilde{n}_{3} \coloneqq \sum_{i=1}^T \bs 1_{(\nicefrac{3}{4}, 1]} ( \tilde{x}_i ) \;.
\end{equation*}
Since, for each $\epsilon \in [-1,1]$ and each $j\in\{1,2,3\}$,
\begin{equation*}
	E^\epsilon(\tilde{n}_{j})
=
	\sum_{i=1}^T P^\epsilon_{x_i} \brb{ (\nicefrac{1}{2},\nicefrac{3}{4}] } \overset{\eqref{eq:simulation}}{=} \sum_{i=1}^T P^\epsilon_{\tilde{x}_i} \brb{(\nicefrac{1}{2},\nicefrac{3}{4}]} = E^{\epsilon}(n_j)\;,   
\end{equation*}
to prove the claim \eqref{eq:Pinsker}, it is enough to prove that, for each $\epsilon \in [-1,1]$,
\begin{equation*}
E^{-\epsilon}(\tilde{n}_{3}) \ge E^{\epsilon}(\tilde{n}_{3}) - c_3 \cdot \epsilon \cdot T \cdot \sqrt{E^\epsilon(\tilde{n}_{1})} \;.
\end{equation*}
We first prove the result when the sequence of randomization seeds is fixed, i.e., we suppose first that $\bar{w}_1, \bar{w}_2, \dots$ are such that $w_1 = \bar{w}_1, w_2 = \bar{w}_2, \dots$.
For each $\epsilon \in [-1,1]$, we consider the associated probability distribution $Q^\epsilon$, defined as the conditional probability distribution $P^\epsilon(\cdot \mid w_1 = \bar{w}_1, w_2 = \bar{w}_2, \dots)$.
For each $t \in \mathbb{N}$, let $I_t \coloneqq \big\{ i\in\{1,\dots,t\} \mid \tilde{x}_i \in (\nicefrac{1}{2}, \nicefrac{3}{4}] \big\}$, and for each $s \in \{1,\dots,t\}$, let
\begin{equation*}
Z_{t,s} \coloneqq \begin{cases}
                    \emptyset & \text{if } s \notin I_t \;,\\
                    \bs 1 (\nicefrac{1}{2} < v_s) & \text{if } s \in I_t \;.
                    \end{cases}
\end{equation*}
Notice that for each $t_1, t_2 \in \mathbb{N}$ and each $s \in \{1,\dots,\min(t_1,t_2)\}$, we have that $Z_{t_1, s} = Z_{t_2, s}$. Then, for each $s \in \mathbb{N}$, it is well defined the random variable $Z_s \coloneqq Z_{t,s}$, where $t \in \mathbb{N}$ is any number $t \ge s$. Define, for each $t \in \mathbb{N}$, the random vector $\bar{Z}_t \coloneqq (Z_1,\dots,Z_t)$. Notice that, given that the sequence of randomization seeds is fixed and that, for each $s \in \mathbb{N}$, we have that $v_s \in \{\nicefrac{1}{4}, \nicefrac{1}{2}, \nicefrac{3}{4}, 1 \}$ (hence, for each $x \in (\nicefrac{1}{2}, \nicefrac{3}{4}]$, it holds that $\bs 1 (\nicefrac{1}{2} < v_s) = \bs 1 (x = v_s)$), the random vector $(\tilde{x}_1, \dots, \tilde{x}_T)$ is measurable with respect to the $\sigma$-algebra generated by $\bar{Z}_{T-1}$.
Hence, for each $\epsilon \in [0,1]$ and each $i \in \{1,\dots, T\}$, we can deduce from Pinsker's inequality (see, e.g., \cite[Lemma 2.5]{tsybakov2008}) that
\begin{equation}
\label{eq:Pinsky}
Q^\epsilon \brb{\tilde{x}_i \in (\nicefrac{3}{4}, 1]} \le Q^{-\epsilon} \brb{\tilde{x}_i \in (\nicefrac{3}{4}, 1]} + \sqrt{\frac{1}{2} \mathcal{D}_{\text{KL}} \brb{Q^\epsilon_{\bar{Z}_{T-1}} \mid \mid Q^{-\epsilon}_{\bar{Z}_{T-1}}} } \;,
\end{equation}
where $\mathcal{D}_{\text{KL}}$ is the Kullback-Leibler divergence.
Now, for each $t \in \mathbb{N}$ and each $\epsilon \in [0,1]$, by the chain rule for Kullback-Leibler divergence (see, e.g., \cite[Theorem 2.5.3]{ThomasM.Cover2006}), we have
\begin{multline}
	\label{eq:chain_rule}
    \mathcal{D}_{\text{KL}} \brb{ Q^\epsilon_{\bar{Z}_{t+1}} \mid \mid Q^{-\epsilon}_{\bar{Z}_{t+1}} }   
= 
    \mathcal{D}_{\text{KL}} \brb{ Q^\epsilon_{(\bar{Z}_{t},Z_{t+1})} \mid \mid Q^{-\epsilon}_{(\bar{Z}_{t},Z_{t+1})} }
\\
\begin{aligned}
&
=
    \mathcal{D}_{\text{KL}} \brb{ Q^\epsilon_{\bar{Z}_{t}} \mid \mid Q^{-\epsilon}_{\bar{Z}_{t}} }
+ \sum_{(\bar{z},z) \in \{\emptyset,0,1\}^t \times \{\emptyset,0,1\} } \log \lrb{ \frac{Q^\epsilon (Z_{t+1} = z \mid \bar{Z}_t = \bar{z})}{Q^{-\epsilon}(Z_{t+1} = z \mid \bar{Z}_t = \bar{z})} } 
\cdot Q^\epsilon \brb{ \bar{Z}_{t} = \bar{z} \cap Z_{t+1} = z } \;.
\end{aligned}
\end{multline}
Notice that, for each $t \in \mathbb{N}$ and each $\epsilon \in [0,1]$ we have
\begin{multline}
\label{eq:conditional_divergence}
\sum_{(\bar{z},z) \in \{\emptyset,0,1\}^t \times \{\emptyset,0,1\} } \log \lrb{ \frac{Q^\epsilon (Z_{t+1} = z \mid \bar{Z}_t = \bar{z})}{Q^{-\epsilon}(Z_{t+1} = z \mid \bar{Z}_t = \bar{z})} } \cdot Q^\epsilon \brb{ \bar{Z}_{t} = \bar{z} \cap Z_{t+1} = z }
\\
\begin{aligned}[b]
&= 
   \sum_{ \substack{(\bar{z},z) \in \{\emptyset,0,1\}^t \times \{\emptyset,0,1\} \\ t+1 \in I_{t+1} } } \log \lrb{ \frac{Q^\epsilon (Z_{t+1} = z \mid \bar{Z}_t = \bar{z})}{Q^{-\epsilon}(Z_{t+1} = z \mid \bar{Z}_t = \bar{z})} } \cdot Q^\epsilon \brb{ \bar{Z}_{t} = \bar{z} \cap Z_{t+1} = z }
\\
&=
	\lrb{ \sum_{ \substack{\bar{z} \in \{\emptyset,0,1\}^t \\ t+1 \in I_{t+1} } } Q^\epsilon (\bar{Z}_t = \bar{z}) }
\cdot \sum_{z \in \{0,1\}} \log \lrb{ \frac{Q^\epsilon \brb{ \bs 1 (\nicefrac{1}{2} < v_{t+1}) = z }}{Q^{-\epsilon}\brb{ \bs 1 (\nicefrac{1}{2} < v_{t+1}) = z } } } \cdot Q^\epsilon(\bs 1 (\nicefrac{1}{2} < v_{t+1}) = z)
\\
&=
	Q^{\epsilon}\brb { \tilde{x}_{t+1} \in (\nicefrac{1}{2}, \nicefrac{3}{4}] }
\cdot \sum_{z \in \{0,1\}}  \log \lrb{ \frac{Q^\epsilon \brb{ \bs 1 (\nicefrac{1}{2} < v_{t+1}) = z }}{Q^{-\epsilon}\brb{ \bs 1 (\nicefrac{1}{2} < v_{t+1}) = z } } } \cdot Q^\epsilon(\bs 1 (\nicefrac{1}{2} < v_{t+1}) = z) \;.
\end{aligned}
\end{multline}
Algebraic calculations show that, for each $t \in \mathbb{N}$ and each $\epsilon \in [0,1]$,
\begin{multline}
\label{eq:logarithmic_part}
\sum_{z \in \{0,1\}} \log \lrb{ \frac{Q^\epsilon \brb{ \bs 1 (\nicefrac{1}{2} < v_{t+1}) = z }}{Q^{-\epsilon}\brb{ \bs 1 (\nicefrac{1}{2} < v_{t+1}) = z } } } \cdot Q^\epsilon(\bs 1 (\nicefrac{1}{2} < v_{t+1}) = z)
\\
\begin{aligned}[b]
&=
	\log \lrb{ \frac{Q^\epsilon \brb{ \frac{1}{2} < v_{t+1} }}{Q^{-\epsilon}\brb{ \frac{1}{2} < v_{t+1} } } } \cdot Q^\epsilon \lrb{ \frac{1}{2} < v_{t+1}}  
+   \log \lrb{ \frac{Q^\epsilon \brb{ \frac{1}{2} \ge v_{t+1} }}{Q^{-\epsilon}\brb{ \frac{1}{2} \ge v_{t+1}} } } \cdot Q^\epsilon \lrb{\frac{1}{2} \ge v_{t+1}}   
\\
&=
    \log \lrb{ \frac{1-a-(1+\epsilon)\cdot b}{1-a-(1-\epsilon)\cdot b} } \cdot \brb{1-a-(1+\epsilon)\cdot b  }
+ \log \lrb{ \frac{a + (1+\epsilon) \cdot b}{a + (1-\epsilon) \cdot b} } \cdot \brb{a + (1+\epsilon) \cdot b}
\\
&\le
	\frac{4 \cdot b^2 \cdot \epsilon ^2}{\brb{1-a-(1-\epsilon) \cdot b} \cdot \brb{a + (1-\epsilon)\cdot b}} \le \frac{4 \cdot b^2 \cdot \epsilon^2}{a\cdot(1-a-2b)} = 2 \cdot c_3^2 \cdot \epsilon ^2 \;. 
\end{aligned}
\end{multline}
Putting \eqref{eq:chain_rule}, \eqref{eq:conditional_divergence} and \eqref{eq:logarithmic_part} together, we obtain that, for each $t \in \mathbb{N}$ and each $\epsilon \in [0,1]$,
\begin{equation}
\label{eq:bound_divergence_1}
	\mathcal{D}_{\text{KL}} \brb{ Q^\epsilon_{\bar{Z}_{t+1}} \mid \mid Q^{-\epsilon}_{\bar{Z}_{t+1}} }
\le
	\mathcal{D}_{\text{KL}} \brb{ Q^\epsilon_{Z_{1}} \mid \mid Q^{-\epsilon}_{Z_{1}} } + 2 \cdot c_3^2 \cdot \epsilon^2 \cdot \sum_{s=1}^t Q^\epsilon \brb{ \tilde{x}_{s+1} \in (\nicefrac{1}{2}, \nicefrac{3}{4}] }\;.
\end{equation}
With the same technique used above, for each $\epsilon \in [0,1]$, we can prove that
\begin{equation}
\label{eq:bound_divergence_2}
\mathcal{D}_{\text{KL}} \brb{ Q^\epsilon_{Z_{1}} \mid \mid Q^{-\epsilon}_{Z_{1}} } \le 2 \cdot c_3^2 \cdot \epsilon^2 \cdot Q^\epsilon \brb{\tilde{x}_1 \in (\nicefrac{1}{2}, \nicefrac{3}{4}]}\;.
\end{equation}
For each $t \in \{1,\dots, T\}$, putting \eqref{eq:bound_divergence_1} and \eqref{eq:bound_divergence_2} together, we obtain
\begin{align}
\label{eq:bound_divergence_3}
	\mathcal{D}_{\text{KL}} \brb{ Q^\epsilon_{\bar{Z}_{t}} \mid \mid Q^{-\epsilon}_{\bar{Z}_{t}} }
&\overset{\eqref{eq:bound_divergence_1} + \eqref{eq:bound_divergence_2} }{\le}
	2 \cdot c_3^2 \cdot \epsilon^2 \cdot \sum_{s=1}^t Q^\epsilon \brb{ \tilde{x}_{s} \in (\nicefrac{1}{2}, \nicefrac{3}{4}] }
\nonumber
\\
&\overset{\phantom{\eqref{eq:bound_divergence_1} + \eqref{eq:bound_divergence_2}} }{\le}
	2 \cdot c_3^2 \cdot \epsilon^2 \cdot E^\epsilon \brb{ \tilde{n}_{1} \mid w_1 = \bar{w}_1, w_2 = \bar{w}_2 ,\dots} \;.
\end{align}
Now, \eqref{eq:Pinsky} and \eqref{eq:bound_divergence_3} imply that, for each $\epsilon \in [0,1]$ and each $i \in \{1,\dots, T\}$,
\begin{equation}
\label{eq:istantaneous_bound}
	Q^\epsilon\brb{\tilde{x}_i \in (\nicefrac{3}{4},1]} \le Q^{-\epsilon} \brb{ \tilde{x}_i \in (\nicefrac{3}{4},1] } + c_3 \cdot \epsilon \cdot \sqrt{ E^\epsilon \brb{ \tilde{n}_{1} \mid w_1 = \bar{w}_1, w_2 = \bar{w}_2, \dots } }\;. 
\end{equation}
Taking the sum of \eqref{eq:istantaneous_bound} over $i \in \{1,\dots,T\}$, we obtain that for each $\epsilon \in [0,1]$,
\begin{multline}
\label{eq:cumulative_bound}
    E^{-\epsilon} \brb{ \tilde{n}_{3} \mid w_1 = \bar{w}_1, w_2 = \bar{w}_2, \dots }
\\
\begin{aligned}
&\ge 
	E^{\epsilon} \brb{ \tilde{n}_{3} \mid w_1 = \bar{w}_1, w_2 = \bar{w}_2, \dots } - c_3 \cdot \epsilon \cdot T \cdot \sqrt{ E^\epsilon \brb{ \tilde{n}_{1} \mid w_1 = \bar{w}_1, w_2 = \bar{w}_2, \dots } }\;.
\end{aligned}
\end{multline}
Now, since the sequence $\bar{w}_1, \bar{w}_2, \dots$ of randomization seeds has been arbitrarily chosen, for each $\epsilon \in [0,1]$, using the fact that $P^\epsilon_{(w_t)_{t \in \mathbb{N}}} = P^{-\epsilon}_{(w_t)_{t \in \mathbb{N}}}$ and Jensen's inequality, we have that
\begin{align*}
    E^{-\epsilon} (\tilde{n}_{3})
&\overset{\phantom{\eqref{eq:iid_seeds}}}{=}
	\int_{[0,1]^\mathbb{N}} E^{-\epsilon} \brb{ \tilde{n}_{3} \mid w_1 = \bar{w}_1, w_2 = \bar{w}_2, \dots }  \mathrm{d} P^{-\epsilon}_{(w_t)_{t \in \mathbb{N}}}(\bar{w}_1, \bar{w}_2, \dots)
\\
&\overset{\eqref{eq:iid_seeds}}{=}
	\int_{[0,1]^\mathbb{N}} E^{-\epsilon} \brb{ \tilde{n}_{3} \mid w_1 = \bar{w}_1, w_2 = \bar{w}_2, \dots }  \mathrm{d} P^{\epsilon}_{(w_t)_{t \in \mathbb{N}}}(\bar{w}_1, \bar{w}_2, \dots)
\\
&\overset{\eqref{eq:cumulative_bound}}{\ge}
	\int_{[0,1]^\mathbb{N}} E^{\epsilon} \brb{ \tilde{n}_{3} \mid w_1 = \bar{w}_1, w_2 = \bar{w}_2, \dots } \mathrm{d} P^{\epsilon}_{(w_t)_{t \in \mathbb{N}}}(\bar{w}_1, \bar{w}_2, \dots)
\\
&\quad-
	c_3 \cdot \epsilon \cdot T \cdot \int_{[0,1]^\mathbb{N}} \sqrt{ E^\epsilon \brb{ \tilde{n}_{1} \mid w_1 = \bar{w}_1, w_2 = \bar{w}_2, \dots } }   \mathrm{d} P^{\epsilon}_{(w_t)_{t \in \mathbb{N}}}(\bar{w}_1, \bar{w}_2, \dots)
\\
\text{(by Jensen) \qquad}&\overset{\phantom{\eqref{eq:iid_seeds}}}{\ge}
		\int_{[0,1]^\mathbb{N}} E^{\epsilon} \brb{ \tilde{n}_{3} \mid w_1 = \bar{w}_1, w_2 = \bar{w}_2, \dots } \mathrm{d} P^{\epsilon}_{(w_t)_{t \in \mathbb{N}}}(\bar{w}_1, \bar{w}_2, \dots)
\\
&\quad-
	c_3 \cdot \epsilon \cdot T \cdot \sqrt{ \int_{[0,1]^\mathbb{N}}  E^\epsilon \brb{ \tilde{n}_{1} \mid w_1 = \bar{w}_1, w_2 = \bar{w}_2, \dots }  \mathrm{d} P^{\epsilon}_{(w_t)_{t \in \mathbb{N}}}(\bar{w}_1, \bar{w}_2, \dots) }
\\
&\overset{\phantom{\eqref{eq:iid_seeds}}}{=}
	E^\epsilon(\tilde{n}_{3}) - c_3 \cdot \epsilon \cdot \sqrt{ E^{\epsilon}(\tilde{n}_{1}) } \;. \qedhere
\end{align*}
\end{proof}

\subsection{Theorem \ref{theo:lower_concave} (Lower bound on regret for the concave case)}

\begin{figure}[t]
    \caption{Construction for proving the lower bound on regret for the concave case}
    \label{fig:lower_bound_concave}
    \begin{center}
      \includegraphics[width=.8\textwidth]{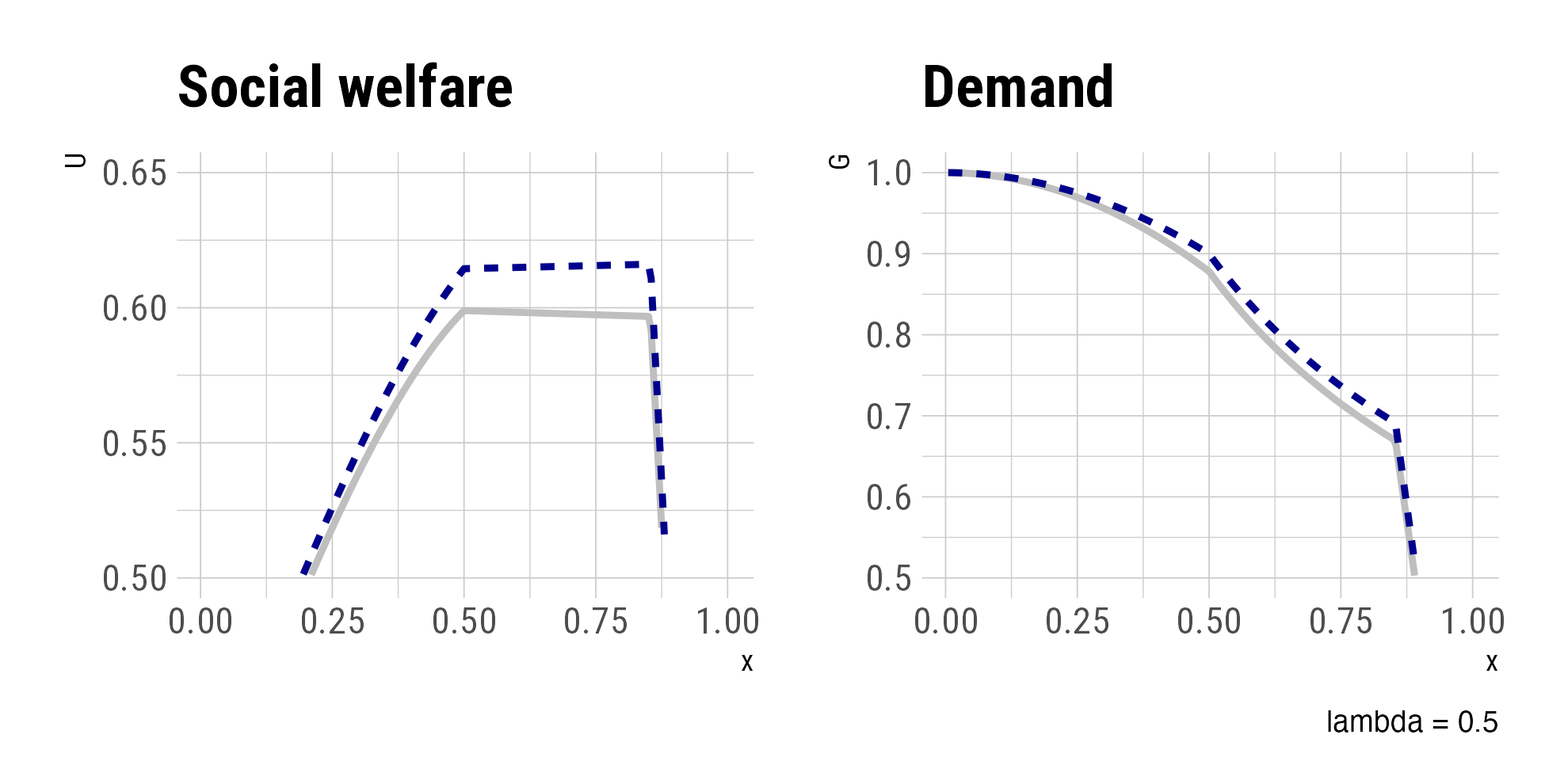}
    \end{center}
  \end{figure}

    \paragraph{Defining a family of distributions for $\wtp$}
Define $\bar{h} \coloneqq \frac{1-\sqrt{1-\lambda}}{2}$ and notice that $0 < \bar{h} < \frac{1}{2}$.
Define $\bar{\eta} \coloneqq \brb{\bar{h} \cdot (1-\bar{h})^{1-\lambda}\cdot(1-\lambda)}^{-1}$ and $\bar{\epsilon} \coloneqq \frac{1}{2} \cdot \min(\bar{\eta}, \frac{2}{3} \cdot 2^{-\lambda})$. For each $\epsilon \in (-\bar{\epsilon}, \bar{\epsilon})$ and each $x \in [0,1]$, define
\[
   f^\epsilon (x) \coloneqq \bar{c} \cdot \lrb{\brb{ 2^{2-\lambda} - 8 \cdot \bar{h}\cdot \epsilon)} \cdot x \cdot \bs 1_{[0,\frac{1}{2})}(x)  + \frac{1}{x^{2-\lambda}} \cdot \bs 1_{[\frac{1}{2},1-\bar{h}]}(x) + (\bar{\eta} + \epsilon) \cdot \bs 1_{(1-\bar{h},1]}(x)} \;,
\]
where $\bar{c}$ is such that $\int_0^1 f^0(x) \mathrm{d}x  = 1$. For each $\epsilon \in (-\bar{\epsilon}, \bar{\epsilon})$, note that $f^\epsilon$ is a density function on $[0,1]$, i.e., a non-negative function whose integral is $1$.
For each $\epsilon \in (-\bar{\epsilon}, \bar{\epsilon})$, let $\mu^\epsilon$ be the probability measure whose density is $f^\epsilon$, and define $\demexp^\epsilon$ and $\swexp^\epsilon$ as the demand function and the expected social welfare associated to $\mu^\epsilon$, respectively (see Figure \ref{fig:lower_bound_concave} for an illustration).

\paragraph{Properties of $\swexp$}
Define also $\bar{x} \coloneqq \frac{1}{2} \cdot \lrb{\frac{1}{2} + (1-\bar{h})} = \frac{3}{4} - \frac{\bar{h}}{2}$ and $\bar{m} \coloneqq \frac{1-\sqrt{1-\lambda}}{8} \cdot (1-\lambda)^{3/2}$.
Notice that, for all $\epsilon \in (-\bar{\epsilon}, \bar{\epsilon})$, we have:
\begin{itemize}
   \item $\swexp^\epsilon$ is continuous and concave.
   \item $\swexp^\epsilon$ is strictly increasing in $[0, \frac{1}{2}]$, linear in $[\frac{1}{2}, 1-\bar{h}]$ with slope $(1-\lambda)\cdot \bar{h} \cdot \epsilon$, and strictly decreasing on $[1-\bar{h},1]$, which in particular implies that the maximum of $\swexp^\epsilon$ is at $1-\bar{h}$ if $\epsilon>0$, and at $\frac{1}{2}$ if $\epsilon < 0$.
   \item If $\epsilon > 0$, then $\swexp^\epsilon(1-\bar{h}) - \max_{x \in [0, \bar{x}]} \swexp^\epsilon(x) = \bar{m} \cdot |\epsilon| = \swexp^{-\epsilon}(\frac{1}{2}) - \max_{x \in [\bar{x}, 1]} \swexp^{-\epsilon}(x)$.
\end{itemize}
Now, consider the sequence of individual valuations $v_1,v_2, \dots \in [0,1]$, and assume that, for each $\epsilon \in (-\bar{\epsilon},\bar{\epsilon})$, when the underlying distribution is $P^\epsilon$, this sequence is i.i.d. (independent of the randomization used by the algorithm) with common distribution $\mu^\epsilon$.
The previous list of properties implies that, for each $\epsilon \in (0,\bar{\epsilon})$ (resp., $\epsilon \in (-\bar{\epsilon},0)$), when the underlying distribution is $P^\epsilon$, the expected instantaneous regret at time $t$ is at least $\bar{m} \cdot |\epsilon|$ if the learner plays in the region $\bar{I} \coloneqq [0, \bar{x}]$ (resp., in the region $\bar{J} \coloneqq (\bar{x} , 1]$).
It follows that, in order not to suffer linear regret, the learner has to discriminate the sign of $\epsilon$.

\paragraph{Intuition for the proof}
Now, the high-level idea is that in order to discriminate the sign of $\epsilon$, the learner needs on the order of $\frac{1}{\epsilon^2}$ observations.
Therefore, for a number of periods on the order of $\frac{1}{\epsilon^2}$, the algorithm is playing ``in the dark,'' and thus suffers an expected regret on the order of $\epsilon  \cdot T$, or, equivalently, on the order of $\sqrt{T}$, when the underlying distribution is between $P^\epsilon$ or $P^{-\epsilon}$.

\paragraph{Defining constants}
We now formalize this idea.
Let
\[
\gamma \coloneqq \lrb{\int_0^{1/2} \lrb{ \frac{16 \bar{h} }{2^{2-\lambda} - 8 \bar{h} \bar{\epsilon}}}^2 f^{-\bar{\epsilon}}(x)\mathrm{d}x + \int_{1-\bar{h}}^1 \lrb{\frac{2}{\bar{\eta} - \bar{\epsilon}}}^2 f^{\bar{\epsilon}}(x) \mathrm{d} x }^{1/2} > 0
\]
Let $\bar{M} > 0$ such that $2 \cdot \sqrt{\frac{\sqrt{2}}{3}\cdot \frac{\gamma \cdot \bar{M}}{\bar{m}}} = 1$.
Let $M \in (0, \bar{M})$ such that $$k \coloneqq \sqrt{\frac{\frac{M}{\bar{m}}}{\frac{\sqrt{2}}{3} \cdot \gamma}} < \bar{\epsilon} \;.$$
From now on, fix a time horizon $T \in \mathbb{N}$ and let $\epsilon \coloneqq \frac{k}{\sqrt{T}}$.
In the following we use the notation $E^\epsilon$ (resp., $E^{-\epsilon}$) to denote the expectation with respect to the probability measure $P^\epsilon$ (resp., $P^{-\epsilon}$).
Let $x_1,x_2,\dots$ be the policies chosen by the algorithm. Note that, since the algorithm bases its decision at time $t$ only on the (partial) knowledge of $v_1, \dots, v_{t-1}$ and some independent randomization, there exists a (measurable) function $\varphi_t \colon [0,1]^{t-1} \to [0,1]$ such that
\[
   E^{\epsilon}\brb{\bs 1 (x_t \in \bar{I}) \mid v_1,\dots,v_{t-1}} = \varphi_t(v_1,\dots,v_{t-1}) = E^{-\epsilon}\brb{\bs 1 (x_t \in \bar{I})\mid v_1,\dots,v_{t-1}}\;.
\]
Then, for each time $t$, it holds
\begin{align*}
   \babs{E^{\epsilon}\brb{\bs 1 (x_t \in \bar{I})} - E^{-\epsilon}\brb{\bs 1 (x_t \in \bar{I})} }
&=
   \babs{E^{\epsilon}\brb{\varphi_t(v_1,\dots,v_{t-1})} - E^{-\epsilon}\brb{\varphi_t(v_1,\dots,v_{t-1})} }
\\
&\le 
   \bigg\| \bigotimes_{s=1}^{t-1} \mu^\epsilon - \bigotimes_{s=1}^{t-1} \mu^{-\epsilon} \bigg\|_{\mathrm{TV}}
=
   (\star)
\end{align*}

\paragraph{Relating choice probabilities for positive and negative $\epsilon$}
By Pinsker's inequality and the fact that the Kullback-Leibler divergence is upper bounded by the $\chi^2$-divergence, it follows that
\begin{align*}
   (\star)
&\le
   \sqrt{\frac{\mathcal{D}_{\mathrm{KL}}\lrb{\bigotimes_{s=1}^{t-1} \mu^{-\epsilon},\bigotimes_{s=1}^{t-1} \mu^{\epsilon}}}{2}}
=
   \sqrt{\frac{(t-1) \cdot \mathcal{D}_{\mathrm{KL}}\lrb{ \mu^{-\epsilon}, \mu^{\epsilon}}}{2}}
\le
   \sqrt{\frac{(t-1) \cdot \mathcal{D}_{\chi^2}\lrb{ \mu^{-\epsilon}, \mu^{\epsilon}}}{2}}
\\
&=
   \sqrt{\frac{t-1}{2} \int_0^1 \labs{ \frac{f^\epsilon(x)}{f^{-\epsilon}(x)} - 1}^2 f^{-\epsilon}(x) \mathrm{d}x}
=
   (\star\star)
\end{align*}
Now, noticing that
\begin{align*}
   \int_0^1 \labs{ \frac{f^\epsilon(x)}{f^{-\epsilon}(x)} - 1}^2 f^{-\epsilon}(x) \mathrm{d}x
&=
   \int_0^{1/2} \lrb{ \frac{16 \bar{h} \epsilon}{2^{2-\lambda} + 8 \bar{h} \epsilon}}^2 f^{-\epsilon}(x)\mathrm{d}x + \int_{1-\bar{h}}^1 \lrb{\frac{2\epsilon}{\bar{\eta} - \epsilon}}^2 f^{-\epsilon}(x) \mathrm{d} x
\\
&\le
   \lrb{\int_0^{1/2} \lrb{ \frac{16 \bar{h} }{2^{2-\lambda} - 8 \bar{h} \bar{\epsilon}}}^2 f^{-\bar{\epsilon}}(x)\mathrm{d}x + \int_{1-\bar{h}}^1 \lrb{\frac{2}{\bar{\eta} - \bar{\epsilon}}}^2 f^{\bar{\epsilon}}(x) \mathrm{d} x } \cdot \epsilon^2
\\
&=
	\gamma^2 \cdot \epsilon^2
\;,
\end{align*}
it follows that
\[
   (\star\star) \le \gamma \cdot \epsilon \cdot \sqrt{\frac{t-1}{2}} \;.
\]
Summing over $t= 1,2,\dots, T$, we obtain
\[
   \labs{E^{\epsilon}\lrb{\sum_{t=1}^T\bs 1 (x_t \in \bar{I})} - E^{-\epsilon}\lrb{\sum_{t=1}^T \bs 1 (x_t \in \bar{I})} } \le \frac{\sqrt{2}}{3} \cdot \gamma \cdot \epsilon \cdot T^{3/2} = \frac{\sqrt{2}}{3} \cdot \gamma \cdot k \cdot T\;.
\]

\paragraph{Upper bound on regret for $\epsilon>0$ implies lower bound on regret for $-\epsilon$.}
Now, suppose that in the scenario determined by $P^\epsilon$ the algorithm suffer a regret $R^\epsilon_T \le M\cdot \sqrt{T}$.
Then
\[
   M\cdot \sqrt{T}
\ge
   R^\epsilon_T
\ge
   \bar{m} \cdot \epsilon \cdot \sum_{t=1}^T E^\epsilon\brb{ \bs 1 (x_t \in \bar{I}) }
=
   \bar{m} \cdot \frac{k}{\sqrt{T}} \cdot \sum_{t=1}^T E^\epsilon\brb{ \bs 1 (x_t \in \bar{I}) } \;.
\]
and rearranging
\[
   \sum_{t=1}^T E^\epsilon\brb{ \bs 1 (x_t \in \bar{I}) }
\le
   \frac{M \cdot T}{\bar{m}\cdot k}
\]
It follows that the expected number of times the algorithm plays in the (correct) region $I$ when the underlying scenario is determined by $P^{-\epsilon}$ is
\begin{align*}
   \sum_{t=1}^T E^{-\epsilon} \brb{ \bs 1 (x_t \in \bar{I}) }
&=
   \lrb{    \sum_{t=1}^T E^{-\epsilon} \brb{ \bs 1 (x_t \in \bar{I}) } -    \sum_{t=1}^T E^{\epsilon} \brb{ \bs 1 (x_t \in \bar{I}) } } +    \sum_{t=1}^T E^{\epsilon} \brb{ \bs 1 (x_t \in \bar{I}) }
\\
&\le
   \lrb{ \frac{\sqrt{2}}{3} \cdot \gamma \cdot k + \frac{M}{\bar{m} \cdot k}} \cdot T  
\end{align*}
The last inequality implies that the expected number of times that the algorithm plays in the (wrong) region $J = I^c$ when the underlying scenario is determined by $P^{-\epsilon}$ is lower bounded by
\[
   \sum_{t=1}^T E^{-\epsilon} \brb{ \bs 1 (x_t \in \bar{J}) } = \sum_{t=1}^T E^{-\epsilon} \brb{ \bs 1 (x_t \notin \bar{I}) } \ge \lrb{ 1 - \lrb{ \frac{\sqrt{2}}{3} \cdot \gamma \cdot k + \frac{M}{\bar{m} \cdot k}}  } \cdot T \;,
\] 
which implies that the regret the algorithm suffers in the scenario determined by $P^{-\epsilon}$ is lower bounded by
\begin{align*}
   R^{-\epsilon}_T
&\ge
   \bar{m} \cdot \epsilon \cdot \sum_{t=1}^T E^{-\epsilon} \lrb{ \bs 1 ( x_t \in J ) }
=
      \bar{m} \cdot \frac{k}{\sqrt{T}} \cdot \sum_{t=1}^T E^{-\epsilon} \lrb{ \bs 1 ( x_t \in J ) }
\\
&\ge
   \bar{m} \cdot \frac{k}{ \sqrt{T}} \cdot \lrb{ 1 - \lrb{ \frac{\sqrt{2}}{3} \cdot \gamma \cdot k + \frac{M}{\bar{m} \cdot k}}  } \cdot T
=
   \bar{m}\cdot k \cdot \lrb{ 1 - 2\sqrt{\frac{\sqrt{2} \cdot \gamma \cdot M}{3 \cdot \bar{m}}}} \cdot \sqrt{T} \;. 
\end{align*}
Putting everything together, any algorithm suffers at least $\min\lrb{M, \bar{m}\cdot k \cdot \lrb{1 - 2 \cdot \sqrt{\frac{\sqrt{2} \cdot \gamma \cdot M}{3 \cdot \bar{m}}}}} \cdot \sqrt{T}$ regret, in at least one scenario between the ones determined by $P^\epsilon$ and $P^{-\epsilon}$. Recalling that our choice of $M$ implies $1 - 2\sqrt{\frac{\sqrt{2} \cdot \gamma \cdot M}{3 \cdot \bar{m}}} >0$, the conclusion follows.

\subsection{Theorem \ref{theo:upper_concave} (Stochastic upper bound on regret of \dyadic)}

For the sake of simplicity, we assume that $\swexp$  admits a unique maximizer $x^\star \in [0,1]$ (the other cases can be treated similarly and, actually, they ended up having better constants in the final regret guarantees).

For each epoch $\tau = 1,2,\dots $, we refer to the three current $l$ (left), $c$ (center) and $r$ (right) points of the corresponding epoch $\tau$ using $l_\tau, c_\tau$ and $r_\tau$, respectively. For any time $t$, the epoch to which the time $t$ belongs is denoted $\tau_t$.
The length of an interval $J$ is denoted $|J|$, while the number of elements in a finite set $A$ is denoted $\#A$.

Consider a family $(v_{x,i})_{x \in [0,1], i \in \mathbb{N}}$ of random variables such that, for each $x \in [0,1]$, the sequence $(v_{x,i})_{i\in \mathbb{N}}$ is i.i.d. with the same distribution as $(v_i)_{i \in \mathbb{N}}$. With these random variables, we can define the auxiliary family $(y_{x,i})_{x \in [0,1], i \in \mathbb{N}} \coloneqq \brb{\bs 1(x \le v_{x,i})}_{x \in [0,1], i \in \mathbb{N}}$.
We assume that, whenever we select a policy $x \in [0,1]$ at time $t$, we observe $\bs 1(x \le v_{x,n_t(x)})$ (recall that $n_t(x) = \sum_{s=1}^t \bs 1 (x_s = x)$) instead of $\bs 1(x \le v_t)$.
This does not change anything in expectation, but will be useful in what follows.

The next lemma states that Algorithm \ref{alg:dyadic} maintains confidence intervals containing the differences of the welfare function (among left, center and right points) with high probability. 

\begin{lem}[Confidence intervals contain true welfare differences with high probability]
\label{lem:high_probability}
There exists a constant $\constone \in (0,20]$ such that, for every time horizon $T$ and any $\delta \in (0,1)$, if the learner runs Algorithm \ref{alg:dyadic} with confidence parameter $\delta$, then the probability of the event
\[
	\mathcal{E}
\coloneqq
	\bigcap_{t=1}^T \Brb{
	\bcb{ \swexp(c_{\tau_t}) - \swexp(l_{\tau_t}) \in J_t(l_{\tau_t},c_{\tau_t})}
	\cap
	\bcb{ \swexp(r_{\tau_t}) - \swexp(c_{\tau_t}) \in J_t(c_{\tau_t},l_{\tau_t})}
	\cap
	\bcb{ \swexp(r_{\tau_t}) - \swexp(l_{\tau_t}) \in J_t(r_{\tau_t},l_{\tau_t})}
	}
\]
is lower bounded by $1 - \constone \cdot T^2 \cdot \delta \;.$
\end{lem}
The proof of this lemma can be found in \ref{proof:high_probability}.

The following lemma establishes the rate of shrinking of the length of the confidence intervals  as the length of an epoch increases.

\begin{lem}[Confidence intervals shrink with epoch length]
\label{lem:confidence_length}
For any $\delta \in (0,1)$, if the learner runs Algorithm \ref{alg:dyadic} with confidence parameter $\delta$ then, for any time $t$,
\begin{equation}
	\max\lrb{ \babs{J_t(l_{\tau_t},c_{\tau_t})},\babs{J_t(c_{\tau_t},r_{\tau_t})}, \babs{J_t(l_{\tau_t},r_{\tau_t})} }
\le
	 \frac{\consttwo}{\sqrt{ t - t_{{\tau_t}-1} }}\;,
\end{equation}
whenever $t - t_{{\tau_t}-1} \ge  \constthree$, where $\constthree = 10$ and $\consttwo = 72 \cdot \sqrt{10} \cdot \lrb{ \sqrt{2 \log(2/\delta)} +4 }$.
\end{lem}
The proof of this lemma can be found in \ref{proof:confidence_length}.

Lemma \ref{lem:high_probability} and Lemma \ref{lem:confidence_length} allow us to prove Theorem \ref{theo:upper_concave}, which closely follows the proof given in \citep{bachoc2022regret}.

\subsubsection{Lemma \ref{lem:high_probability} (Confidence intervals contain true welfare differences with high probability)}
\label{proof:high_probability}
\begin{proof}
    For each $n\in \mathbb{N}$, let $\mathcal{D}_n \coloneqq \{ k \cdot 2^{-n}  \mid k \in \mathbb Z \}$, let $\mathcal{D}^\star_n \coloneqq \{x_{n,1},\dots, x_{n,10}\} \subset \mathcal{D}_n$ such that
    \[
        x_{n,1}
    <
        \dots
    <
        x_{n,5}
    \le
        x^\star
    \le
        x_{n,6}
    <
        \dots
    <
        x_{n,10}
    \]
    and $x_{n,j+1} - x_{n,j} \le 2^{-n}$, for all $j\in \{1,\dots,9\}$.
    Define $\mathcal{D} \coloneqq \bigcup_{n=1}^T \mathcal{D}^\star_n \cap (0,1)$.
    Consider the following events
    \small
    \begin{align*}
        \mathcal{E}'
    &\coloneqq
        \bigcap_{\substack{n,t\in \{1,\dots,T\}\\j\in \{1,\dots,10\}}}
        \lcb{ 
            \labs{ \frac{1}{t} \sum_{s=1}^t y_{x_{n,j},s} - \demexp(x_{n,j}) }
            \le
            \sqrt{\frac{1}{2t} \log\lrb{\frac{2}{\delta}}}
        }
    \\
        \mathcal{E}''
    &\coloneqq
        \bigcap_{\substack{n\in \{1,\dots,T\}\\m \in \{1,\dots,\lfloor \log_2(T) \rfloor\}\\j\in \{1,\dots,9\}}}
        \lcb{ 
            \labs{ \frac{1}{2^m} \sum_{i=1}^{2^m-1} y_{x_{n,j}+\frac{i}{2^{n+m}},1} - \frac{1}{x_{n,j+1}-x_{n,j}} \cdot \int_{x_{n,j}}^{x_{n,j+1}} \demexp(x)\mathrm{d}x }
            \le
            \sqrt{\frac{1}{2 \cdot 2^{m}} \log\lrb{\frac{2}{\delta}}} + \frac{2}{2^{m}}
        }    
    \end{align*}
    \normalsize
    and note that $\mathcal{E} \subset \mathcal{E}' \cap \mathcal{E}''$, since, in the event $\mathcal{E}' \cap \mathcal{E}''$, Algorithm \ref{alg:dyadic} will query only points in $\mathcal{D}^\star$, given that it uses only a subset of the estimates in the definition of $\mathcal{E'}$ and $\mathcal{E''}$ to build its own estimates (in particular, due to the ties breaking rules, to estimate the integral terms it will only use the \emph{first} query of the relevant dyadic points).
    Now, notice that for each $n \in \{1,\dots,n\}$, each $m \in \{1,\dots,\lfloor \log_2(T) \rfloor\}$ and each $j \in \{1,\dots,9\}$ we have
    \begin{align*}
    &
    \lcb{ 
        \labs{ \frac{1}{2^m} \sum_{i=1}^{2^m-1} y_{x_{n,j}+\frac{i}{2^{n+m}},1} - \frac{1}{x_{n,j+1}-x_{n,j}} \cdot \int_{x_{n,j}}^{x_{n,j+1}} \demexp(x)\mathrm{d}x }
            >
            \sqrt{\frac{1}{2\cdot 2^m} \log\lrb{\frac{2}{\delta}}} + \frac{2}{2^m}
        }
    \\
    &\quad\subset
        \lcb{ 
        \labs{ \frac{1}{2^m} \sum_{i=1}^{2^m-1} y_{x_{n,j}+\frac{i}{2^{n+m}},1} - \frac{1}{2^m} \sum_{i=1}^{2^m-1} \demexp\lrb{x_{n,j}+\frac{i}{2^{n+m}}} }
            >
            \sqrt{\frac{1}{2\cdot 2^m} \log\lrb{\frac{2}{\delta}}}
        }
    \\
    &\qquad\cup
        \lcb{ 
        \labs{ \frac{1}{2^m} \sum_{i=1}^{2^m-1} \demexp\lrb{x_{n,j}+\frac{i}{2^{n+m}}} - \frac{1}{x_{n,j+1}-x_{n,j}} \cdot \int_{x_{n,j}}^{x_{n,j+1}} \demexp(x)\mathrm{d}x }
            >
            \frac{2}{2^m}
        }
    \\
    &\quad=
        \lcb{ 
        \labs{ \frac{1}{2^m} \sum_{i=1}^{2^m-1} y_{x_{n,j}+\frac{i}{2^{n+m}},1} - \frac{1}{2^m} \sum_{i=1}^{2^m-1} \demexp\lrb{x_{n,j}+\frac{i}{2^{n+m}}} }
            >
            \sqrt{\frac{1}{2\cdot 2^m} \log\lrb{\frac{2}{\delta}}}
        }
    \end{align*}
    where the last equality follows from
    \begin{align*}
        &\labs{ \frac{1}{2^m} \sum_{i=1}^{2^m-1} \demexp\lrb{x_{n,j}+\frac{i}{2^{n+m}}} - \frac{1}{x_{n,j+1}-x_{n,j}} \cdot \int_{x_{n,j}}^{x_{n,j+1}} \demexp(x)\mathrm{d}x }
    \\
    \quad&\le
        \sum_{i=1}^{2^m-1} \int_{x_{n,j}+\frac{i-1}{2^{n+m}}}^{x_{n,j}+\frac{i}{2^{n+m}}} \lrb{\demexp(x)- \demexp\lrb{x_{n,j}+\frac{i}{2^{n+m}}}} \mathrm{d}x
    +
    \frac{1}{2^{m}}
    \\
    \quad&\le
        \sum_{i=1}^{2^m-1} \int_{x_{n,j}+\frac{i-1}{2^{n+m}}}^{x_{n,j}+\frac{i}{2^{n+m}}} \lrb{\demexp\lrb{x_{n,j}+\frac{i-1}{2^{n+m}}}- \demexp\lrb{x_{n,j}+\frac{i}{2^{n+m}}}} \mathrm{d}x
    +
    \frac{1}{2^{m}}
    \\
    \quad&\le
        \frac{1}{2^{m}} \cdot \lrb{\demexp\lrb{x_{n,j}}- \demexp\lrb{x_{n,j+1}} } + \frac{1}{2^{m}}
    \le
        \frac{2}{2^{m}}
    \end{align*} 
    By De Morgan's laws, a union bound and Hoeffding's inequality, we have 
    $ P (\mathcal{E}^c) \le  P \brb{(\mathcal{E}')^c}+ P \brb{(\mathcal{E}'')^c} \le 20 \cdot T^2\cdot \delta$.
    \end{proof}

\subsubsection{Lemma \ref{lem:confidence_length} (Confidence intervals shrink with epoch length)}
\label{proof:confidence_length}

We break the proof of Lemma \ref{lem:confidence_length} in several steps. Let $d_1,d_2,d_3,d_4,d_5>0$ be constants. For each $k \in \{1,2,3\}$, define
\[
	f_k : \{0,1,2,\dots\}\to [0,+\infty], \qquad n\mapsto \frac{d_k}{\sqrt{n}}
\]
and for each $k \in \{4,5\}$ define
\[
	f_k : \{0,1,2,\dots\}\to [0,+\infty], \qquad n\mapsto \frac{d_4}{\sqrt{2^{\lfloor \log_2(n+1) \rfloor } -1 }} + \frac{d_5}{2^{\lfloor \log_2(n+1) \rfloor}}\;,
\]
with the usual convention that $a/0 = +\infty$, for any $a> 0$.
Suppose that $m_1(0), m_2(0), m_3(0)$, $m_4(0), m_5(0) \in \{0,1,2,\dots\}$ and consider Algorithm \ref{alg:shrinking}.
\begin{algorithm}[t]
    \caption{Index selection}
    \label{alg:shrinking}
    \begin{algorithmic}[1]
        \FOR{$s=1,2,\dots$}
			\STATE Let $k_s = \min\lrb{\operatorname{argmax}_{k \in [5]} f_k \brb{m_k (s-1)}}$
			\STATE $m_{k_s} (s) = m_{k_s} (s-1) + 1$
			\FOR{$i \in [5] \backslash \{k_s\}$}
				\STATE $m_{i}(s) = m_{i}(s-1)$
			\ENDFOR 
        \ENDFOR 
    \end{algorithmic}
\end{algorithm}

The following lemma holds.

\begin{lem}
\label{lem:at_least_one}
	Consider Algorithm \ref{alg:shrinking} and the notation defined therein. For each $s \in \mathbb{N}$ there exists an index $i \in [5]$ for which $m_i(s) \ge \lceil s/5 \rceil$.
\end{lem}
\begin{proof}
	Let $s \in \mathbb{N}$ and suppose by contradiction that for each $k \in [5]$ it holds that $m_k(s) < s/5$. Then
\[
	s \le \sum_{k=1}^5 m_k(s) \le 5 \cdot \max_{k \in[5]} m_k(s) < 5 \cdot \frac{s}{5} = s\;, 
\]
which is a contradiction. It follows that there exists $k \in [5]$ for which $m_k(s) \ge s/5$, which also implies $m_k(s) \ge \lceil s/5 \rceil$. Given that $s$ was arbitrarily chosen, the conclusion follows.
\end{proof}

Notice that, for each $n \in \{0,1,2,\dots\}$, we have
\[
	\frac{d_4}{\sqrt{n}} \le \frac{d_4}{\sqrt{2^{\lfloor \log_2(n+1) \rfloor } -1 }} \le \frac{2 d_4}{\sqrt{n}}
\]
and
\[
	0\le \frac{d_5}{\sqrt{2^{\lfloor \log_2(n+1) \rfloor } }} \le \frac{2 d_5}{n} \;,
\]
which implies that, for each $k \in [5]$ and each $n \in \{0,1,2,\dots\}$
\[
	\frac{d_k}{\sqrt{n}} \le f_k(n) \le \frac{D_k}{\sqrt{n}}
\]
where $D_1=d_1, D_2=d_2, D_3=d_3, D_4 = D_5 = 2(d_4+d_5)$.

The following lemma holds.

\begin{lem}
\label{lem:comparison}
Consider Algorithm \ref{alg:shrinking} and the notation defined therein.
For any $i,j \in [5]$ and any $s \in\mathbb{N}$ it holds
\[
	m_i(s) \ge \lrb{\frac{d_i}{D_j}}^2 (m_j(s)-1) \;.
\]
\end{lem}
\begin{proof}
Let $i,j \in [5]$. Suppose by contradiction that the conclusion does not hold. Then there exists a smallest $s \in \{0,1,2,\dots\}$ for which
\[
	m_i(s) < \lrb{\frac{d_i}{D_j}}^2 (m_j(s)-1) \;,
\]
which we call $s_0$. Notice that $s_0 \neq 0$. Then, the fact that
\[
	m_i(s_0-1) \ge \lrb{\frac{d_i}{D_j}}^2 (m_j(s_0-1)-1) \;,
\]
implies that at time $s_0$ the algorithm selected $k_{s_0} = j$, which in turn implies that $m_i(s_0-1) = m_i(s_0)$ and $m_j(s_0 - 1) = m_j(s_0)-1$.
It follows that
\[
	\lrb{\frac{d_i}{D_j}}^2 m_j(s_0-1)
=
	\lrb{\frac{d_i}{D_j}}^2 \brb{m_j(s_0)-1}
>
	m_i(s_0)
=
	m_i(s_0-1)\;,
\]
Rearranging, we get
\[
	m_j(s_0-1) > \lrb{\frac{D_j}{d_i}}^2 m_i(s_0-1)\;.
\]
from which it follows that
\[
	f_j\brb{m_j(s_0-1)}
\le
	\frac{D_j}{\sqrt{m_j(s_0-1)}}
<
	\frac{d_i}{\sqrt{m_i(s_0-1)}}
\le
	f_i\brb{m_i(t_0-1)}\;.
\]
This last inequality implies that at time $s_0$ the algorithm should have chosen the index $i$ and not the index $j$, which is a contradiction.
\end{proof}

Combining the last two lemmas we can prove the following result.

\begin{lem}
\label{lem:index_selection_upper_bound}
Consider Algorithm \ref{alg:shrinking} and the notation defined therein. Then, for any $s \ge 5$ it holds that
\[
	\max_{k \in [5]} f_k\brb{m_k(s)} \le \frac{D}{\sqrt{s-5}}
\]
where $D = \sqrt{5} \cdot \brb{\max_{j \in[5]} D_j} \cdot \brb{\max_{k \in [5]} \frac{D_k}{d_k}}$.
\end{lem}
\begin{proof}
Let $s \ge 5$. Pick $j \in [5]$ such that $m_j(s) \ge \lceil s/5 \rceil$ (which does exist by Lemma \ref{lem:at_least_one}). Then, by Lemma \ref{lem:comparison}
\begin{align*}
	\max_{k \in [5]} f_k\brb{m_k(s)}
&\le
	\max_{k \in [5]} \frac{D_k}{\sqrt{m_k(s)}}
\le
	\max_{k \in [5]} \frac{D_k}{\sqrt{\lrb{\frac{d_k}{D_j}}^2\brb{m_j(s)-1}}}
\\
&=
	D_j \cdot \max_{k \in [5]} \lrb{\frac{D_k}{d_k}} \frac{1}{\sqrt{m_j(s)-1}}
\le
	D_j \cdot \max_{k \in [5]} \lrb{\frac{D_k}{d_k}} \frac{1}{\sqrt{\lceil s/5\rceil-1}}
\le
	\frac{D}{\sqrt{s-5}} \;. \qedhere
\end{align*}
\end{proof}

We are now ready for the proof of Lemma \ref{lem:confidence_length}.

\begin{proof}[Proof of Lemma \ref{lem:confidence_length}]
It is enough to notice that Algorithm \ref{alg:dyadic} with confidence parameter $\delta \in (0,1)$ relies, inside each epoch,  on the same routine given by Algorithm \ref{alg:shrinking} with $d_1 = l \cdot \sqrt{\frac{\log(2/\delta)}{2}}, d_2 = c \cdot \sqrt{\frac{\log(2/\delta)}{2}}, d_3 = r \cdot \sqrt{\frac{\log(2/\delta)}{2}}, d_4 = \lambda \cdot (c-l) \cdot \sqrt{\frac{\log(2/\delta)}{2}}, d_5 = 2 \cdot \lambda \cdot (c-l)$, with the convention that $l$ correspond to $1$, $c$ corresponds to $2$, $r$ corresponds to $3$, $(l,c)$ corresponds to $4$ and $(c,r)$ corresponds to $5$, the correspondence between times is given by $s = t-t_{\tau_{t}-1}$, and, for each $s \in \{0,1,2,\dots\}$, $m_1(s) = n_{s+t_{\tau_{t}-1}}(l)$, $m_2(s) = n_{s+t_{\tau_{t}-1}}(c)$, $m_3(s) = n_{s+t_{\tau_{t}-1}}(r)$, $m_4(s) = \sum_{i \le s+t_{\tau_{t}-1}} \bs 1 \brb{x_i \in (l,c)}$, $m_5(s) = \sum_{i \le s+t_{\tau_{t}-1}} \bs 1 \brb{x_i \in (c,r)}$. With these conventions, in Lemma \ref{lem:index_selection_upper_bound} we have that $D \le 9 \cdot \sqrt{5} \cdot \lrb{ \sqrt{2 \log(2/\delta)} + 4}$ and, for example (the other cases can be proved analogously)
\begin{align*}
	\babs{J_t(l_{\tau_t},r_{\tau_t})} 
&\le
	2 \cdot \lrb{ \Gamma_{t}(r)+\Gamma_{t}(l)+\Gamma_{t}(l,c)+\Gamma_{t}(c,r)}
\le
	2 \cdot 4 \cdot \max_{k \in [5]} f_k\brb{m_k(s)}
\\
&\le
	8 \cdot \frac{D}{\sqrt{t-t_{\tau_{t}-1}-5}}
\le
	\frac{\consttwo}{\sqrt{2}} \cdot \frac{1}{\sqrt{t-t_{\tau_{t}-1}-5}}
\le
	\frac{\consttwo}{\sqrt{t-t_{\tau_{t}-1}}}\;
\end{align*}
where in the last inequality we used the fact that $t-t_{\tau_{t}-1} \ge 10$.
\end{proof}

\subsubsection{Theorem \ref{theo:upper_concave} (Completing the proof)}

\begin{proof}[Proof of Theorem \ref{theo:upper_concave}]
    Define $\tau_T$ as the last epoch, $t_0 = 0$ and (if not already defined) $t_{\tau_{T}} = T$ .
    
    Due to Lemma \ref{lem:high_probability}, we may (and do!) assume that for each $t \in \{1,\dots,T\}$ it holds
    \[
        \brb{ \swexp(c_{\tau_t}) - \swexp(l_{\tau_t}) \in J_t(l_{\tau_t},c_{\tau_t})}
        \land
        \brb{ \swexp(r_{\tau_t}) - \swexp(c_{\tau_t}) \in J_t(c_{\tau_t},l_{\tau_t})}
        \land
        \brb{ \swexp(r_{\tau_t}) - \swexp(l_{\tau_t}) \in J_t(r_{\tau_t},l_{\tau_t})}\;.
    \]
    This is because, given our choice $\delta = \frac{1}{T^{5/2}}$, assuming these conditions costs us in the expected regret a further additive term which is no greater than $T \cdot \constone \cdot T^2 \cdot \delta = \constone \cdot \sqrt{T}$.
    
    Under these assumptions, notice that for each $\tau \in [\tau_T]$ we have that $x^\star \in I_\tau$. In fact, if the confidence intervals are guaranteed to contain the corresponding differences in the expected welfare, every time Algorithm \ref{alg:dyadic} shrinks the active interval is because all the discarded points are guaranteed to be suboptimal.
    
    For each epoch $\tau \in \{1,\dots,\tau_T\}$, define
    \[
        B_\tau \coloneqq (t_\tau-1) - t_{\tau-1}\;.
    \]
    Now, for each epoch $\tau \in \{1,\dots,\tau_T\}$ if $B_\tau \ge \constthree$, then
    \[
        \max_{x \in [ l_\tau, r_\tau ]} \brb{ \swexp(x^\star) - \swexp(x) } \le 2 \cdot \consttwo \cdot \sqrt{\frac{1}{ B_\tau }} \;.
    \]
    In fact, assume that $x^\star > r_\tau$ (the other cases have similar proofs). Then, leveraging concavity, and recalling that $\inf \brb{J_{t_{\tau}-1}(l_\tau,r_\tau)} < 0$ and that $x^\star \in I_\tau$ (which implies $\frac{x^\star-l_\tau}{r_\tau-l_\tau}\le 2$), we have
    \begin{align*}
        \max_{x \in [l_\tau, r_\tau ]} \brb{ \swexp(x^\star) - \swexp(x) }
    &=
        \swexp(x^\star) - \swexp(l_\tau)
    =
        \frac{\swexp(x^\star) - \swexp(r_\tau)}{x^\star - r_\tau} (x^\star - r_\tau) + \swexp(r_\tau) - \swexp(l_\tau)
    \\
    &\le
        \frac{\swexp(r_\tau) - \swexp(l_\tau)}{r_\tau - l_\tau} (x^\star - r_\tau) + \swexp(r_\tau) - \swexp(l_\tau)
    =
        \frac{x^\star-l_\tau}{r_\tau-l_\tau}\cdot \brb{\swexp(r_\tau) - \swexp(l_\tau)} 
    \\
    &\le
        2 \cdot \brb{\swexp(r_\tau) - \swexp(l_\tau)}
    \le
        2 \cdot \sup(J_{t_{\tau}-1}(l_\tau,r_\tau))
    \le
        2 \cdot \labs{J_{t_{\tau}-1}(l_\tau,r_\tau)}
    \\
    &\le
        2 \cdot \consttwo \cdot \sqrt{\frac{ 1}{B_\tau }} \;,
    \end{align*}
    where the final inequality follows by Lemma \ref{lem:confidence_length}.
    
    Let $\tau^\star$ be the first epoch from which it holds $x^\star \in [l_\tau, r_\tau]$.
    If $\tau^\star \ge 2$, then for each $\tau \in \{2,\dots,\tau^\star-1\}$ it holds that
    \[
       \max_{x \in [l_\tau, r_\tau] } \brb{\swexp(x^\star)-\swexp(x)} \le \frac{3}{4} \cdot \max_{x \in [l_{\tau-1}, r_{\tau-1}] } \brb{\swexp(x^\star)-\swexp(x)}\;.
    \]
    In fact, either for all $\tau \in \{1,\dots, \tau^\star - 1\}$ it holds that $r_\tau < x^\star$, or for all $\tau \in \{1,\dots, \tau^\star - 1\}$ it holds that $l_\tau > x^\star$. In the first case, for all $\tau \in \{1,\dots, \tau^\star - 1\}$, leveraging concavity and recalling that $x^\star \in I_\tau$ (which implies $\frac{x^\star-l_{\tau}}{x^\star-l_{\tau-1}}\le \frac{3}{4}$), we have
    
    \begin{align*}
       \max_{x \in [l_\tau, r_\tau] } \brb{\swexp(x^\star)-\swexp(x)}
    &=
        \swexp(x^\star)-\swexp(l_\tau)
    =
        \frac{\swexp(x^\star)-\swexp(l_\tau)}{x^\star-l_\tau} \cdot(x^\star-l_\tau)
    \\
    &\le
        \frac{\swexp(x^\star)-\swexp(l_{\tau-1})}{x^\star-l_{\tau-1}} \cdot(x^\star-l_{\tau})
    \\
    \le
           \frac{3}{4}\cdot\brb{\swexp(x^\star)-\swexp(l_{\tau-1})}
    \\
    &
    =
        \frac{3}{4} \cdot \max_{x \in [l_{\tau-1}, r_{\tau-1}] } \brb{\swexp(x^\star)-\swexp(x)} \;,
    \end{align*}
    while the second case can be deduced analogously.
    
    For each $m \in \mathbb{N}$, let $A_m \coloneqq \bcb{ x\in (0,1) : \exists k \in \{1,\dots,2^m-1\}, x = k/2^m }$ be the dyadic mesh in $(0,1)$ of index $m$.
    For any epoch $\tau \in \mathbb{N}$, let $m_\tau \coloneqq - \log_2 ( c_\tau - l_\tau )$ be the index of the dyadic mesh in $(0,1)$ at epoch $\tau$ of Algorithm \ref{alg:dyadic} (note that $m_\tau \ge 2$ for all $\tau \in \mathbb{N}$ because Algorithm \ref{alg:dyadic} begins with a step-size of $1/4$).
    
    Let $m^\star \coloneqq \min \bcb{m \in \mathbb{N} : \#\brb{ A_m \cap (0,x^\star] } \ge 4 \text{ and } \#\brb{ A_m \cap [x^\star, 1) } \ge 4 }$ be the smallest index of the dyadic mesh in $(0,1)$ such that there are at least 4 points of the dyadic mesh in $(0,1)$ to the right and to the left of $x^\star$.
    For each $m \ge m^\star$ let $x_1^m < x_2^m < x_3^m < x_4^m \le x^\star$ be the four points of $A_m \cap (0,x^\star]$ closest to $x^\star$ and $x^\star \le x_5^m < x_6^m < x_7^m < x_8^m$ be the four points of $A_m \cap [x^\star, 1)$ closest to $x^\star$.
    Observe that, for all epochs $\tau \ge \tau^\star + 3$, Algorithm \ref{alg:dyadic} selects policies only in the closed interval $[x_1^{m_\tau}, x_8^{m_\tau}]$.
    Observe further that, for each $m \ge m^\star + 1$, it holds
    \[
        \max_{x \in [x_1^m,x_8^m]} \brb{\swexp(x^\star)-\swexp(x)} \le \frac{4}{7} \cdot \max_{x \in [x_1^{m-1},x_8^{m-1}]} \brb{\swexp(x^\star)-\swexp(x)}\;.
    \]
    In fact, either $\max_{x \in [x_1^m,x_8^m]} \brb{\swexp(x^\star)-\swexp(x)} = \swexp(x^\star)-\swexp(x_1^m) $ or $\max_{x \in [x_1^m,x_8^m]} \brb{\swexp(x^\star)-\swexp(x)} = \swexp(x^\star)-\swexp(x_8^m) $. In the first case, leveraging concavity and observing that $\frac{x^\star-x_1^{m}}{x^\star-x_1^{m-1}}\le \frac{4}{7}$, we have
    \begin{align*}
        \max_{x \in [x_1^m,x_8^m]} \brb{\swexp(x^\star)-\swexp(x)}
    &=
        \swexp(x^\star)-\swexp(x_1^m)
    =
        \frac{\swexp(x^\star)-\swexp(x_1^m)}{x^\star-x_1^m}\cdot (x^\star-x_1^m)
    \\
    &\le
        \frac{\swexp(x^\star)-\swexp(x_1^{m-1})}{x^\star-x_1^{m-1}}\cdot (x^\star-x_1^m)
    \le
        \frac{4}{7} \cdot \brb{\swexp(x^\star)-\swexp(x_1^{m-1})}
    \\
    &\le
        \frac{4}{7} \cdot \max_{x \in [x_1^{m-1},x_8^{m-1}]} \brb{\swexp(x^\star)-\swexp(x)}\;.
    \end{align*}
    The second case can be worked out similarly.
    
    Define $\ts \coloneqq \bfl{ 4 + 2 \log_{\nicefrac43}(\sqrt{T}) }$ so that
    \[
        \lrb{\frac34}^{ \lfl{ \frac{\ts -1}{2} } }
    =
        \lrb{\frac34}^{ \lfl{ \frac{\bfl{ 4 + 2 \log_{\nicefrac43}(\sqrt{T}) } -1}{2} } }
    \le
        \lrb{\frac34} ^ { \log_{\nicefrac43}(\sqrt{T}) }
    =
        \frac{1}{\sqrt{T}} \;.
    \]
    Assume that $\ts < \tau^\star$ and $\tau^\star + 2 + \ts < \tau_T$ (the other cases can be treated analogously, omitting terms which are not there anymore). 
    Then, the expected regret can be decomposed as follows:
    \begin{multline*}
        \sum_{t=1}^T \brb{\swexp( x^\star ) - \swexp(x_t)}
    =
        \sum_{\tau=1}^{\ts} \sum_{t = t_{\tau-1} +1}^{t_\tau} \brb{\swexp( x^\star ) - \swexp(x_t)}
        +
        \sum_{\tau = \ts+1}^{\tau^\star -1} \sum_{t = t_{\tau-1} +1}^{t_\tau} \brb{\swexp( x^\star ) - \swexp(x_t)}
    \\
        +
        \sum_{\tau = \tau^\star}^{\tau^\star+2} \sum_{t = t_{\tau-1} +1}^{t_\tau} \brb{\swexp( x^\star ) - \swexp(x_t)}
        +
        \sum_{\tau = \tau^\star+3}^{\tau^\star+2+\ts} \sum_{t = t_{\tau-1} +1}^{t_\tau} \brb{\swexp( x^\star ) - \swexp(x_t)}
        +
        \sum_{\tau = \tau^\star+3+\ts}^{\tau_T} \sum_{t = t_{\tau-1} +1}^{t_\tau} \brb{\swexp( x^\star ) - \swexp(x_t)}.
    \end{multline*}
    We analyze these five terms individually.
    
    For the first one, we further split the sum into two terms, depending on whether or not $B_\tau \coloneqq t_\tau - 1  - t_{\tau-1} \ge \constthree$. 
    Recalling that for each $\tau \in \{1,\dots,\tau_T\}$ and for each $t \in \{t_{\tau-1}+1,\dots,t_{\tau}\}$ Algorithm \ref{alg:dyadic} selects the policy $x_t$ in the closed interval $[l_\tau,r_\tau]$, we have that
    \begin{align*}
        \sum_{\substack{\tau=1\\B_\tau \ge \constthree}}^{\ts} \sum_{t = t_{\tau-1} +1}^{t_\tau} \brb{\swexp( x^\star ) - \swexp(x_t)}
    &
    \le
        \sum_{\substack{\tau=1\\B_\tau \ge \constthree}}^{\ts} (B_\tau+1) \cdot \max_{x \in [l_\tau, r_\tau ]} \brb{ \swexp(x^\star) - \swexp(x) }
    \\
    &\le
        \sum_{\substack{\tau=1\\B_\tau \ge \constthree}}^{\ts} (B_\tau+1) \cdot 2 \cdot \consttwo \cdot \sqrt{\frac{ \log(2/\delta)}{B_\tau }}
    \\
    &
    \le
        4 \cdot \consttwo  \cdot \sum_{\substack{\tau=1\\B_\tau \ge \constthree}}^{\ts}  \sqrt{B_{\tau}}
    \le
        4 \cdot \consttwo \cdot \ts \cdot \sqrt{T}.
    \end{align*}
    On the other hand, we also have that
    \[
        \sum_{\substack{\tau=1\\B_\tau \le (\constthree-1)}}^{\ts} \sum_{t = t_{\tau-1} +1}^{t_\tau} \brb{\swexp( x^\star ) - \swexp(x_t)}
    \le
        (\constthree-1) \sum_{\tau=0}^{\infty} \brb{ \nicefrac34 }^\tau
    =
        4 \cdot (\constthree-1).
    \]
    Thus, the first term is upper bounded by $ 4 \cdot \consttwo \cdot \ts \cdot \sqrt{T} + 4 \cdot (\constthree-1)$. 
    
    For the second term, leveraging the definition of $\ts$, we obtain
    \begin{align*}
        \sum_{\tau = \ts+1}^{\tau^\star -1} \sum_{t = t_{\tau-1} +1}^{t_\tau} \brb{\swexp( x^\star ) - \swexp(x_t)}
    &
    \le
        \sum_{\tau = \ts+1}^{\tau^\star -1} \sum_{t = t_{\tau-1} +1}^{t_\tau} \brb{ \nicefrac34 }^{\tau-1}
    \le
        \brb{ \nicefrac34 }^{\ts-1} \cdot \sum_{\tau = \ts+1}^{\tau^\star -1} \sum_{t = t_{\tau-1} +1}^{t_\tau} 1
    \\
    &\le
        \brb{ \nicefrac34 }^{\bfl{\frac{\ts-1}{2}}} \cdot \sum_{\tau = \ts+1}^{\tau^\star -1} \sum_{t = t_{\tau-1} +1}^{t_\tau} 1
    \le
        \sqrt{T}.
    \end{align*}
    For the third term, we further split the sum into two terms, depending on whether or not $B_\tau \ge \constthree$.
    Proceeding exactly as for the first term, we obtain
    \[
        \sum_{\tau = \tau^\star}^{\tau^\star+2} \sum_{t = t_{\tau-1} +1}^{t_\tau} \brb{\swexp( x^\star ) - \swexp(x_t)}
    \le
        3 \cdot 4 \cdot \consttwo \cdot \sqrt{T} + 3 \cdot (\constthree-1).
    \]
    For the fourth term, we split again the sum into two terms, depending on whether or not $B_\tau \ge \constthree$.
    If $B_\tau \ge \constthree$, proceeding exactly as for the corresponding part of the first term, we obtain
    \[
        \sum_{\substack{\tau=\tau^\star+3\\B_\tau \ge \constthree}}^{\tau^\star+2+\ts} \sum_{t = t_{\tau-1} +1}^{t_\tau} \brb{\swexp( x^\star ) - \swexp(x_t)}
    \le
        4 \cdot \consttwo \cdot \ts \cdot \sqrt{T}.
    \]
    Instead, if $B_\tau \le (\constthree-1)$, we get
    \begin{align*}
        \sum_{\substack{\tau=\tau^\star+3\\B_\tau \le (\constthree-1)}}^{\tau^\star+2+\ts} \sum_{t = t_{\tau-1} +1}^{t_\tau} \brb{\swexp( x^\star ) - \swexp(x_t)}
    &\le
        (\constthree-1) \cdot \sum_{\substack{\tau=\tau^\star+3\\B_\tau \le (\constthree-1)}}^{\tau^\star+2+\ts} \max_{x\in [l_\tau, r_\tau ]} \brb{\swexp( x^\star ) - \swexp(x)}
    \\
    &\le
        (\constthree-1) \cdot \sum_{\substack{\tau=\tau^\star+3\\B_\tau \le (\constthree-1)}}^{\tau^\star+2+\ts} \max_{x \in \lsb{ x^{m_\tau}_1, x^{m_\tau}_8 }} \brb{\swexp( x^\star ) - \swexp(x)}
    \\
    &\le
        2 \cdot (\constthree-1) \cdot \sum_{\tau= 0}^{\infty} \brb{\nicefrac47}^\tau
    \\
    &\le
        \frac{14}{3} \cdot (\constthree-1) \;.
    \end{align*}
    For the last term, we have
    \begin{align*}
        \sum_{\tau=\tau^\star+3+\ts}^{\tau_T} \sum_{t = t_{\tau-1} +1}^{t_\tau} \brb{\swexp( x^\star ) - \swexp(x_t)}
    &
    \le
        \sum_{\tau=\tau^\star+3+\ts}^{\tau_T} \sum_{t = t_{\tau-1} +1}^{t_\tau} \max_{x \in \lsb{ x^{m_\tau}_1, x^{m_\tau}_8 }} \brb{\swexp( x^\star ) - \swexp(x)}
    \\
    &
    \le
        \sum_{\tau=\tau^\star+3+\ts}^{\tau_T} \sum_{t = t_{\tau-1} +1}^{t_\tau} \brb{ \nicefrac47 }^{\lfl{ \frac{\tau-(\tau^\star+3) -1}{2} }}
    \\
    &\le
        \brb{ \nicefrac34 }^{\lfl{ \frac{\ts-1}{2} }} \sum_{\tau=\tau^\star+3+\ts}^{\tau_T} \sum_{t = t_{\tau-1} +1}^{t_\tau} 1
    \le
        \sqrt{T}\;.
    \end{align*}
    Putting everything together, and recalling the definition of $\ts$, the conclusion follows.
    \end{proof}

    \subsection{Theorem \ref{theo:upper_exp3_income} (Upper bound on regret of Tempered Exp3 for  Optimal Income Taxation)}

We prove this result by \textbf{reduction to our baseline model}, as analyzed in Section \ref{sec:regret_bounds}.
Assume that $\mathcal W = \{w^1,\dots,w^H\}$ with $0=w^1<w^2<\dots<w^H\le 1$. For each tax bracket $[\wage^h,\wage^{h+1})$, \textit{Tempered Exp3 for  Optimal Income Taxation} essentially reduces to a separate instance of \textit{\algexp}.
Denote 
\bals
    \sw_i^h(\mathbf{\pol}(\cdot)) &= \sw_i(\mathbf{\pol}(\cdot))  \cdot \bs 1(\left\lfloor \wage_i\right\rfloor = \wage^h),&
    \swcum^h_i (\mathbf{\pol}(\cdot)) &= \sum_{j\leq i} \sw_j^h(\mathbf{\pol}(\cdot)),\\
    \swcum^h_i &= \sum_{j\leq i} \sw_j^h(\mathbf{\pol}_j(\cdot)),&
    T^h &= \sum_{i\leq T} \bs 1(\left\lfloor \wage_i\right\rfloor = \wage^h),\\
    \mathcal R^h_T &= \sup_{\mathbf{\pol}( \cdot ) \in \mathcal{X_W}} E\left [ \swcum^h_T(\mathbf{\pol}( \cdot ))- \swcum^h_T \Big| \{\wtp_i\}_{i=1}^T, \{\wage_i\}_{i=1}^T \right ].
\eals
It is immediate that
$$
\mathcal R_T = \sum_h \mathcal R^h_T \text{, and } T= \sum_h T^h.
$$

Assume for a moment that the \textbf{upper bound} on regret of Theorem \ref{theo:upper_exp3} (with $\lambda$ replaced by $1$) \textbf{applies to each instance} (tax bracket) $h$, separately.
That is, assume that 
$$\mathcal R^h_T \leq \left ( \gamma   + \eta  \cdot (e-2) 
\tfrac{K+1}{K} \cdot \left (\tfrac{2K+1}{6} + \tfrac{1}{\gamma}  \right )
+ \tfrac{1}{K} \right ) \cdot \mathbf{T^h}
+ \tfrac{\log(K+1)}{\eta}.$$
Then it follows that
$$\mathcal R_T \leq \left ( \gamma   + \eta  \cdot (e-2) 
\tfrac{K+1}{K} \cdot \left (\tfrac{2K+1}{6} + \tfrac{1}{\gamma}  \right )
+ \tfrac{1}{K} \right ) \cdot \mathbf{T}
+ \tfrac{\mathbf{H} \cdot \log(K+1)}{\eta},$$
and the claims of Theorem \ref{theo:upper_exp3_income} are immediate.\\

It remains to show that indeed the upper bound on regret of Theorem \ref{theo:upper_exp3} applies to each instance (tax bracket) $h$.
For any given pair of sequences $\{\wtp_i\}_{i=1}^T, \{\wage_i\}_{i=1}^T $, consider the subsequence of observations $i$ for which $\left\lfloor \wage_i\right\rfloor = \wage^h$.
Along this subsequence, the policy choice reduces to the choice of a tax rate $\pol_i = \mathbf{\pol}_i(\wage^h) \in \mathcal X$, and the algorithm \textit{Tempered Exp3 for  Optimal Income Taxation} reduces to an instance of the algorithm \textit{\algexp}, with the following \textbf{modifications}:
\begin{enumerate}
    \item Estimated demand $\widehat \dem_i(\pol, \wage^h)$ is multiplied by an additional factor $w_i \in [0,1]$.
    \item Estimated social welfare $\widehat \swcum_{i+1}(\pol, \wage^h)$ is updated with a term for private welfare that includes a time-varying welfare weight $\weight(\wage_i) \leq 1$, rather than a fixed weight $\lambda$.
\end{enumerate}
We need to verify that, with these modifications, the following key claims in the proof of Theorem \ref{theo:upper_exp3} continue to hold:
\begin{enumerate}
    \item Unbiasedness: $\widehat \swcum_{i}(\pol, \wage^h)$ is an unbiased estimator of $\tilde  \swcum_{i}(\pol, \wage^h)$, for a suitably discretized version of cumulative social welfare. 
    (Step 2 of the original proof.)
    In the present setting, discretization requires substituting $\tilde \wtp_i$ for $\wtp_i$, where 
    $\tilde \wtp_i = \min \{\pol \in \mathcal X:\; \wage_i(1-\pol) \geq \wtp_i\}$.
    
    \item Bounded support: $\widehat \sw_{i}(\pol, \wage^h)<\tfrac{K+1}{\gamma}$, where 
    $$\widehat \sw_{i}(\pol, \wage^h) = \pol \cdot  \widehat \dem_i(\pol, \wage^h)
    + \frac{\weight(\wage_i)}{K}  \cdot \sum_{\pol' \in \mathcal X, \pol'>\pol } \widehat \dem_{i}(\pol', \wage^h).$$ 
    (Step 4 of the original proof.)
    
    \item Bounded second moment of $\widehat \sw_{i}(\pol, \wage^h)$:
    \bals
    E_i\left [ \widehat \sw_{i}(\pol, \wage^h) ^2 \right ] 
    &\leq  \frac{\pol^2 }{p_i(x|w^h)} 
  +\left ( \frac{1}{K} \right )^2 \cdot \sum_{\pol' \in \mathcal X, \pol'>\pol } \frac{1}{p_i(x'|w^h)},   
  \eals
  (Step 6 of the original proof.)
\end{enumerate}
Unbiasedness follows as before. 
 To show bounded support, as well as the bound on the second moment, note that we can rewrite
 $$\widehat \sw_{i}(\pol, \wage^h) = 
 \left(\pol  \cdot \bs 1(\pol_i=\pol) + \frac{\weight(\wage_i)}{K} \cdot  \bs 1(\pol_i>\pol)\right) \cdot \frac{ \out_i \cdot \wage_i}{p_{i}(\pol_i | \wage^h)}.$$ 
Recall that $\pol, \weight(\wage_i)$, $\wage_i$, and $\out_i$ are all bounded above by $1$, and that $p_i(x|w^h) \geq \frac{\gamma}{K+1}$.
Bounded support and the bound on the second moment follow.
 The remaining steps of the proof are as before.

\end{document}